\documentclass[onecolumn]{pasj00}
\SetRunningHead{Nuclear starbursts over a wide luminosity range of Seyfert galaxies}{Oi, Imanishi \& Imase}

\usepackage{times}
\usepackage{multicol}

\begin{document}

\title{Comparison of AGN and Nuclear Starburst Activity in Seyfert 1 and 2 Galaxies over a Wide Luminosity Range Based on Near-infrared 2--4 $\mu$m Spectroscopy}
\author{Nagisa \textsc{Oi}\altaffilmark{1,2,3}}
\email{nagisa.oi@nao.ac.jp}
\author{Masatoshi \textsc{Imanishi}\altaffilmark{2,1,3} \& Keisuke \textsc{Imase}\altaffilmark{1,2}}
\altaffiltext{1}{Department of Astronomical Science,
 The Graduate University for Advanced Studies (Sokendai),
 Mitaka, Tokyo 181-8588, Japan}
\altaffiltext{2}{National Astronomical Observatory of Japan,
 Mitaka, Tokyo 181-8588, Japan}
\altaffiltext{3}{Present Address; Subaru Telescope, 650, North A'ohoku Place, Hilo, HI 96720, USA}


\maketitle

\begin{abstract}
We present near-infrared K- (1.9--2.5$\mu$m) and L- (2.8--4.2$\mu$m) band
 spectroscopy of 22 Seyfert nuclei.
We use two methods to investigate the presence of nuclear starbursts:
 (1) the Polycyclic Aromatic Hydrocarbon (PAH) emission feature at
 $\lambda_{rest}$ = 3.3 $\mu$m in the rest frame of L-band spectrum (a
 starburst indicator) and (2) the CO absorption feature at $\lambda_{rest}$ =
 2.3--2.4 $\mu$m in the rest frame of the K-band spectrum, originating in the
 CO molecule.
We clearly detected the 3.3 $\mu$m PAH emission features in five objects and
 the CO absorption features in 17 objects.
Seyfert 2 galaxies tend to show bluer $K-L$ colors compared with Seyfert 1
 galaxies.
We interpret the discrepancy as resulting from relative strength of stellar
 emission because AGN emission is affected by dust extinction.
The 3.3 $\mu$m PAH emission luminosity ($L_{3.3PAH}$) distributions for the
 Seyfert 1s and Seyfert 2s are very similar when normalized to the AGN power.
Star-formation rates estimated from $L_{3.3PAH}$ could be large enough to
 inflate the dusty torus by supernova explosion.
We find that $L_{3.3PAH}$ positively correlates with N-band luminosity with
 small aperture over a wide AGN luminosity range, and is independent of
 physical area we probed.
The results suggest that nuclear region has a concentration of star formation
 and the star formation would control AGN activity.
\end{abstract}

\begin{multicols}{2}

\section{Introduction}
A galaxy that exhibits bright emission in the nuclear region is termed an 
 Active Galactic Nucleus (AGN).
The origin of the radiative energy is putative release of gravitational energy
 through accretion of the interstellar medium from the host galaxy onto a
 central super-massive black hole (SMBH) with a mass of
 $>$10$^6{\rm M}_{\odot}$.
Starbursts are one of the candidates for transporting material onto the center
 \citep{WN02}, so studying the correlation between AGNs and starbursts is
 crucial to understanding the origin of AGN activity.
Seyfert galaxies are the second most numerous AGN class in the local
 universe\footnote{It assumes that LINERs are powered by AGN \citep{HoFS97}}.
They are classified into two types: type 1 Seyfert galaxies show broad and
 narrow optical emission lines, and type 2 Seyfert galaxies show narrow
 emission lines only.
The difference is explained by the AGN unification paradigm \citep{ant93} in
 which molecular gas and dust with a torus-shaped structure (the so-called
 dusty torus) surrounds the central SMBH.
The intrinsic AGN properties of the two types of Seyfert galaxies are assumed
 to be the identical.
The dusty torus plays various roles in observed spectra:
 (1) obscures the AGN for a Seyfert 2 galaxy and
 (2) absorbs the energetic radiation from the AGN and re-radiates it as
 infrared dust emission for both types of Seyfert galaxies.
Because the torus is composed of significant amounts of molecular gas and dust,
 it is a natural site for star formation.
In particular, the outer portion of the torus is thought to be an ideal site
 for star formation because star formation can be suppressed in the inner
 portion by tidal forces from the SMBH.
In fact, it was argued that such starburst activity could occur in the dusty
 torus in the nuclear region ($<$100pc) of a Seyfert galaxy
 \citep{WN02, Ima03}.
We call such starbursts in Seyfert nuclei "nuclear starbursts".
\citet{WN02} showed that the nuclear starbursts in the dusty torus can produce
 an inflated turbulent torus around the central SMBH.
However, with limited spatial resolution, it is difficult to detect the
 emission from the nuclear starbursts as they are diluted by strong AGN
 emission.
Here, we discern these emission and estimate the magnitudes of starburst
 activities using infrared K-band (1.9--2.5$\mu$m) and L-band (2.8--4.2$\mu$m)
 spectra.

Polycyclic Aromatic Hydrocarbon (PAH) emission at around 3.3 $\mu$m in the
 L-band is a powerful tool for distinguishing starburst emission from AGN
 emission.
PAH molecules are widely distributed throughout interstellar space
 \citep{tanaka96}.
The PAH molecules are not destroyed and are excited by non-ionizing UV photons
 from stars and then emit the line at 3.3 $\mu$m in the Photo-Dissociation
 Region (PDR) for starburst \citep{sellg81}, whereas the PAH molecules are
 destroyed by X-rays \citep{voit92} from AGN.
Observationally, a normal starburst galaxy, which consists of HII regions,
 PDRs, and molecular gas (e.g. M82 and NGC253), shows the 3.3 $\mu$m PAH
 emission \citep{tok91, ID00}, whereas a pure AGN shows only PAH-free
 continuum emission from larger-sized AGN-heated hot dust grains
 \citep{Moorwood86, ID00} because an AGN emits strong X-rays in addition to UV.
Additionally, because the 3.3 $\mu$m PAH emission feature is very strong,
 it is detectable even if the starbursts are weak.

Nuclear starbursts in Seyfert galaxies can also be investigated through
 infrared K-band spectra.
The CO($\Delta$v=2--0) absorption feature at $\lambda_{rest}$=2.3--2.4 $\mu$m
 is caused by the CO molecule in the photosphere of cool stellar population
 (i.e. red giants and supergiants) \citep{Fro78, doyon94}.
Empirically, starburst galaxies exhibit a substantially deeper
 CO($\Delta$v=2--0) feature than do quiescent spirals galaxies
 \citep{Fro78, ridg94}, and AGNs do not exhibit this absorption feature.

Furthermore dust extinction is much lower in the K- and L-bands than at shorter
 wavelengths, $A_{K}\sim 0.062A_{V}$ and $A_{L}\sim 0.031A_{V}$, respectively
 \citep{Nishi08, Nishi09}, so that the quantitative uncertainty of the dust
 extinction correction is significantly reduced compared with shorter
 wavelengths (e.g. UV, optical).
Thus, K- and L-band spectra enable us to estimate the relative contributions
 of AGN and starbursts emission to an observed spectrum in a quantitatively
 reliable manner \citep{Oli99, Iva00, Rod03, Ima03, IA04, IW04, davies07}.

The above authors studied mainly nearby Seyfert galaxies with high infrared
 luminosities (L$_{IR}\gtrsim$10$^{44}$ergs s$^{-1}$), and found a correlation
 between the AGN activity and nuclear starburst activity
 \citep{Ima03, IA04, IW04}.
However, it is unclear whether the correlation holds over a wide range of
 AGN luminosity.
\citet{Haas05} argued that the height of the torus is smaller in lower
 luminosity Seyfert galaxies compared with higher luminosity Seyfert galaxies,
 suggesting that nuclear starbursts are weaker in low-luminosity AGNs.
In contrast, \citet{balla08} predicted that parsec-scale starbursts would be
 associated with lower luminosity AGNs.
\citet{KW08} predicted that the ratio of the scatter in the AGN to starburst
 luminosity would increase in low-luminosity AGNs.
To resolve this issue, it is necessary to study Seyfert galaxies over a wide
 AGN luminosity range. 
Thus, we performed K- and L-band spectroscopy of a large sample of Seyfert
 galaxies with luminosity ranging from low to high.
Throughout the paper, $H_0=75$km s$^{-1}$Mpc$^{-1}$, $\Omega_{M}=$0.3,
 $\Omega_{\Lambda}=$0.7 are adopted.

\section{Target Selection}
\citet{Ima03}, \citet{IA04}, and \citet{IW04} studied Seyfert galaxies taken
 from the CfA \citep{HB92} and 12 $\mu$m \citep{rush93} samples.
The CfA and 12 $\mu$m samples were selected through optical spectroscopy of
 large numbers of galaxies limited by optical and $IRAS$ 12 $\mu$m fluxes,
 respectively.
The CfA sample is usually regarded as a complete sample of optically selected
 Seyfert galaxies.
However, \citet{HU01} argued that the CfA sample is likely to be complete only
 for relatively bright objects.

Therefore we added objects from the Palomar sample \citep{HoFS95} in this
 paper.
Seyfert galaxies in the Palomar sample are selected through optical
 spectroscopy of nearby, bright ($B_T\le$12.5 mag), northern (declination
 $>$0$^\circ$) galaxies covering a wider AGN luminosity range. 
Our aim is to confirm the relationship between AGN and nuclear starburst
 activity over a wide AGN luminosity range, paying particular attention to
 low-luminosity AGN.
To be easily observable from Mauna Kea, Hawaii (our observing site; latitude
 $\sim$20$^\circ$), the declinations of Seyfert galaxies were limited to
 greater than $-$30$^\circ$.
Owing to the telescope limit on the IRTF 3.0-m telescope used in this study, a
 restriction on declination of less than 68$^\circ$ was also applied.
In total, 13 objects in the Palomar Sample (Table \ref{tb:info}) were observed.
We also added some additional re-observed Seyfert galaxies to our targets.
These galaxies are those studied by \citet{Ima03}, \citet{IA04} and
 \citet{IW04}, in which PAH emission was barely discernible, but with less
 than 3$\sigma$, as reported in these papers (NGC262, NGC513, Mrk993,
 MCG--3-58-7, F03450+0055, NGC931 and NGC5548).
Moreover, we also included 0152+06 in the CfA sample and MCG--2-8-39 in the 12
 $\mu$m sample in our targets.
Both of them meet the declination criteria of these paper but were not
 observed in previous papers.
To summarize, 13 objects from the Palomar sample, seven re-observed objects
 discussed in previous papers, and two new objects from CfA and 12 $\mu$m
 samples are included in this paper. 

Six of the 22 targets are Seyfert 1 galaxies and the remaining 16 targets are
 Seyfert 2 galaxies.
Infrared (8-1000 $\mu$m) luminosity of our targets is
 log$L_{IR}$=42.5--44.8 ergs s$^{-1}$ for the Palomar sample and
 log$L_{IR}$=43.6--44.8 ergs s$^{-1}$ for the CfA and 12$\mu$m samples
 (see a caption of Table \ref{tb:info} for the definition of $L_{IR}$).
Detailed information on the targets is summarized in Table \ref{tb:info}.
We define type 1--1.5 and type 1.8--2 Seyfert galaxies as Seyfert 1 and
 Seyfert 2 galaxies, respectively.
Mrk993 was classified as a Seyfert 2 galaxy in the CfA sample, but later
 reclassified as a Seyfert 1.5 galaxy \citep{OM93}.
We adopt the latter classification and thus classified it as a Seyfert 1
 galaxy, although it was regarded as a Seyfert 2 galaxy by \citet{IA04}.

\section{Observations \& Data Analysis}
Table \ref{tb:Observing Log} summarizes our observation log.
All targets were observed with SpeX \citep{ray03} on the IRTF 3.0-m
 telescope on Mauna Kea, Hawaii.
SpeX has a 1.9--4.2$\mu$m cross-dispersed spectroscopic mode (LXD1.9 mode),
 enabling us to observe K- and L-band spectra simultaneously.
Because the seeing in $K$ measured with the SpeX guiding/imaging mode was about
 0$\farcs$4--0$\farcs$8 (FWHM) throughout the observations (except on 26 and 27
 March 2009), we consistently used a 0$\farcs$8-wide slit.
On 26 and 27 March 2009, the seeing was over 1$\farcs$0 at K, so we used a
 1$\farcs$6-wide slit during those observations.
A standard telescope nodding technique (ABBA pattern) with a throw of
 7$\farcs$5 was employed along the slit to subtract background emission.
The physical scale probed by our slit spectroscopy is 24--500 pc taking into
 account the redshift, z = 0.002--0.032, and slit width.
We observed A-, F- and G-type main-sequence stars
 (Table \ref{tb:Observing Log}) as standard stars.
These were used to correct for transmission through the Earth's atmosphere.
We observed the standard stars at airmasses similar to the targets
 ($<$0.1 difference), before and after the observation of each target to better
 correct for possible time variations in the Earth atmosphere.

The spectra were reduced using standard IRAF tasks.\footnote{IRAF is distributed by 
the National Optical Astronomy Observatories, operated by the Association of Univers
ities for Research in Astronomy (AURA), Inc., under cooperative agreement with the N
ational Science Foundation.}
Frames taken with an A (B) beam were subtracted from frames taken with a B (A)
 beam, and then median images were taken and divided by a spectroscopic flat
 image.
Known bad pixels and remaining pixels hit by cosmic rays were replaced with
 interpolated values from the surrounding pixels.
The Seyfert galaxies and standard-star spectra were then extracted by
 integrating signals over 1$\farcs$3--2$\farcs$5 along the slit.
The one-dimensional spectrum was extracted by fitting the signals in each
 spectral order in the two-dimensional image.
We used the wavelength-dependent nature of transmission through the Earth's
 atmosphere for our wavelength calibration.
The object spectra were divided by spectra of standard stars to remove the
 effects of atmospheric absorption and then multiplied by the spectra of
 blackbodies with temperatures corresponding to those of the individual
 standard stars (Table \ref{tb:Observing Log}).
Flux calibration was performed based on the signal detected inside our slit,
 taking into account the differences in exposure time between our targets and
 the standard stars.
K- and L-band magnitudes for the standard stars were estimated from their
 V-band magnitudes, adopting appropriate $V-K$ and $K-L$ colors for each
 individual standard star stellar type \citep{tok00}.
A strong methane absorption line appears $\lambda_{rest}$ = 3.315$\mu$m in the
 Earth's atmosphere.
Although this could be corrected by standard stars in theory, it introduces a
 relatively strong amount of noise.
Thus, we excluded a few data points at around $\lambda_{rest}$ = 3.315 $\mu$m,
 whose transmission were less than 40\% of nearby signals without being
 affected by the absorption line.

\section{Results}
\subsection{L-band spectrum}
\label{sec:L-band spectrum}

Figure \ref{fig:L-band_spectra} shows flux-calibrated 3.0--3.6 $\mu$m slit
 spectra in the L-band.
Although the full L-band (2.8--4.2 $\mu$m) wavelength range is covered by SpeX,
 both ends of the L-band are noisy because of poor transmission through the
 Earth's atmosphere ($\lambda < 3.0\mu$m) and high background emission from the
 Earth's atmosphere ($\lambda>3.6\mu$m).
We removed those data points for the discussions in this paper because they
 were not necessary for our scientific aims.
In these plots, some objects show a clear excess at 3.29 $\mu$m$\times$(1+z)
 whose wavelength corresponds to the redshifted 3.3 $\mu$m PAH emission
 feature.
We regard the features as detected when at least two consecutive data points
 at around 3.29 $\mu$m$\times$(1+z) are higher than 1.5 times the scatter of
 the continuum level.
Using this criterion, objects with detectable PAH emission are marked with
 "3.3 $\mu$m PAH", and PAH non-detected objects are marked with
 "3.3 $\mu$m PAH(?)" in the figure.
We fitted the emission line to a Gaussian function by setting the
 normalization, full width of half maximum (FWHM) and central wavelength of
 the emission as free parameters, and then estimated the flux ($f_{3.3PAH}$),
 luminosity ($L_{3.3PAH}$) and rest-frame equivalent width ($EW_{3.3PAH}$) of
 the 3.3 $\mu$m PAH emission.
For objects with no clear PAH emission features, we estimated upper limits by
 fitting the emission line with a Gaussian function whose height, central
 wavelength and FWHM were fixed as 3 $\sigma$, where 1 $\sigma$ was a
 dispersion of its continuum emission, 3.29$\times$(1+z)$\mu$m and
 0.02$\times$(1+z)$\mu$m \citep{tok91}, respectively.
The strengths of the 3.3 $\mu$m PAH emission features are summarized in Table
 \ref{tb:results}.

The 3.3 $\mu$m PAH emission feature is detected in $\sim$ 23\% (5/22) of the
 observed Seyfert galaxies (NGC6764, 5194, 5273, F03450+0055 and MCG-3-58-7).
NGC6764 shows a strong PAH emission feature, and its rest-frame equivalent
 width ($EW_{3.3PAH}$) is as high as those of starburst galaxies
 ($\sim$100 nm; \cite{ID00}).
Although four objects (NGC5194, NGC5273, F03450+0055 and MCG-3-58-7) show
 detectable 3.3 $\mu$m PAH emission, the $EW_{3.3PAH}$ values are significantly
 smaller than $\sim$100 nm.
The other 17 objects do not display detectable PAH features.

\subsection{K-band spectrum}
\label{sec:K-band spectrum}

Flux-calibrated 2.0--2.5 $\mu$m slit spectra in the K-band are shown in Figure
 \ref{fig:K-band_spectra}.
Most of the spectra show a flux depression at $\lambda_{obs} > 2.3\mu$m in the
 observed frame, which is attributed to the CO ($\Delta$v=2--0) molecular
 absorption feature.
The strength of the CO absorption feature was quantified by a spectroscopic
 $CO_{spec}$ index defined by \citet{doyon94} as follows:
\begin{eqnarray}
  CO_{spec}\equiv -2.5log_{10}\langle R_{2.36}\rangle, \nonumber
\end{eqnarray}
where $\langle R_{2.36}\rangle$ is an average of actual signal at
 $\lambda_{rest}$ = 2.31--2.40 $\mu$m divided by a power-law continuum
($F_{\lambda}=\alpha \times \lambda^{\beta}$) extrapolated from shorter
 wavelengths.
As some strong emission features appear at around 2$\mu$m in K-band,
 such as H$_2$ 1--0 S(1) ($\lambda_{rest}=2.122\mu$m),
 Br$\gamma$ ($\lambda_{rest}=2.166\mu$m) and H$_2$ 1--0 S(0)
 ($\lambda_{rest}=2.223\mu$m), we used data points at
 $\lambda_{rest}=2.1-2.29\mu$m to fit the continuum level after excluding these
 emission lines.
The continuum level is shown as a solid line in
 Figure \ref{fig:K-band_spectra}.
The $CO_{spec}$ of all our targets are given in Table \ref{tb:results}.
The $CO_{spec}$ values of cool stars are typically 0.2--0.3
 \citep{doyon94}.
For objects with a clear CO absorption feature ($CO_{spec}>$ 0.02),
 "CO absorption" signature is written in Figure \ref{fig:K-band_spectra}.
About 77 \% (17/22) of observed Seyfert galaxies show clear CO absorption
 features.
Figure \ref{fig:COhist_lirgs} is a histogram of the $CO_{spec}$ of Luminous
 Infrared Galaxies (LIRGs: their luminosities are generally explained by
 starbursts) measured by \citet{ridg94}.
It suggests that their typical value is also $\sim$ 0.2--0.3.
If the properties of nuclear starbursts in Seyfert galaxies are similar to
 those of starbursts in LIRGs, the $CO_{spec}$ of the starbursts in Seyfert
 galaxies is also expected to be 0.2--0.3.
However, the observed $CO_{spec}$ is smaller than the typical value for a LIRG
 because featureless AGN emission is superposed in the K-band spectrum.

\section{Discussion}
\subsection{$K-L$ color}
\label{sec:$K-L$ color}

Because we took K- and L-band spectra simultaneously, possible slit loss caused
 by IRTF/SpeX tracking error is similar and sky conditions are the same for the
 K- and L-bands.
Therefore, the derived $K-L$ colors are very reliable.
The emission from normal stars ($>$2000K) in the L-band is much weaker
 than in the K-band, whereas hot ($\sim$1000K) dust heated by
 an AGN emits strongly in both the K- and L-bands.
Thus the $K-L$ colors become smaller (bluer) with increasing relative
 contribution from stellar emission to the total emission of nuclear region.
Indeed, intrinsic $K-L$ colors of Seyfert 1 galaxies are typically 1--2 mag
 \citep{alo03}\footnote{K$^{\prime}$ was used in \cite{alo03}; we assumed the differ
ence between K$^{\prime}$ and K was negligible.},
 and those of normal spiral galaxies are $K-L<0.4$ \citep{wil84}.
Therefore, the color enables us to distinguish starburst-important Seyfert
 galaxies from AGN-dominant ones.
The nuclear $K-L$ colors derived from our slit spectra are summarized in
 column (6) of Table \ref{tb:results}.

For the combined sample used in this paper, taken from \citet{Ima03},
 \citet{IA04} and \citet{IW04}, we compared the $K-L$ colors of Seyfert 1 and
 2 galaxies (Figure \ref{fig:K-Lhist}).
Most of the Seyfert 1 galaxies are distributed in the $K-L=$ 1--2 mag range,
 whereas the colors of almost half of Seyfert 2 galaxies are bluer than the
 Seyfert 1 galaxies ($K-L<1$ mag).
It shows that stellar contamination is relatively larger in Seyfert 2s than
 Seyfert 1s in the observed total emission.
Two possibilities could explain the result.
First, AGN emission is strongly affected by the extinction of dusty torus
 compared with nuclear starburst emission.
In the AGN unified model, as the AGN emission comes from the central region,
 the emission from a Seyfert 2 galaxy passes through the dusty torus and is
 more strongly absorbed by the dusty torus than those from a Seyfert 1 galaxy.
So, observed AGN emission of Seyfert 2 galaxies would be relatively weak in
 comparison to Seyfert 1 galaxies.
Whereas, if the nuclear starbursts are occurring in the outer region of the
 dusty torus, then the absorption effect would not be much difference between
 these types of Seyfert galaxies.
In this case, the starburst contribution to the total emission would be
 relatively significant in Seyfert 2s when compared with Seyfert 1s and then
 the $K-L$ colors of Seyfert 2s would be bluer than those of Seyfert 1s.
Another possible explanation is the starbursts in the central region in Seyfert
 2 galaxies could be more active than those in Seyfert 1 galaxies.
If strong nuclear starbursts inflate the dusty torus more thickly than weaker
 starbursts, then a Seyfert galaxy with active starbursts is much more likely
 to be observed as type 2, and the color of it would be bluer than one with
 weak starbursts. 
However, as we will discuss in \S\ref{sec:IRAS 12 and 25 um luminosity},
 we find no clear difference between Seyfert 1 and 2 galaxies in quantitatively
 measured 3.3 $\mu$m PAH luminosity.
So, it is unlikely that the difference of the $K-L$ colors between Seyfert 1s
 and 2s is caused by the intrinsic difference of starburst activities in
 them.
Therefore we suggest that the bluer $K-L$ colors of Seyfert 2s than Seyfert 1s
 are due to the obscuration of AGN emission and to relative contamination of
 the stellar emission in Seyfert 2 galaxies.
\citet{kotil92} and \citet{alo96} deconvolved surface brightness profile of
 Seyfert 1s and 2s within 3-arcsec aperture into a combination of a nuclear
 source (non-stellar component), a bulge component and a disk component, and
 suggested that the stellar contamination to the total emission is larger in
 Seyfert 2s than in Seyfert 1s.

Figures \ref{fig:KL-PAHCO}a and \ref{fig:KL-PAHCO}b show the $K-L$ color
 versus the $EW_{3.3PAH}$ and $CO_{spec}$ of the sample.
These figures show that both the $EW_{3.3PAH}$ and $CO_{spec}$ decrease with
 reddening $K-L$ color.
Thus they are useful tracers for estimating the starburst contribution to the
 nuclear spectra inside our slit.
The trends of Seyfert 2 galaxies in Figure \ref{fig:KL-PAHCO} are similar to
 those for Seyfert 1 galaxies.
In Figure \ref{fig:KL-PAHCO}b, however, some fraction of Seyfert 2 galaxies in
 our sample reach to $CO_{spec}$ = 0.2--0.3, which is closer to those of LIRGs,
 in comparison to Seyfert 1 galaxies.
This is good agreement with the suggestion above.
In other words, the $CO_{spec}$ of some fraction of Seyfert 2 galaxies are
 not strongly diluted by AGN emission.

\section{Comparison of nuclear starbursts with AGN power}
\subsubsection{$IRAS$ 12 and 25 $\mu$m luminosity}
\label{sec:IRAS 12 and 25 um luminosity}

We compare the nuclear starburst activity with AGN activity over a wide AGN
 luminosity range.
The nuclear starburst activity is reasonably quantifiable from the observed
 3.3 $\mu$m PAH emission luminosity inside our slit spectrum.
Meanwhile \citet{alo03} showed that the obscuring effects of the AGN emission
 for Seyfert galaxies became insignificant at longer than 10 $\mu$m.
$IRAS$ 60 and 100 $\mu$m luminosities contain more contamination of emission
 from star formation in the host galaxy than do $IRAS$ 12 and 25 $\mu$m
 luminosities \citep{SpM89, rod97, alo02}, so we use $IRAS$ 12 and 25 $\mu$m luminos
ities as good tracers
 of AGN power in Seyfert galaxies.
A comparison between the $IRAS$ 12 and 25 $\mu$m luminosities and the
 observed nuclear 3.3 $\mu$m PAH emission luminosity is shown in Figures
 \ref{fig:AGNpowervsPAHlumin}a and \ref{fig:AGNpowervsPAHlumin}b, respectively.
In these figures, we combined with the data of four sources observed by
 \citet{Rod03} using the 0$\farcs$8 slit width of IRTF/SpeX.

We see no clear difference in the $L_{3.3PAH}$ between the two types of
 Seyfert galaxies when normalized to the AGN power.
This means that we see no evidence of the possibility that Seyfert 2 galaxies
 tend to show intrinsically stronger starburst activities than do Seyfert 1
 galaxies, a possibility discussed in \S\ref{sec:$K-L$ color}.
In Figure \ref{fig:AGNpowervsPAHlumin}, the nuclear 3.3 $\mu$m PAH luminosities
 for our objects combined with previous data are
 $L_{3.3PAH}\sim 10^{38}$--10$^{42}$ ergs s$^{-1}$ .
In starburst-dominated galaxies, the 3.3 $\mu$m PAH-to-far-infrared
 (40--500 $\mu$m) luminosity ratios
 ($L_{3.3PAH}/L_{FIR}$) are $\sim1\times 10^{-3}$ \citep{mou90}.
The nuclear star-formation rates (SFR) are calculated by 
SFR M$_{\odot}$ yr$^{-1}$
 = ($L_{{\rm FIR}}/2.2\times 10^{43}$ ergs s$^{-1}$) \citep{kenn98}.
So the SFR of our sample could be up to 4.5--450$\times10^{-3}{\rm M}_{\odot}$
 yr$^{-1}$.
\citet{Wada09} showed that a supernova rate (SNR) of 5.4 to 540 $\times$
 10$^{-5}$ yr$^{-1}$, corresponding to
 7.7--770$\times$10$^{-3}{\rm M}_{\odot}$ yr$^{-1}$ (SNR $\simeq$ 0.007 SFR),
 can produce torus heights of as large as 10 pc at the outer side $\simeq$ 5
 pc.
Hence, the star-formation rate of almost our entire sample could be high
 enough to create a geometrically thick dusty torus.
We should note, however, that we have only upper limits on the nuclear 3.3
 $\mu$m PAH luminosities for many objects in our sample, so it is likely that
 the sample contains Seyfert galaxies that do not have a thick dusty torus.

The figures show that the nuclear starburst luminosity decreases with
 decreasing AGN power and that the trend dose not change over a wide AGN
 luminosity range.
The ratios of starburst activity to AGN activities for both Seyfert 1 and 2
 galaxies show the same levels of scatter and we find no obvious difference
 between them.
We apply the generalized Kendall rank correlation statistics provided in the
 Astronomy Survival Analysis package (ASURV: \cite{iso86}) (which handles data
 with upper limits) to Figures \ref{fig:AGNpowervsPAHlumin}a and
 \ref{fig:AGNpowervsPAHlumin}b.
The probabilities that a correlation is not present for the figures are both
 found to be $\sim$0.00\%.
The results show that there are tight correlation between 3.3 $\mu$m PAH
 luminosity in central region and $IRAS$ 12 and 25 $\mu$m luminosities.

\subsubsection{Ground-based N-band luminosity}
\label{sec:Ground-based N-band luminosity}

Although the contamination from star formation activity in $IRAS$ 12 and 25
 $\mu$m is less than in the $IRAS$ 60 and 100 $\mu$m, it could be that the
 $IRAS$ 12 and 25 $\mu$m data still contain a significant amount of emission
 from star-forming activity of host galaxies in our low luminosity AGN sample,
 given the large aperture of $IRAS$ 12 $\mu$m ($0\farcm75\times4\farcm5$) and
 25 $\mu$m ($0\farcm75\times4\farcm6$).
\citet{ramos09} compared small aperture unresolved nuclear 10 $\mu$m emission
 (0$\farcs$4--0$\farcs$5) and 20 $\mu$m emission (0$\farcs$5--0$\farcs$6) with
 the $IRAS$ 12 $\mu$m and 25 $\mu$m emission, respectively, and found the large
 aperture data are largely contaminated by starlight. 
So it is likely that the $IRAS$ 12 $\mu$m and 25 $\mu$m emission is
 significantly contaminated by star formation in host galaxies.
To reduce the possible ambiguity, we should compare the 3.3 $\mu$m PAH
 emission with an AGN indicator measured with a smaller aperture.
The N-band ($\lambda_0$=10.78$\mu$m, $\Delta\lambda$=5.7$\mu$m) is observable
 from the ground, and its luminosity is also thought to be a good tracer of
 AGN power \citep{alo02, ramos09, lev09}.
\citet{gorjian04} presented N-band photometric data for the central regions
 of Seyfert galaxies.
They used the Palomar 5-m telescope with a 1$\farcs$5 aperture. 
Only 41\% (9/22) of our objects were measured by \citet{gorjian04}.
We combine these with data from \citet{IW04}, which is shown in Figure
 \ref{fig:N-band_AGNpowervsPAHlumin}.
Although the number of low-luminosity AGNs is small and their 3.3 $\mu$m PAH
 emission luminosities are only upper limits, the three low luminosity Seyfert
 galaxies are distributed around a line extrapolated from the stronger AGN
 activity region.
We also apply the generalized Kendall rank correlation statistics to Figure
 \ref{fig:N-band_AGNpowervsPAHlumin}.
The uncorrelated probability is $\sim$0.03\% for the figure.
These results mean that the correlation between the luminosities of the nuclear
 starbursts detected inside the slit spectra and central AGNs is statistically
 confirmed in Seyfert galaxies for all observed quantities.

We found the relation between the luminosities of 3.3 $\mu$m PAH emission
 within our slit width and of N-band.
However it is worried that the $L_{3.3PAH}$ is affected by physical aperture
 size probed by the slits.
It means that we are concerned about the possibility that the $L_{3.3PAH}$ of
 the most luminous object is due to the largest area we observed.
If star formation traced by $L_{3.3PAH}$ is spatially extending in a wide area
 of host galaxy, then the $L_{3.3PAH}$ becomes higher with lager physical
 scale.
To explore this, we compare the $L_{3.3PAH}$ with physical area of each source
 (Figure \ref{fig:surfacevsL3.3}).
The figure clearly shows that not all high $L_{3.3PAH}$ objects are wide
 physical area coverage.
That is to say, the star formation would be occurring in the central region
 inside the slits and not spatially extend.

\subsection{Trigger of accretion onto SMBH}
We find no evidence that the ratio of starbursts to AGN luminosity
 deviates upward in low-luminosity AGNs, as predicted by \citet{KW08}, or that
 nuclear starburst activities increase with decreasing AGN activities as
 predicted by \citet{balla08}.
Furthermore, we find no clear change in the correlations over a wide AGN
 luminosity range.
This is explained if the main mechanism that connects the AGN and nuclear
 starbursts is the same at each luminosity range.
A model shown in Wada \& Norman (2002; see also \cite{Wada09}), which predicts
 the enhancement of a mass accretion rate onto a central SMBH owing to
 increased turbulence of molecular gas in the torus caused by nuclear
 starbursts, is one possible scenario that could account for the luminosity
 correlation.

\subsection{The origin of the $CO_{spec}$ index}
\label{The origin of the $CO_{spec}$ index}

We calculated stellar luminosity ($L_{Kstellar}$) from $CO_{spec}$.
If nuclear starbursts with properties similar to star-formation-dominated LIRGs
 appear inside our slit and the entire K-band emission comes from stars, then
 the detected $CO_{spec}$ must be 0.2--0.3.
We assumed the original $CO_{spec}$ of the cool star was 0.25.
When the contributions of stellar emission to total emission are 100\%, the
 average signal at $\lambda_{rest}$ = 2.31--2.4$ \mu$m is reduced by about
 20.5\% compared with an extrapolation from a shorter wavelength.
When the contribution of AGN emission to the total emission is 100 \%, the
 absorption feature does not appear.
When the contributions of both stellar and AGN emission are equal, the
 average signal at $\lambda_{rest}$ = 2.31--2.4$ \mu$m should be 10.25\%
 reduced from the shorter wavelength, and the $CO_{spec}$ should be 0.117.
In the case that the $CO_{spec}$ is 0.1 or 0.15, the contributions of stellar
 emission to total emission should be 43\% or 63\%, respectively.
So, we can estimate the $L_{Kstellar}$.
Figure \ref{fig:LkstellarvsPAHlumin} compares the $L_{Kstellar}$ and the 3.3
 $\mu$m PAH emission luminosity in the nuclear region detected inside our slit
 spectra combined with Fig. 5 of \citet{IA04}.
Although they plotted these results for Seyfert 2 galaxies only, we included
 in Figure \ref{fig:LkstellarvsPAHlumin} Seyfert 1 galaxies whose $CO_{spec}$
 were derived in \citet{IW04}.

The K-band to infrared luminosity ratios of starburst-dominated LIRGs were
 estimated to be $L_{K}/L_{IR(8-1000 \mu {\rm m})}\sim 10^{-1.6\pm0.2}$
 \citep{goldader97}. If the nuclear starbursts in Seyfert galaxies have
 properties similar to the starburst-dominated LIRGs, then the same relation
 should hold.
The 3.3 $\mu$m PAH to infrared luminosity ratios in starbursts were found to be
 $L_{3.3PAH}/L_{FIR(40-500 \mu {\rm m})}\sim 10^{-3}$ \citep{mou90}.
Assuming $L_{IR}\sim L_{FIR}$ for starbursts, the luminosity of the 3.3 $\mu$m
 PAH emission to the K-band stellar luminosity ratios are expected to be
 $L_{3.3PAH}/L_{Kstellar}\sim 10^{-1.4}$, if both luminosities trace
 the same starbursts.
The value is shown as a solid line in Figure \ref{fig:LkstellarvsPAHlumin}.
Most of our Seyfert 1 galaxies and about half of the Seyfert 2 galaxies plotted
 in the figure are distributed around the line.
However, the remaining half of the Seyfert 2 galaxies are distributed under
 the line.
Given that in Figure \ref{fig:AGNpowervsPAHlumin}, the $L_{3.3PAH}$
 distribution of both Seyfert 1 and 2 galaxies is similar within their
 dispersions and no obvious difference is seen between them, the discrepancy
 must be caused by $L_{Kstellar}$.
As we discussed above, the $L_{Kstellar}$ is estimated from the $CO_{spec}$
 value, which is related to the fraction of stellar to AGN emission.
Nuclear starbursts are expected to occur at the outer part of the dusty
 torus and the AGN radiation of type 2 Seyfert galaxy is significantly
 attenuated by the dusty torus.
Therefore, the $CO_{spec}$ is likely to increase and thus increase
 $L_{Kstellar}$ in Seyfert 2 galaxies.
In contrast, because $L_{3.3PAH}$ is estimated from the flux of 3.3 $\mu$m
 PAH emission, PAH-estimated starburst luminosity is not affected by the
 absorption of AGN emission.
In summary, we suggest that because $L_{Kstellar}$ represents a relative value
 of stellar to AGN emission and $L_{3.3PAH}$ represents an absolute value of
 stellar emission, it is possible that $L_{Kstellar}$ is overluminous and that
 the ratio of $L_{3.3PAH}/L_{Kstellar}$ decreases in Seyfert 2 galaxies.

The other possibility is that the signatures of stellar emission detected in
 the K-band spectra are significantly contaminated by old bulge stars in the
 nuclear part of the host galaxy, whereas the 3.3 $\mu$m PAH emission
 originates in spatially unresolved (smaller than subarcsecond) nuclear
 starbursts in the dusty torus because old stars do not have enough
 PAH-exciting UV photons.
\citet{Iva00} and \citet{IA04} suggested that the emission of old stars in the
 host galaxy mainly produces the K-band spectra.
However, most of the Seyfert 1 and some fraction of the Seyfert 2 galaxies have
 $L_{3.3PAH}/L_{Kstellar}$ ratios similar to the value expected
 when the origin of both luminosities is the same.
Therefore, although the old stars in host galaxy may contribute to the observed
 K-band spectra of some fraction of Seyfert galaxies, it is not likely that old
 stars dominate the K-band spectra in all objects.

\section{Conclusion}
The results of an infrared K- and L-band spectroscopic study of 13
 Seyfert galaxies from the Palomar sample, and seven Seyfert galaxies with
 non-detectable 3.3 $\mu$m PAH emission discussed in previous papers and two
 new Seyfert galaxies from CfA and 12$\mu$m samples are presented.
Our L-band spectroscopic method successfully detected the 3.3 $\mu$m PAH
 emission in $\sim$ 23 \% (=5/22) of the observed Seyfert galaxies.
Also, our K-band spectroscopic method showed $\sim$ 77 \% (=17/22) of Seyfert
 nuclei in our sample have clear CO absorption feature.
We examined the relationship between AGN activity and nuclear starburst
 activity over a wide AGN luminosity range using our spectra together on
 previously published spectra \citep{Ima03, IA04, IW04}.
Our conclusions are summarized as follows:
\begin{enumerate}
\item The $K-L$ colors of Seyfert 2 galaxies are widely distributed toward the
  blue, compared with those of Seyfert 1 galaxies.
  This implies either that the dusty torus absorbs the AGN emission of Seyfert
  2 galaxies from the central region or that Seyfert 2 galaxies tend to have
  stronger nuclear starbursts than do Seyfert 1 galaxies.
  As the $L_{3.3PAH}$ of Seyfert 1 and 2 galaxies do not differ significantly,
  we have interpreted this as the result of the effect that the AGN emission of
  Seyfert 2s is absorbed by dust of torus and then the stellar emission
  becomes relatively larger. 
  
\item The $L_{3.3PAH}$ shows the same range in Seyfert 1 and 2 galaxies.
  The star-formation rates of our sample are up to 4.5--450
  $\times10^{-3}{\rm M}_{\odot}$ yr$^{-1}$, which could be enough to swell the
  dusty torus by supernova explosion; in short, it is geometrically thick,
  although many objects we provided are only with upper limits on the 3.3
  $\mu$m PAH luminosities.

\item The $L_{3.3PAH}$ correlates with mid-infrared (N-band) luminosity with
  small aperture (= tracing AGN activities), and their luminosity ratio does
  not vary significantly over a wide AGN luminosity range.
  Moreover the $L_{3.3PAH}$ is independent of physical scale we probed with
  slits.
  Therefore the $L_{3.3PAH}$ (i.e. star formation) would concentrate on the
  central region, and the nuclear star formation would induce the accretion
  onto the SMBH and encourage the AGN activity.
  In this work, we find no evidence that nuclear starbursts are stronger in
  lower-luminosity AGN.
  
\item We suggest that the 3.3 $\mu$m PAH emission luminosity and K-band stellar
  luminosity ($L_{Kstellar}$) originate in the same phenomenon.
  However, for some fraction of Seyfert 2 galaxies, $L_{Kstellar}$ is likely to
  be overestimated because of flux attenuation of AGN emission caused by a
  dusty torus and contaminated by old stars in the spheroid of the host galaxy.
\end{enumerate}

\section*{acknowledgments}
We would like to thank the staff at the IRTF 3.0-m telescope for their help
 with observations.
We thank C. Packham, N. Kawakatu, H. Ando, N. Arimoto, and T. Kodama for useful
 discussions.
M.I. was supported by grants-in-aid for scientific research (19740109).
This work was supported in part by The Graduate University for Advanced Studies
 (Sokendai).
This research made use of the NASA/IPAC Extragalactic Database (NED),
 which is operated by the Jet Propulsion Laboratory, California Institute of
 Technology, and the SIMBAD database, operated at CDS, Strasbourg, France.

\end{multicols}

\clearpage


Col.(1): object name. a, b, c are reference;
 a)Palomar sample \citep{HU01}, b)CfA \citep{HB92}, c)12 $\mu$m \citep{rush93}.
Col.(2): redshift.
Col.(3): luminosity distance in Mpc.
Col.(4): AGN class.
Col.(5)-(8): flux at 12$\mu$m, 25$\mu$m, 60$\mu$m, and 100$\mu$m in Jy,
 respectively from $IRAS$ faint source catalog.
Col.(9): ground-based 10.8$\mu$m photometric data, observed with the Palomar
 5-m telescope using a $1\farcs5$ aperture \citep{gorjian04}.
NGC4579 was observed also by \citet{horst09} using VLT/VISIR with a
 $1\farcs27$ small aperture. 
Col.(10): decimal logarithm of infrared (8-1000$\mu$m) luminosity in
 ergs s$^{-1}$ calculated with $L_{IR}\equiv 2.17\times 10^{39}\times
 D_L(\rm{Mpc})^{2} \times $(13.48$\times f_{12}$+5.16$\times
 f_{25}$+2.58$\times f_{60}$+$f_{100}$) ergs s$^{-1}$ \citep{SM96}.
\begin{longtable}{*{10}{r}}
\caption{Target information}
\label{tb:info}
\hline
\hline
Object&z&D$_{\rm L}$&Class&$f_{12}$&$f_{25}$&$f_{60}$&$f_{100}$&$f_{10.8}$&$logL_{IR
}$ \\
(1)&(2)&(3)&(4)&(5)&(6)&(7)&(8)&(9)&(10)\\
\hline
\multicolumn{10}{c}{Palomar sample} \\ \hline
NGC676           & 0.0050 & 20.2 & Sy2 &    0.04 &    0.06 &  0.27 &   0.80 & ... & 
42.3 \\
NGC1167          & 0.0165 & 66.8 & Sy2 & $<$0.03 & $<$0.04 &  0.12 &   1.05 & ... & 
43.3 \\
NGC1275          & 0.0176 & 71.2 & Sy1 &    0.93 &    3.02 &  7.09 &   7.60 & ... & 
44.8 \\
NGC1358          & 0.0134 & 54.3 & Sy2 & $<$0.08 & $<$0.12 &  0.38 &   0.93 & ... & 
43.4 \\
NGC3982$^{b, c}$ & 0.0035 & 14.2 & Sy2 &    0.47 &    0.97 &  7.18 &  16.20 & $<$0.1
1 & 43.3 \\
NGC4258$^b$      & 0.0015 &  6.2 & Sy2 &    2.25 &    2.81 &  2.81 &  78.40 & ... & 
43.0 \\
NGC4579$^{b, c}$ & 0.0050 & 20.3 & Sy2 &    1.12 &    0.78 &  5.93 &  21.40 &0.08(0.
06$^*$) & 43.7 \\
NGC5194$^{b, c}$ & 0.0020 &  8.0 & Sy2 &    7.21 &    9.56 & 97.40 & 221.00 & $<$0.0
7 & 43.9 \\
NGC5273$^b$      & 0.0036 & 14.4 & Sy1 &    0.12 &    0.29 &  0.90 &  1.560 & ... & 
42.5 \\
NGC6764          & 0.0080 & 32.4 & Sy2 &    0.54 &    1.33 &  6.62 &  12.40 & ... & 
44.0 \\
NGC6951          & 0.0048 & 19.1 & Sy2 &    1.34 &    2.16 & 16.20 &  41.80 & ... & 
44.0 \\
NGC7479$^b$      & 0.0079 & 32.0 & Sy2 &    1.37 &    3.86 & 14.90 &  26.70 & ... & 
44.4\\
NGC7743$^b$      & 0.0057 & 22.9 & Sy2 &    0.10 &    0.18 &  0.92 &   3.40 & ... & 
43.0\\
\hline
\multicolumn{10}{c}{CfA \& 12 $\mu$m} \\ \hline
NGC262$^c$        & 0.0151 &  61.3 & Sy2 &    0.31 &    0.84 & 1.29 &    1.55 &    0
.12 & 44.0\\
NGC513$^c$        & 0.0195 &  79.4 & Sy2 &    0.17 &    0.28 & 1.94 &    4.05 & $<$0
.10 & 44.2\\
NGC931$^c$        & 0.0167 &  67.5 & Sy1 &    0.61 &    0.61 & 1.32 &    4.55 &    0
.21 & 44.3\\
NGC5548$^{a,b,c}$ & 0.0172 &  67.6 & Sy1 &    0.40 &    0.77 & 1.07 &    1.61 &    0
.29 & 44.1\\
Mrk993$^{b, c}$   & 0.0155 &  62.8 & Sy1 & $<$0.13 & $<$0.13 & 0.30 &    1.32 & ... 
& 43.6\\
MCG-2-8-39$^c$    & 0.0299 & 122.3 & Sy2 &    0.20 &    0.48 & 0.51 & $<$1.42 & ... 
& 44.4\\
MCG-3-58-7$^c$    & 0.0320 & 131.1 & Sy2 &    0.28 &    0.80 & 2.42 &    3.36 &    0
.20 & 44.8\\
F03450+0055$^c$   & 0.0310 & 126.9 & Sy1 &    0.28 &    0.51 & 0.47 & $<$3.24 &    0
.10 & 44.6\\
0152+06$^b$       & 0.0174 &  70.6 & Sy2 & $<$0.08 & $<$0.16 & 0.50 &    1.15 & ... 
& 43.7\\
\hline
\endlastfoot
\end{longtable}

\clearpage


Col.(1): object name.
Col.(2): observing date in UT.
Col.(3): net on-source exposure time in minutes.
Col.(4): slit width in arcsec.
Col.(5): name of standard stars.
HR962 was observed as the standard star for F03450+0055.
However, because it was saturated in some part of the K-band spectrum,
 we used HR145, which was observed on the same day, as the K-band standard star
 for F03450+0055.
Col.(6): stellar spectral type.
Col.(7)-(9): V, K and L-band magnitude.
Col.(10): effective temperature in K.

\begin{longtable}{*{10}{r}}
\caption{Observing Log}
\label{tb:Observing Log}
\hline
\hline
Object & Date & Itime & Slit & Std star & Type & V & K & L & $T_{eff}$ \\
(1) & (2) & (3) & (4) &(5) & (6) & (7) & (8) & (9) & (10)\\
\hline
\multicolumn{10}{c}{Palomar sample} \\ \hline
NGC676  & 2009 Oct. 27 &  48 & 0.8 & HD13043  & G2V & 6.87 & 5.39 & 5.34 & 5830 \\
NGC1167 & 2009 Oct. 26 &  70 & 1.6 & HD19600  & A0V & 6.42 & 6.32 & 6.32 & 9480 \\
NGC1275 & 2009 Oct. 27 &  44 & 0.8 & HD20995  & A0V & 5.61 & 5.61 & 5.61 & 9480 \\
NGC1358 & 2009 Oct. 28 & 120 & 0.8 & HD21019  & G2V & 6.20 & 4.42 & 4.37 & 5830 \\
NGC3982 & 2005 Apr. 29 &  68 & 0.8 & HR4496   & G8V & 5.33 & 3.53 & 3.47 & 5430 \\
NGC4258 & 2005 Apr. 27 &  80 & 0.8 & HR4572   & F6V & 6.62 & 5.41 & 5.37 & 6385 \\
NGC4579 & 2005 Apr. 28 &  40 & 0.8 & HR4708   & F8V & 6.40 & 5.05 & 5.01 & 6130 \\
NGC5194 & 2005 May. 1  &  60 & 0.8 & HR4845   & G0V & 5.95 & 4.54 & 4.49 & 5930 \\
NGC5273 & 2005 Apr. 29 &  60 & 0.8 & HR4845   & G0V & 5.95 & 4.54 & 4.49 & 5930 \\
NGC6764 & 2005 May. 1  &  60 & 0.8 & HR7294   & G4V & 6.57 & 5.04 & 4.99 & 5740 \\
NGC6951 & 2005 Apr. 30 & 100 & 0.8 & HR7783   & G3V & 5.93 & 4.44 & 4.39 & 5785 \\
NGC7479 & 2009 Oct. 26 &  56 & 0.8 & HD217577 & G2V & 8.66 & 7.12 & 7.07 & 5830 \\
NGC7743 & 2009 Oct. 27 &  75 & 1.6 & HD377    & G2V & 7.60 & 6.12 & 6.07 & 5830 \\
\hline
\multicolumn{10}{c}{CfA \& 12 $\mu$m } \\ \hline
NGC262      & 2007 Aug. 31 &  80 & 0.8 & HR410  & F7V  & 6.31 & 4.99 & 4.95 & 6240 \\
NGC513      & 2007 Aug. 27 & 120 & 0.8 & HR410  & F7V  & 6.31 & 4.99 & 4.95 & 6240 \\ 
NGC931      & 2007 Aug. 26 &  96 & 0.8 & HR720  & G0V  & 5.89 & 4.48 & 4.43 & 5930 \\
NGC5548     & 2005 Apr. 29 &  20 & 0.8 & HR5346 & F8V  & 6.25 & 4.90 & 4.85 & 6130 \\ 
Mrk993      & 2007 Aug. 31 & 112 & 0.8 & HR410  & F7V  & 6.31 & 4.99 & 4.95 & 6240 \\
MCG-2-8-39  & 2007 Aug. 30 & 136 & 0.8 & HR784  & F6V  & 5.78 & 4.57 & 4.53 & 6385 \\
MCG-3-58-7  & 2007 Aug. 30 &  80 & 0.8 & HR8457 & F6V  & 6.09 & 4.88 & 4.84 & 6385 \\
F03450+0055 & 2007 Aug. 28 & 112 & 0.8 & HR962  & F8V  & 5.06 & 3.71 & 3.67 & 6130 \\
            &              &     &     & HR145  & F7V  & 6.41 & 5.09 & 5.05 & 6240 \\
0152+06     & 2007 Aug. 29 & 120 & 0.8 & HR508  & G3Va & 6.27 & 4.78 & 4.73 & 5785 \\
\hline
\endlastfoot
\end{longtable}

\clearpage


Col.(1): object name.
Col.(2): nuclear 3.3$\mu$m PAH emission flux
 in $10^{-14}$ergs s$^{-1}$cm$^{-2}$.
Col.(3): decimal logarithm of nuclear 3.3$\mu$m PAH emission luminosity
 in ergs s$^{-1}$.
Col.(4): rest-frame equivalent width of 3.3$\mu$m PAH emission line in nm.
Col.(5): spectroscopic $CO_{spec}$ index at $\lambda_{rest}$=2.3-2.4 $\mu$m in
 the rest frame.
 The definition appears in the text \S \ref{sec:K-band spectrum}.
Col.(6): $K-L$ color (mag) measured in this work.
 K- and L-band magnitudes were derived by integrating the flux from
 1.97--2.38$\mu$m and 3.26--3.83$\mu$m, respectively.
Col.(7): nuclear star-formation rate in ${\rm M}_\odot$ $yr^{-1}$ inside our
 slit sizes.

\begin{longtable}{*{7}{r}}
\caption{Properties of the nuclear 3.3 $\mu$m PAH emission and CO absorption feature
s}
\label{tb:results}
\hline
\hline 
Object & $f_{3.3PAH}$ &$log(L_{3.3PAH})$ & $EW_{3.3PAH}$ &$CO_{spec}$ & $K-L$ &SFR \\
 & ($\times 10^{14}$ergs s$^{-1}$cm$^{-2}$) & (ergs s$^{-1}$) & (nm) &  & (mag) & (M$_{\odot}$ yr$^{-1}$)\\
(1) & (2) & (3) & (4) & (5) & (6) & (7)\\ 
\hline
\multicolumn{7}{c}{Palomar sample} \\ \hline
NGC676      & $<$11.58 & $<$39.73 & $<$10.79 & 0.20 & $-$0.08 & $<$0.24 \\
NGC1167     & $<$ 7.85 & $<$40.60 & $<$18.02 & 0.23 &    0.14 & $<$1.80 \\
NGC1275     & $<$31.58 & $<$41.26 & $<$ 4.26 & 0    &    1.79 & $<$8.22 \\
NGC1358     & $<$ 9.21 & $<$40.49 & $<$21.53 & 0.18 & $-$0.04 & $<$1.39 \\
NGC3982     & $<$ 6.17 & $<$39.15 & $<$28.42 & 0.16 &    0.27 & $<$0.06 \\
NGC4258     & $<$13.57 & $<$38.77 & $<$ 4.17 & 0.16 &    0.95 & $<$0.03 \\
NGC4579     & $<$ 3.22 & $<$39.17 & $<$ 1.37 & 0.18 &    0.61 & $<$0.07 \\
NGC5194     &     6.99 &    38.71 &    12.08 & 0.20 &    0.14 & 0.02 \\
NGC5273     &     8.55 &    39.30 &     9.57 & 0.11 &    0.72 & 0.09 \\
NGC6764     &    40.50 &    40.68 &    91.72 & 0.22 &    0.54 & 2.18 \\
NGC6951     & $<$ 6.05 & $<$39.40 & $<$19.64 & 0.24 &    0.15 & $<$0.11 \\ 
NGC7479     & $<$16.53 & $<$40.28 & $<$10.77 & 0.10 &    1.59 & $<$0.87 \\
NGC7743     & $<$ 4.39 & $<$39.42 & $<$ 4.76 & 0.25 &    0.28 & $<$0.12 \\
\hline
\multicolumn{7}{c}{CfA \& 12 $\mu$m} \\
\hline
NGC262      & $<$13.01 & $<$40.74 & $<$ 1.94 & 0.02 & 1.98 & $<$2.51 \\
NGC513      & $<$ 4.35 & $<$40.49 & $<$ 5.92 & 0.11 & 1.11 & $<$1.41 \\
NGC931      & $<$11.15 & $<$40.76 & $<$ 1.34 & 0    & 1.74 & $<$2.61 \\
NGC5548     & $<$14.63 & $<$40.88 & $<$ 5.09 & 0.01 & 1.54 & $<$3.43 \\
Mrk993      & $<$ 4.72 & $<$40.32 & $<$10.98 & 0.16 & 0.57 & $<$0.96 \\
MCG-2-8-39  & $<$ 6.24 & $<$41.02 & $<$21.46 & 0.16 & 1.20 & $<$4.80 \\
MCG-3-58-7  &    10.64 &    41.32 &     1.96 & 0.02 & 1.84 & 9.40 \\
F03450+0055 &     3.36 &    40.79 &     0.56 & 0.01 & 1.65 & 2.78 \\
0152+06     & $<$ 2.35 & $<$40.12 & $<$ 6.70 & 0    & 1.02 & $<$0.60 \\
\hline
\endlastfoot
\end{longtable}

\clearpage


\begin{figure*}
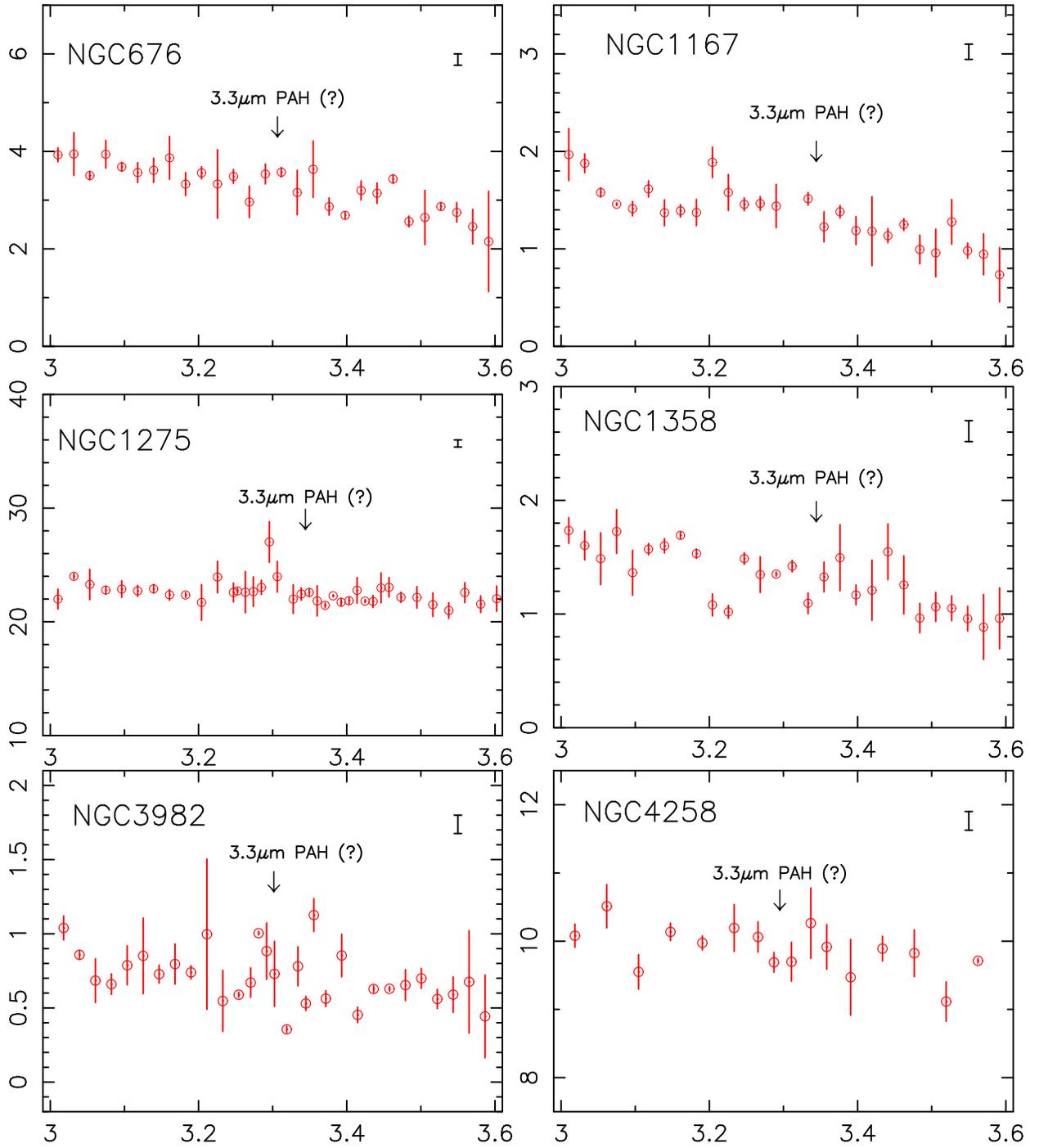

  \begin{center}
\includegraphics[scale=.4, angle=-90]{N676L.ps}    
\includegraphics[scale=.4, angle=-90]{N1167L.ps}    \\
\includegraphics[scale=.4, angle=-90]{N1275L.ps}    
\includegraphics[scale=.4, angle=-90]{N1358L.ps}    \\
\includegraphics[scale=.4, angle=-90]{N3982L.ps}    
\includegraphics[scale=.4, angle=-90]{N4258L.ps}    \\
\caption{Infrared 3.0-3.6$\mu$m slit spectra (L-band) of 22 Seyfert nuclei.
 The abscissa and ordinate are the observed wavelength in $\mu$m and
 $F_{\lambda}$ in $10^{-15}\rm{Wm}^{-2}\mu \rm{m}^{-1}$, respectively.
 The length of the bar at the upper right of each figure indicates the
 1$\sigma$ error for the continuum.
 The phrase "3.3 $\mu$m PAH" ("3.3 $\mu$m PAH(?)") indicates the detectable
 (undetectable) 3.3 $\mu$m PAH emission.
 The best-fit Gaussian profile is shown for objects with detected 3.3 $\mu$m
 PAH emission.
 For our 3.3 $\mu$m PAH detection criteria, please refer to
 \S \ref{sec:L-band spectrum} in the text.}
\label{fig:L-band_spectra}
\end{center}
\end{figure*}

\setcounter{figure}{0}
\begin{figure*}
\begin{center}
\includegraphics[scale=.4, angle=-90]{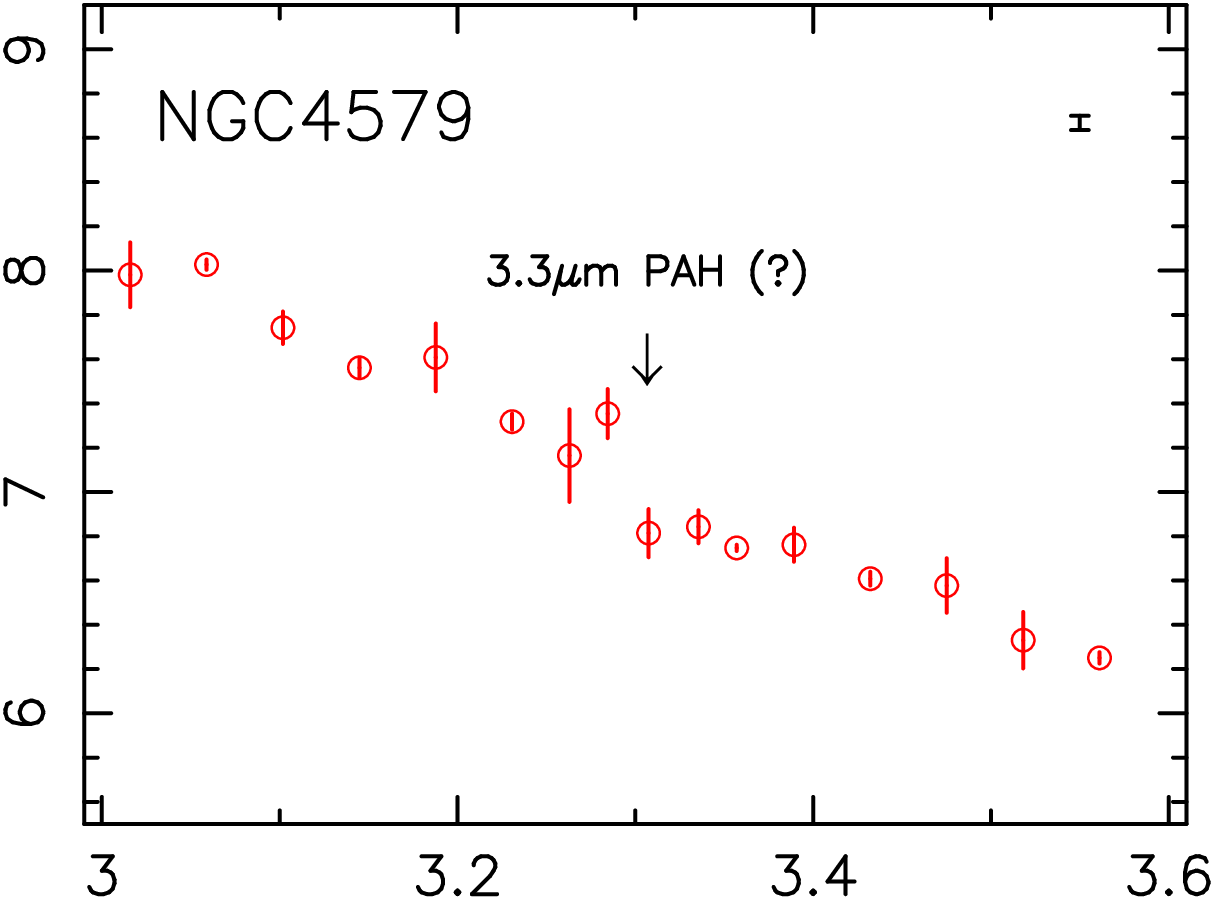}    
\includegraphics[scale=.4, angle=-90]{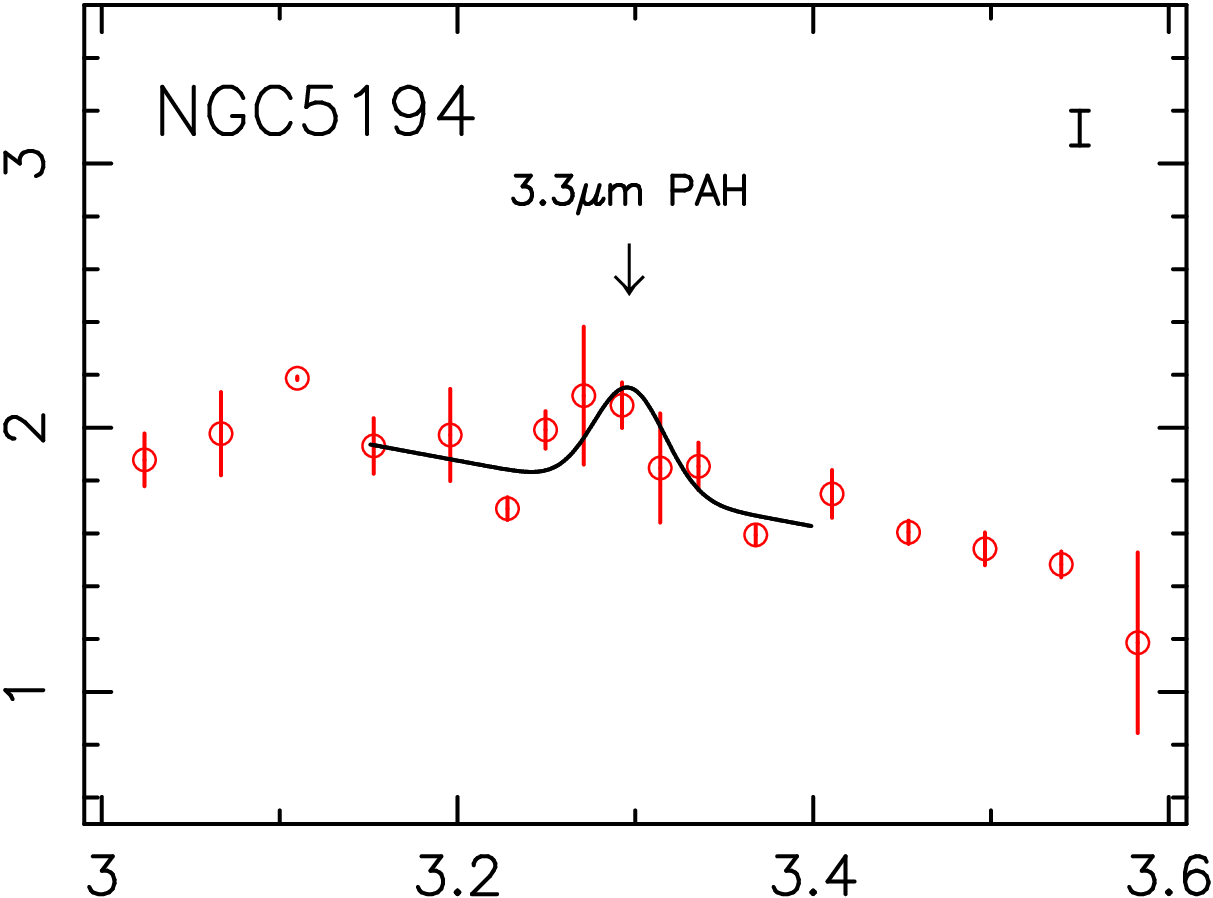}    
\includegraphics[scale=.4, angle=-90]{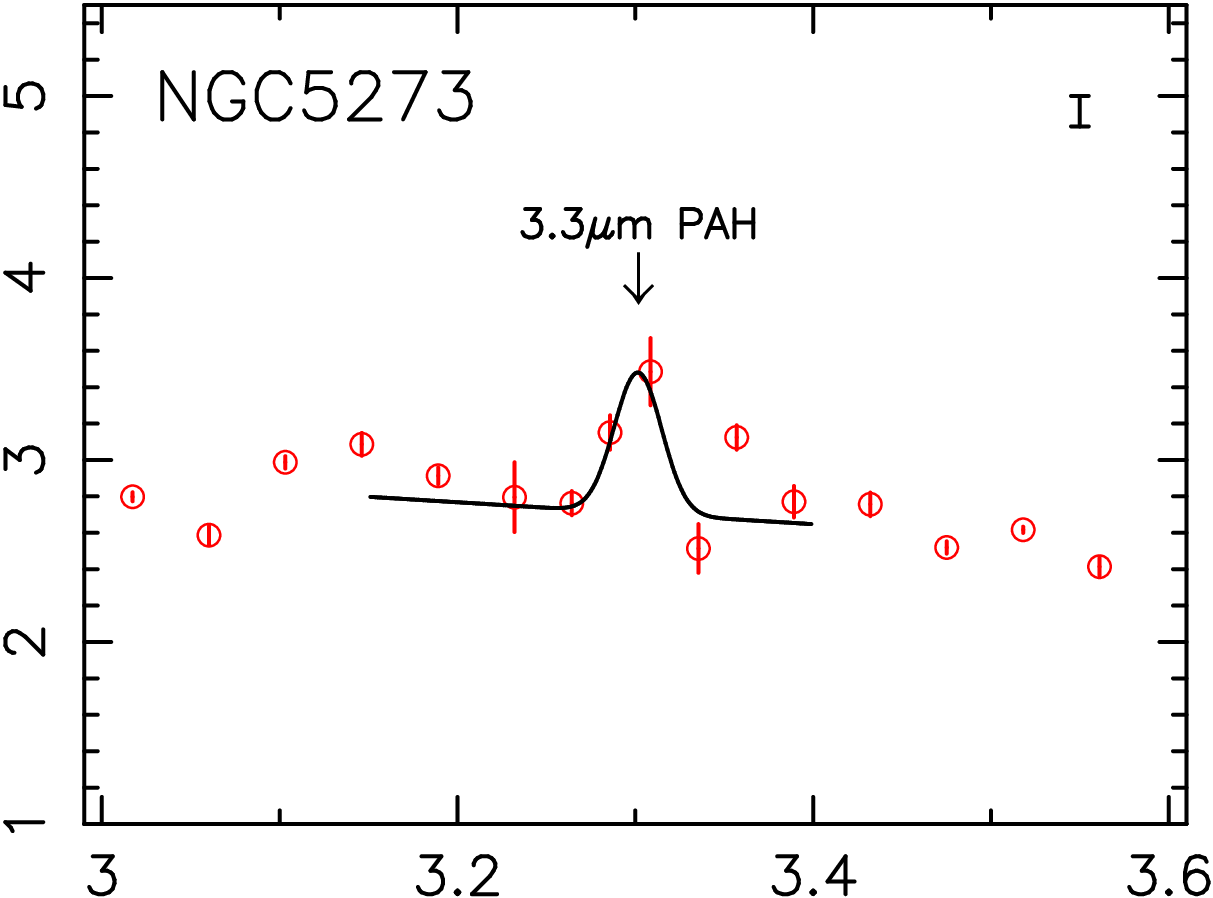}    
\includegraphics[scale=.4, angle=-90]{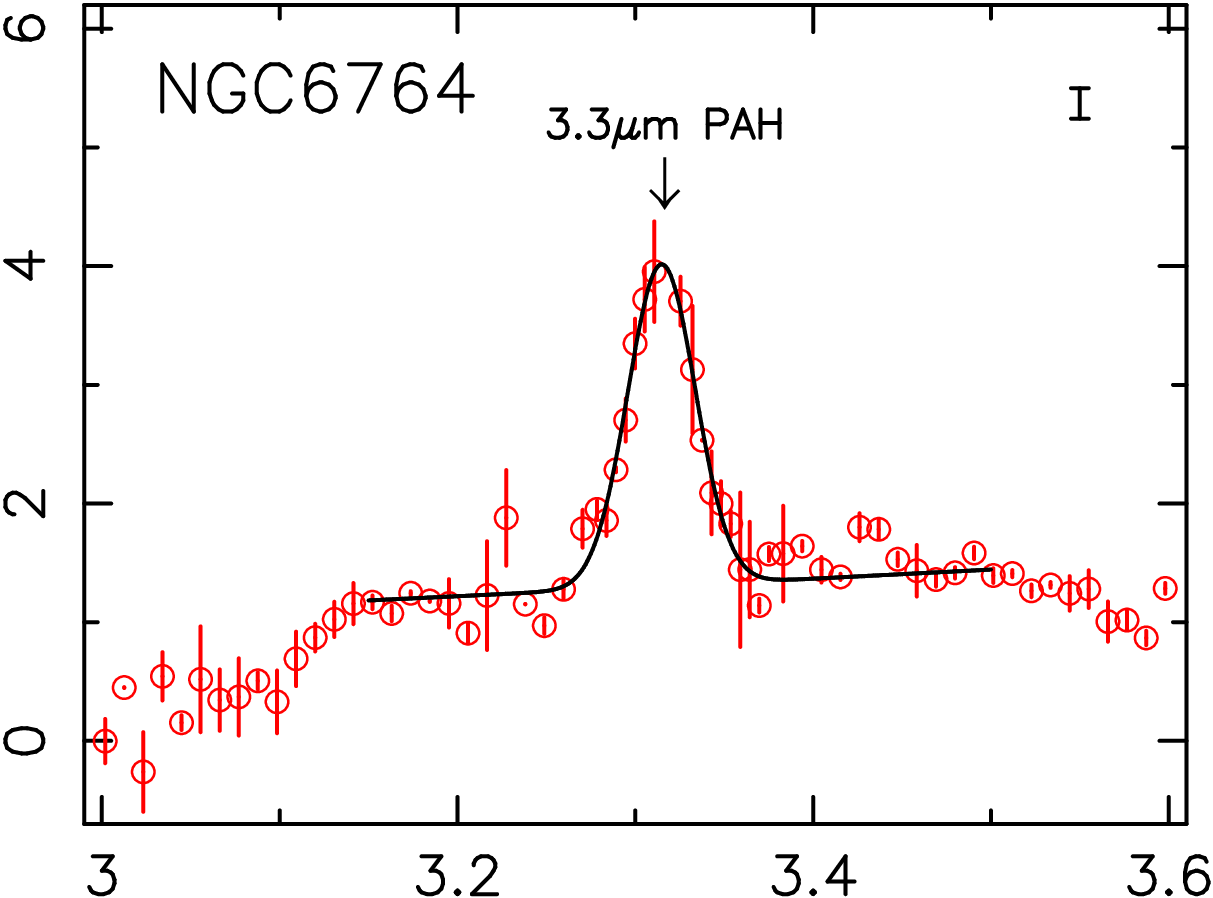}    
\includegraphics[scale=.4, angle=-90]{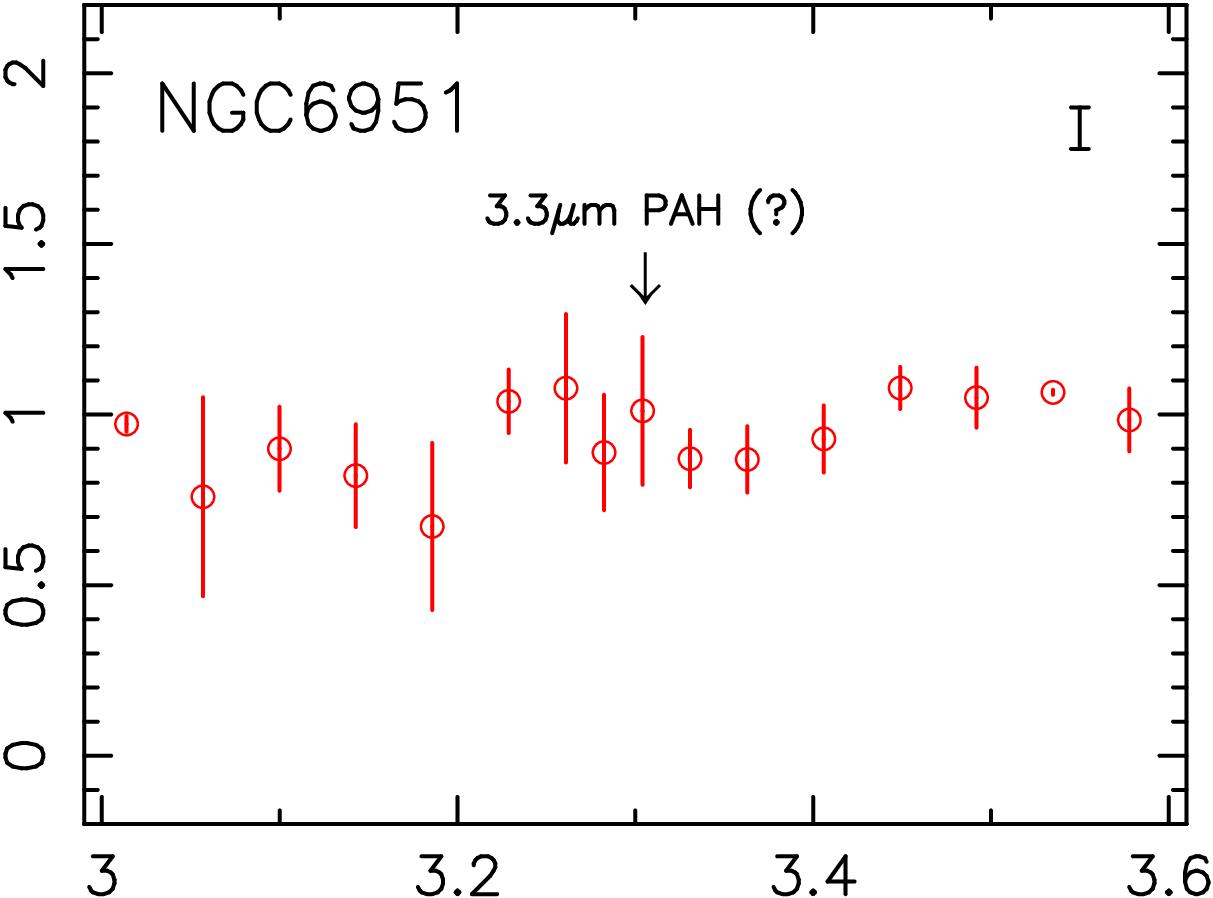}    
\includegraphics[scale=.4, angle=-90]{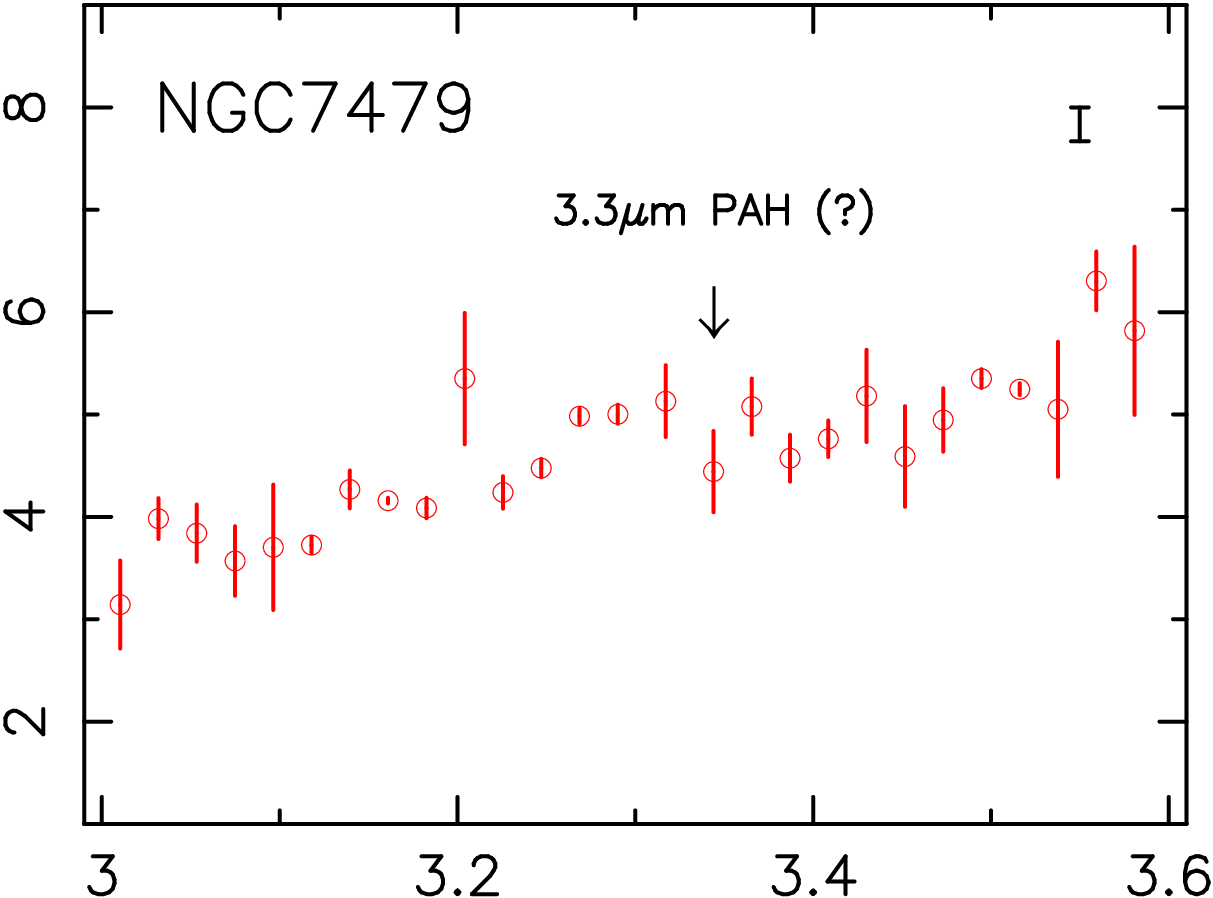}    
\includegraphics[scale=.4, angle=-90]{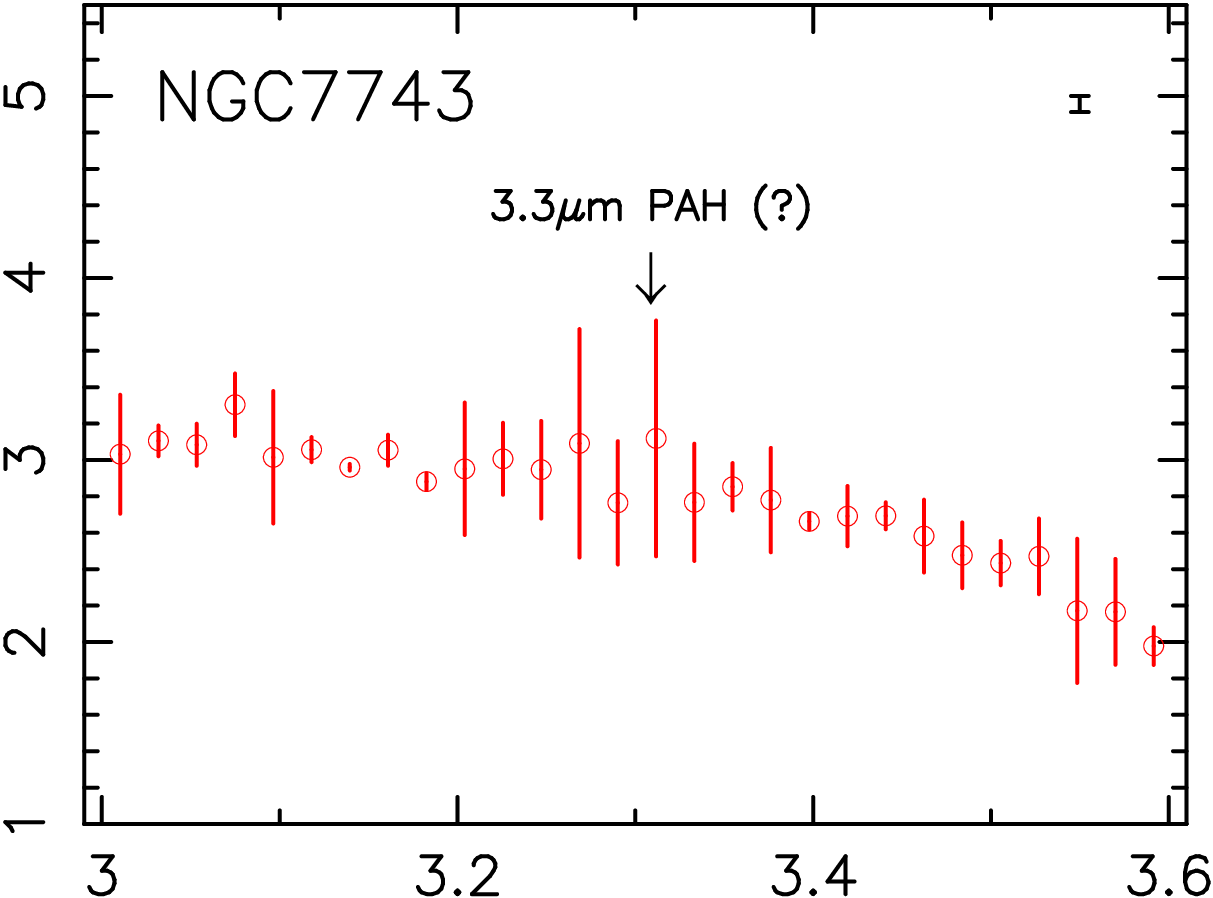}    
\includegraphics[scale=.4, angle=-90]{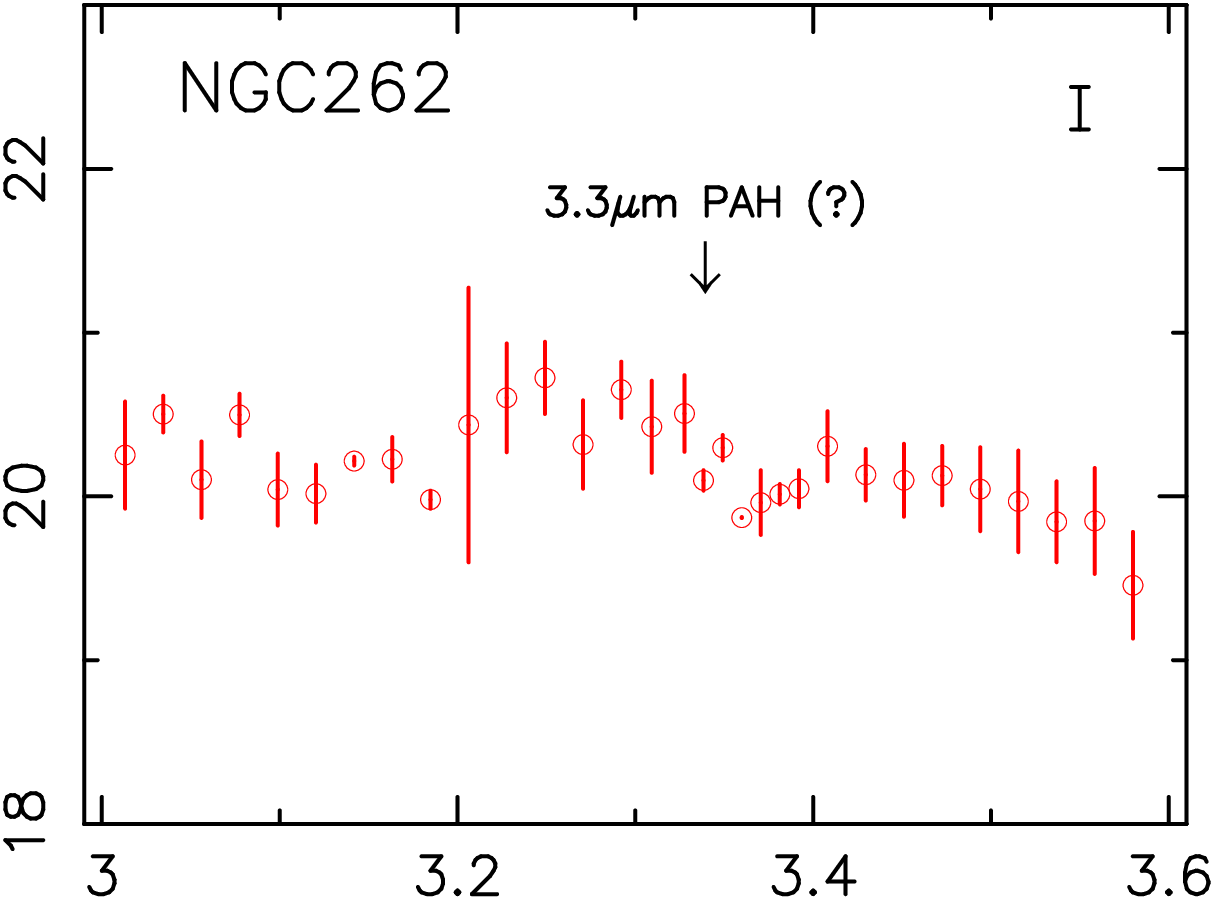}    
\caption{--Continued}
\end{center}
\end{figure*}

\setcounter{figure}{0}
\begin{figure*}
\begin{center}
\includegraphics[scale=.4, angle=-90]{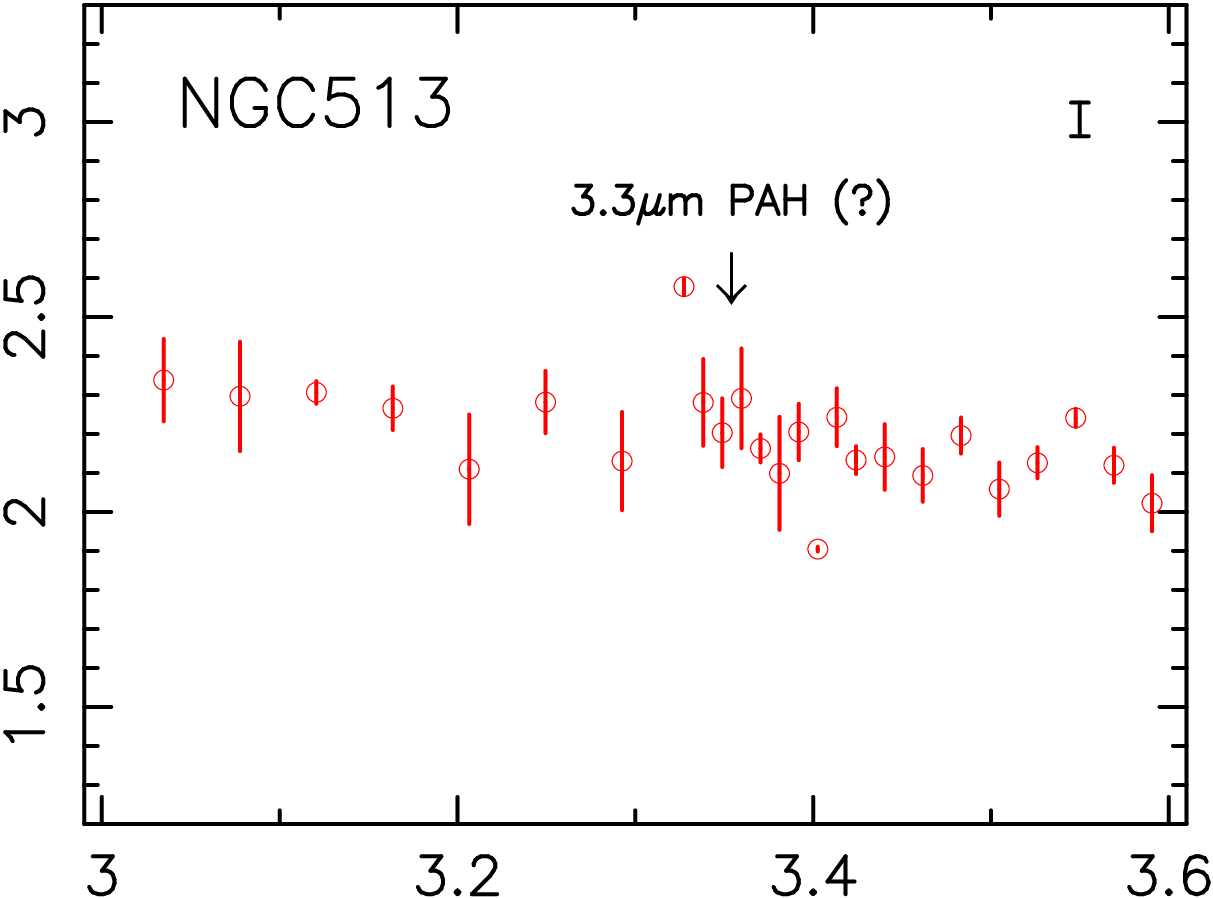}    
\includegraphics[scale=.4, angle=-90]{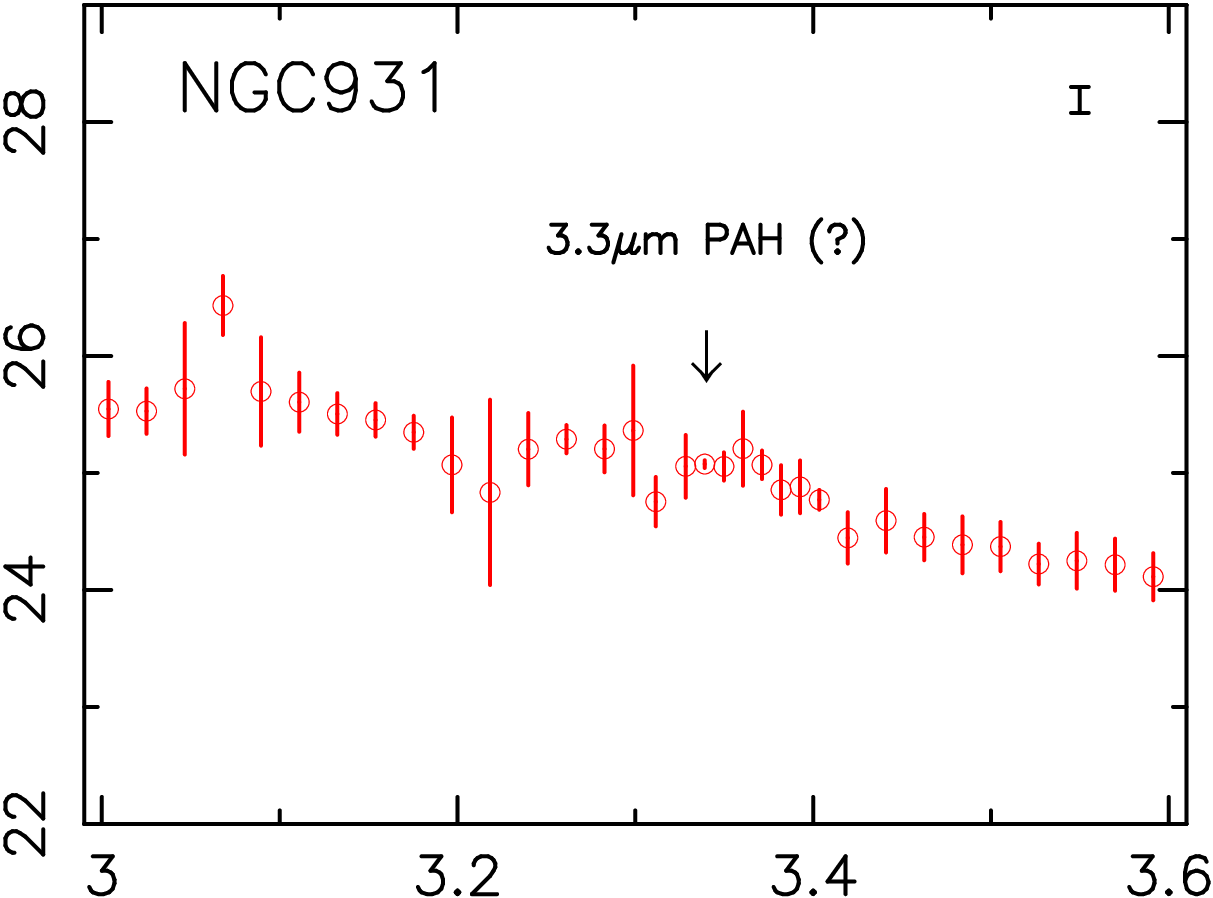}    
\includegraphics[scale=.4, angle=-90]{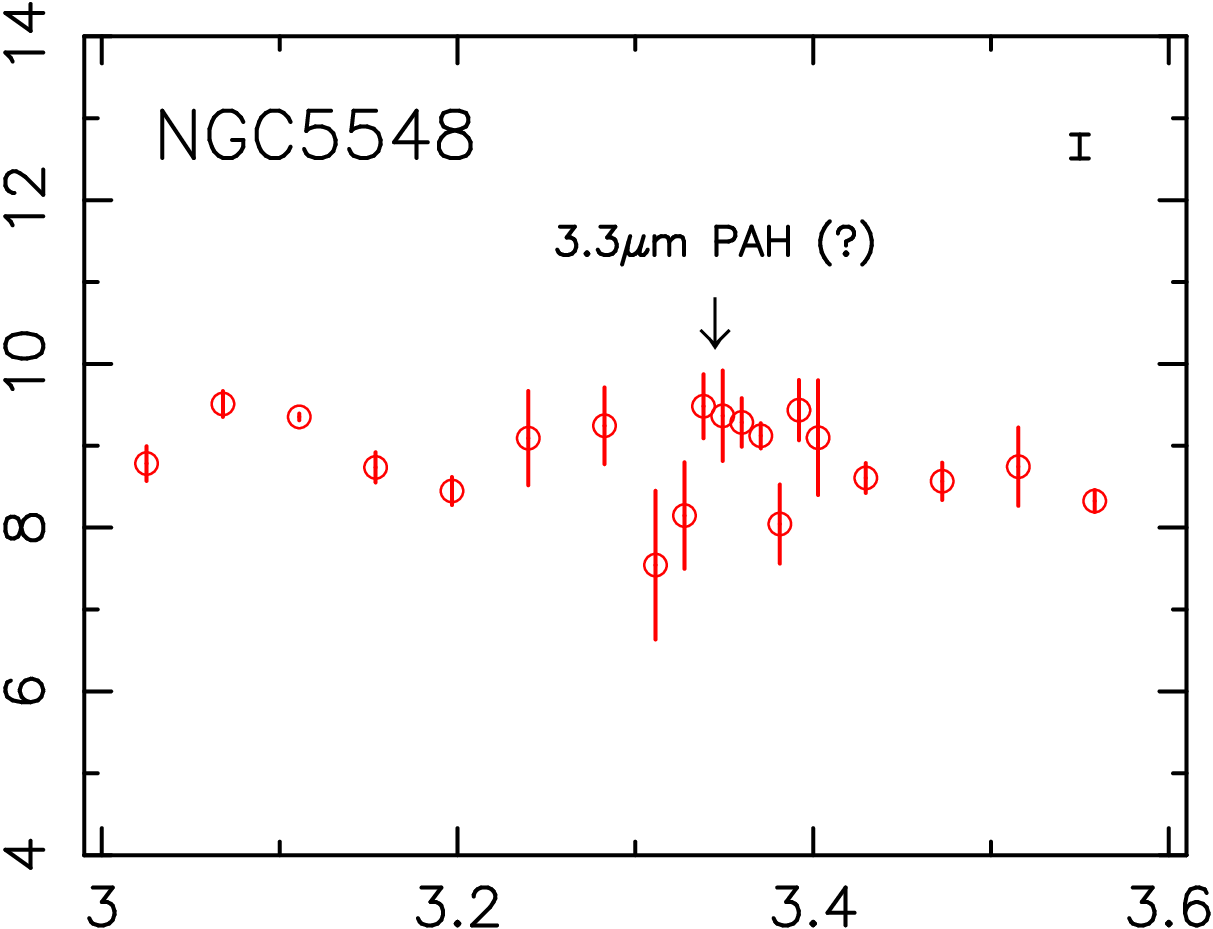}    
\includegraphics[scale=.4, angle=-90]{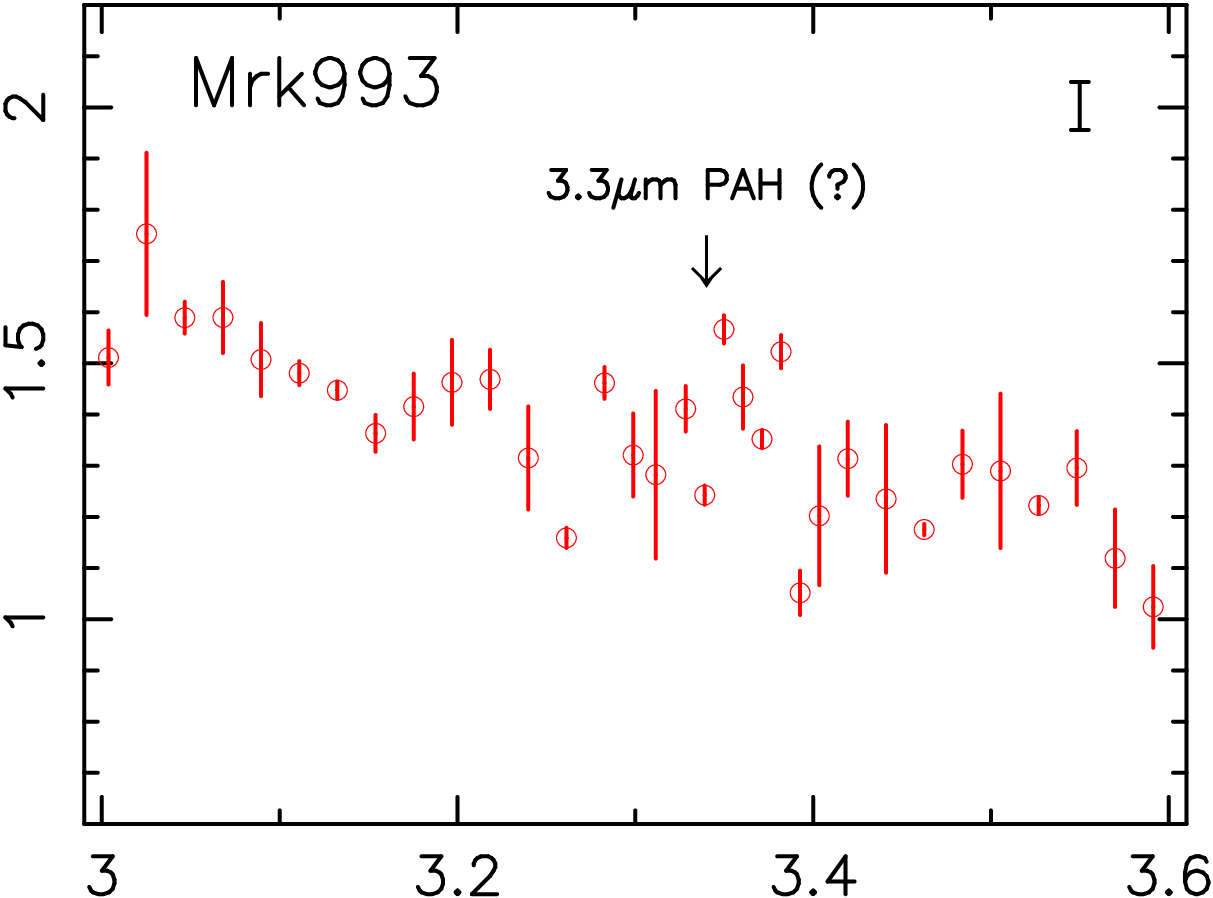}    
\includegraphics[scale=.4, angle=-90]{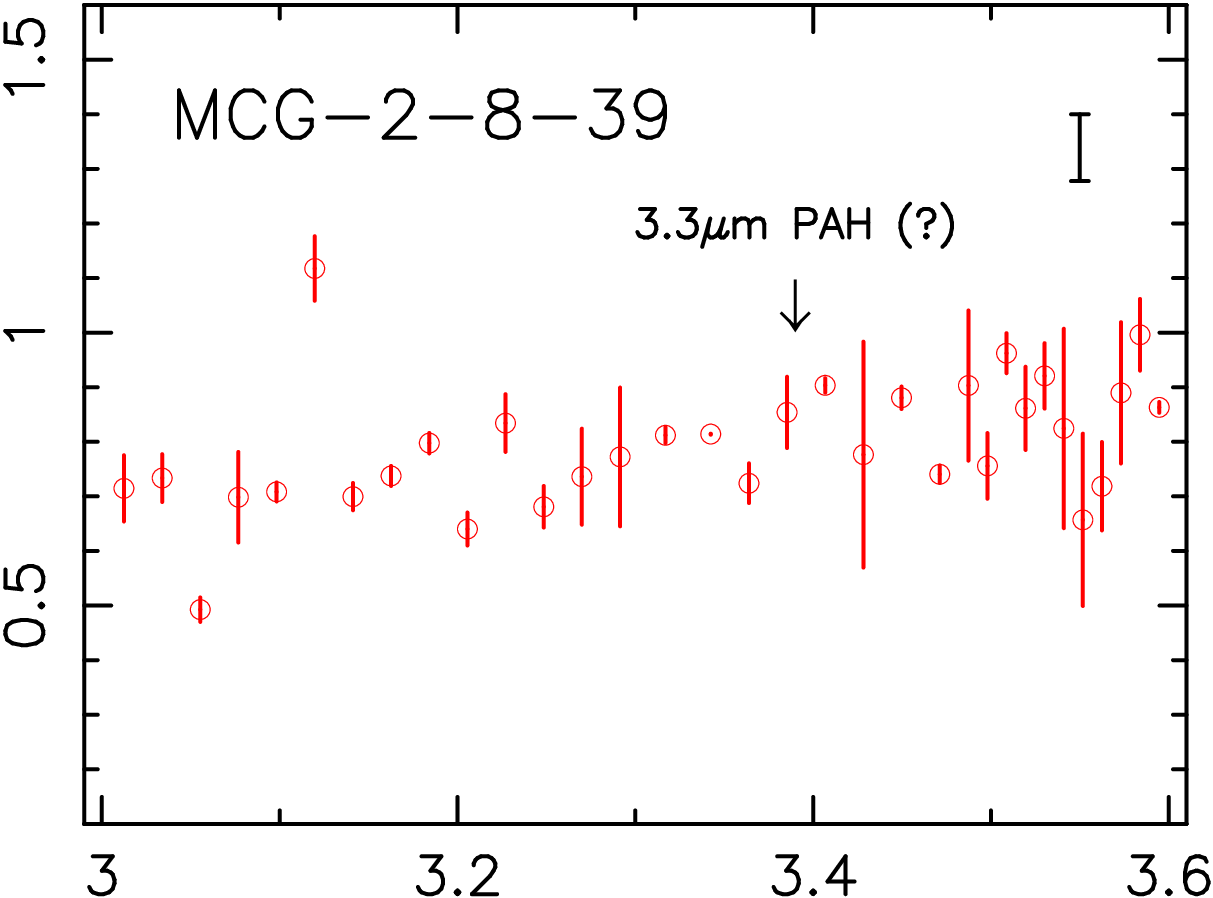}    
\includegraphics[scale=.4, angle=-90]{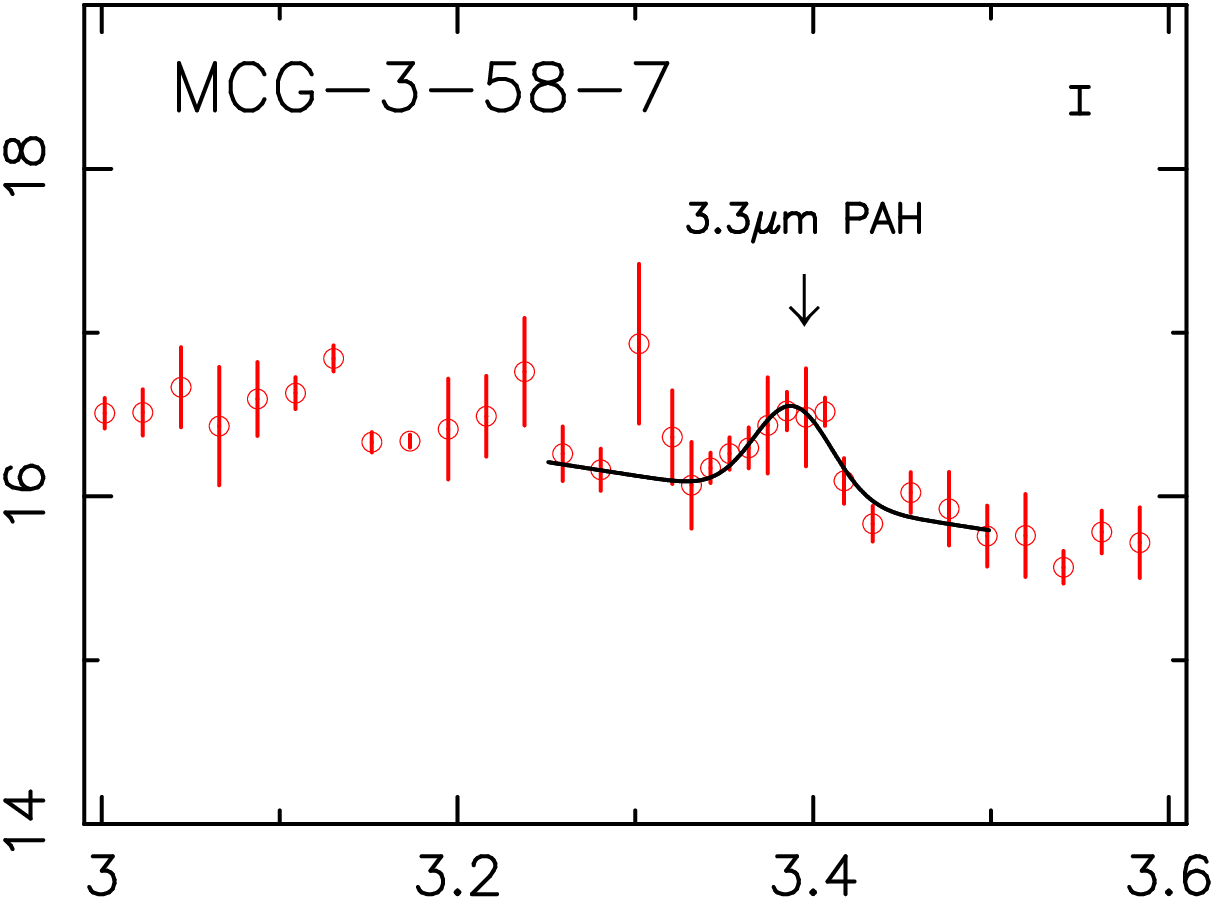}    
\includegraphics[scale=.4, angle=-90]{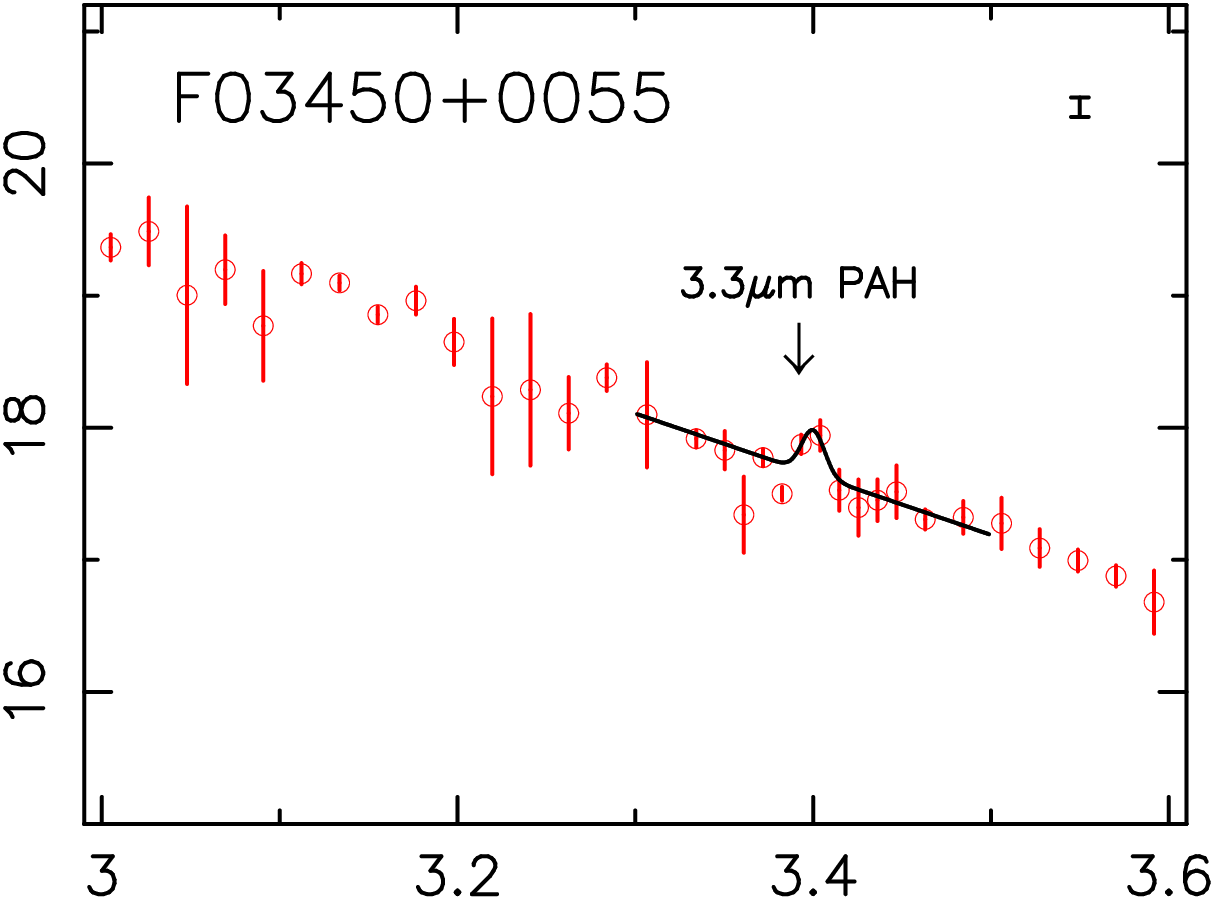}    
\includegraphics[scale=.4, angle=-90]{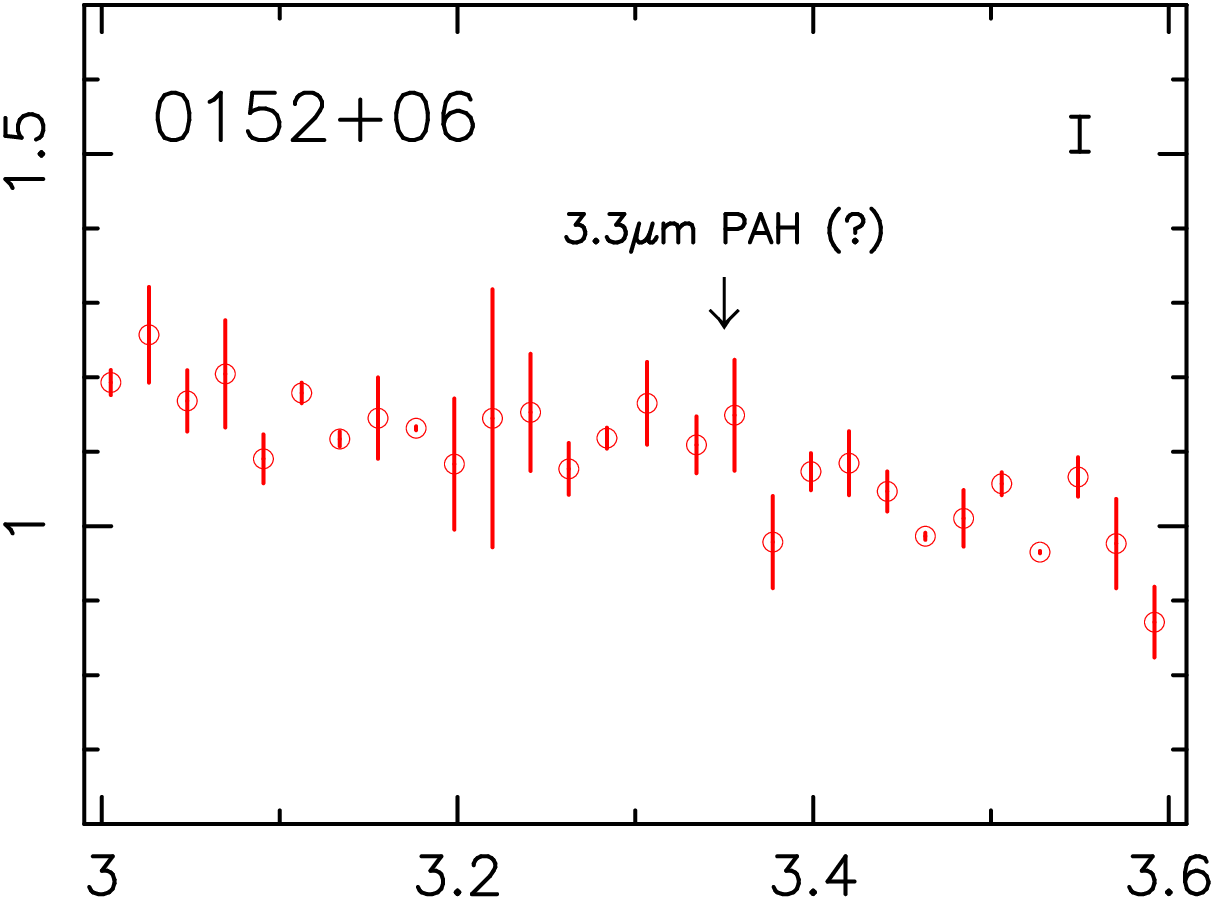}    
\caption{--Continued}
\end{center}
\end{figure*}

\clearpage


\begin{figure*}
  \begin{center}
\includegraphics[scale=.4, angle=-90]{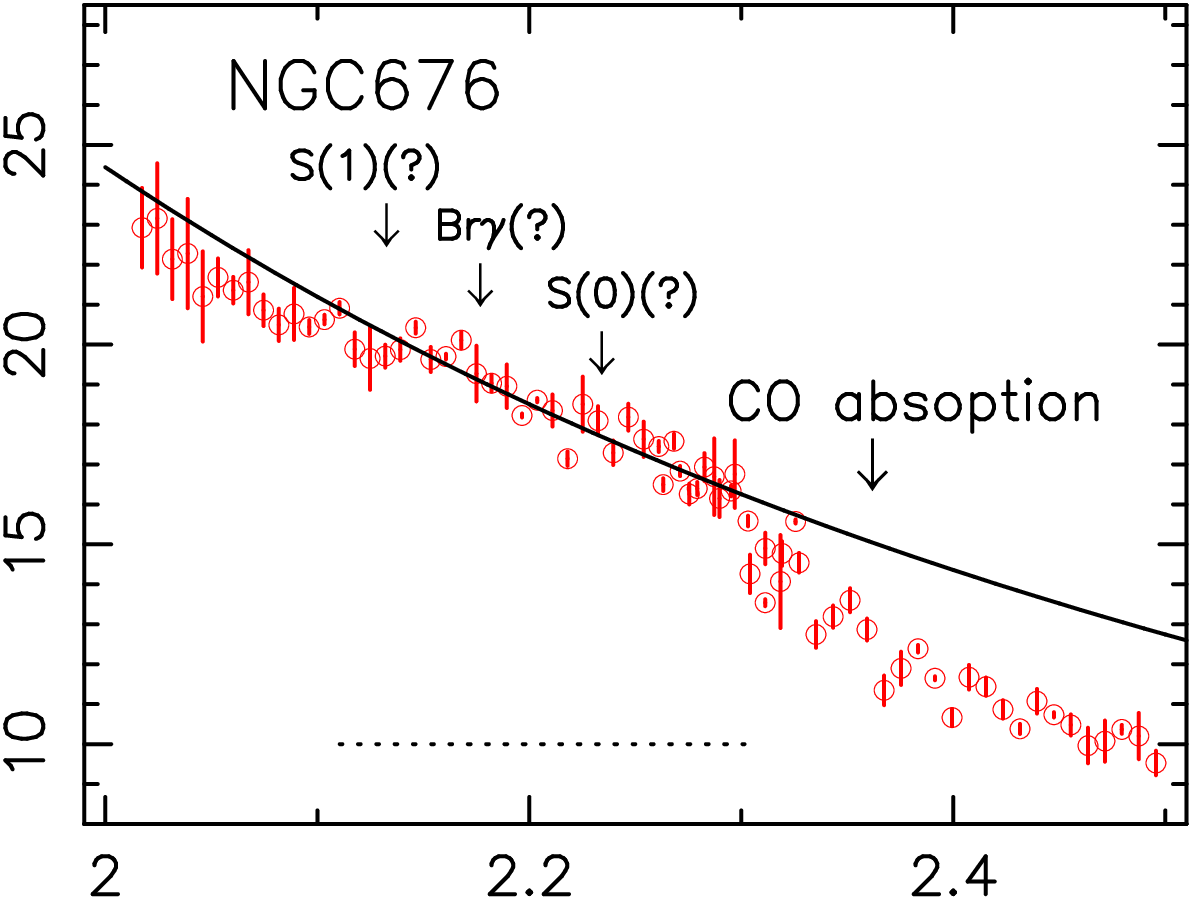}    
\includegraphics[scale=.4, angle=-90]{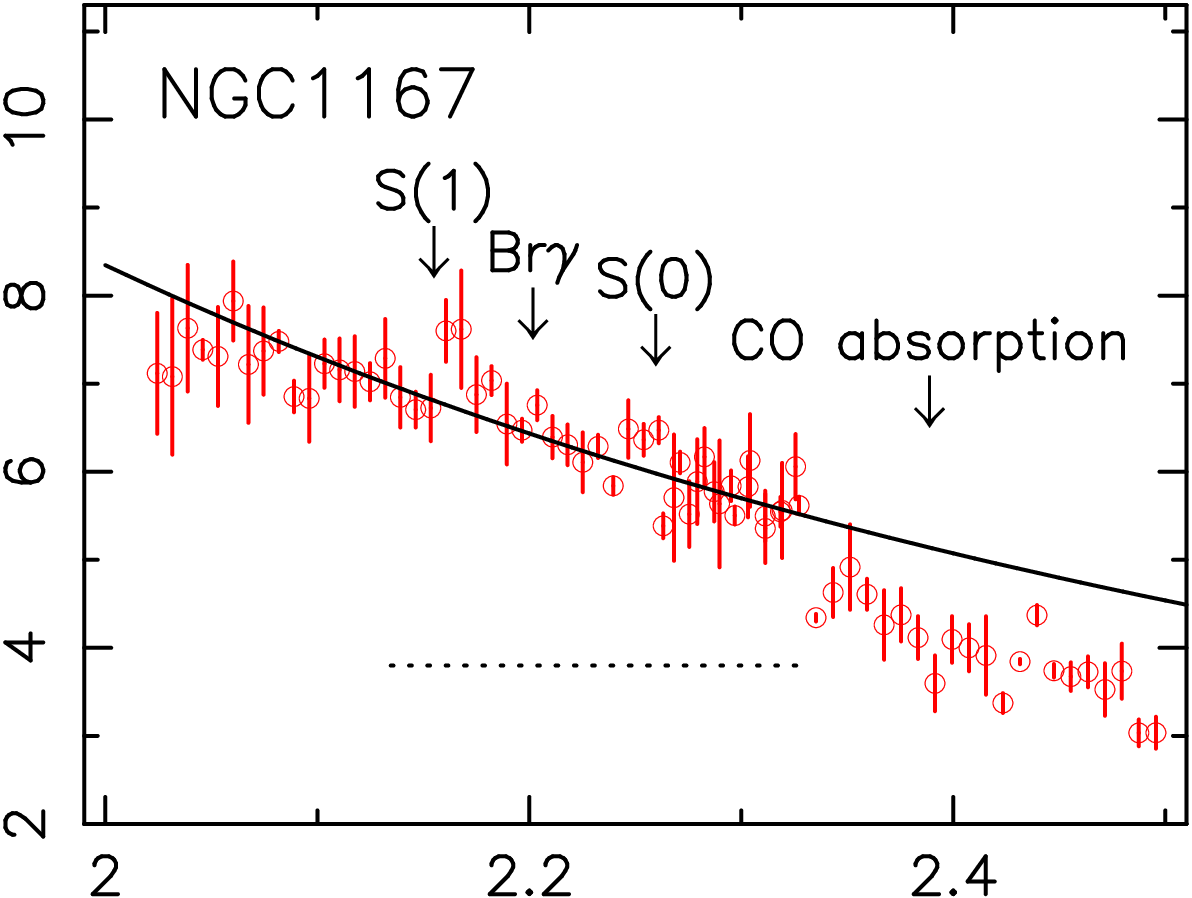}    \\
\includegraphics[scale=.4, angle=-90]{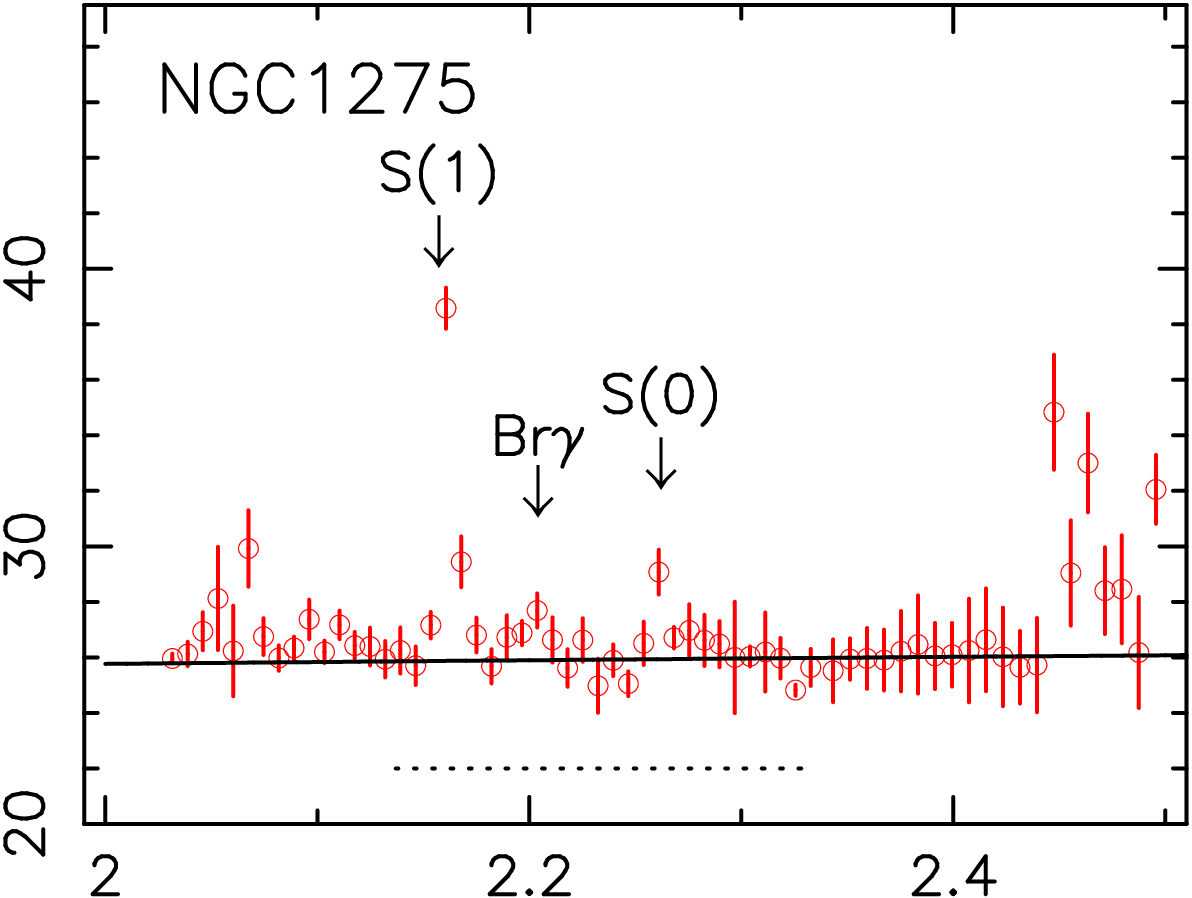}    
\includegraphics[scale=.4, angle=-90]{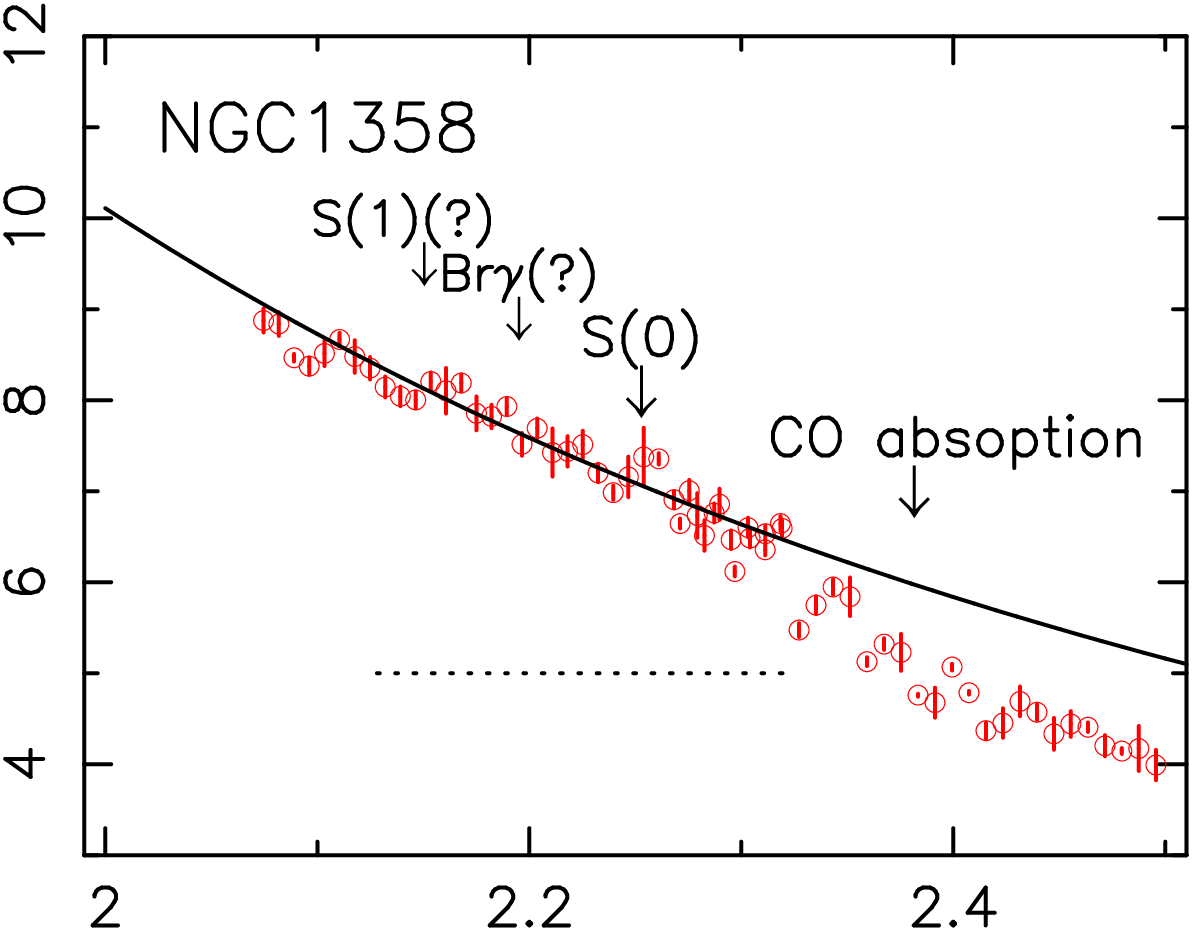}    \\
\includegraphics[scale=.4, angle=-90]{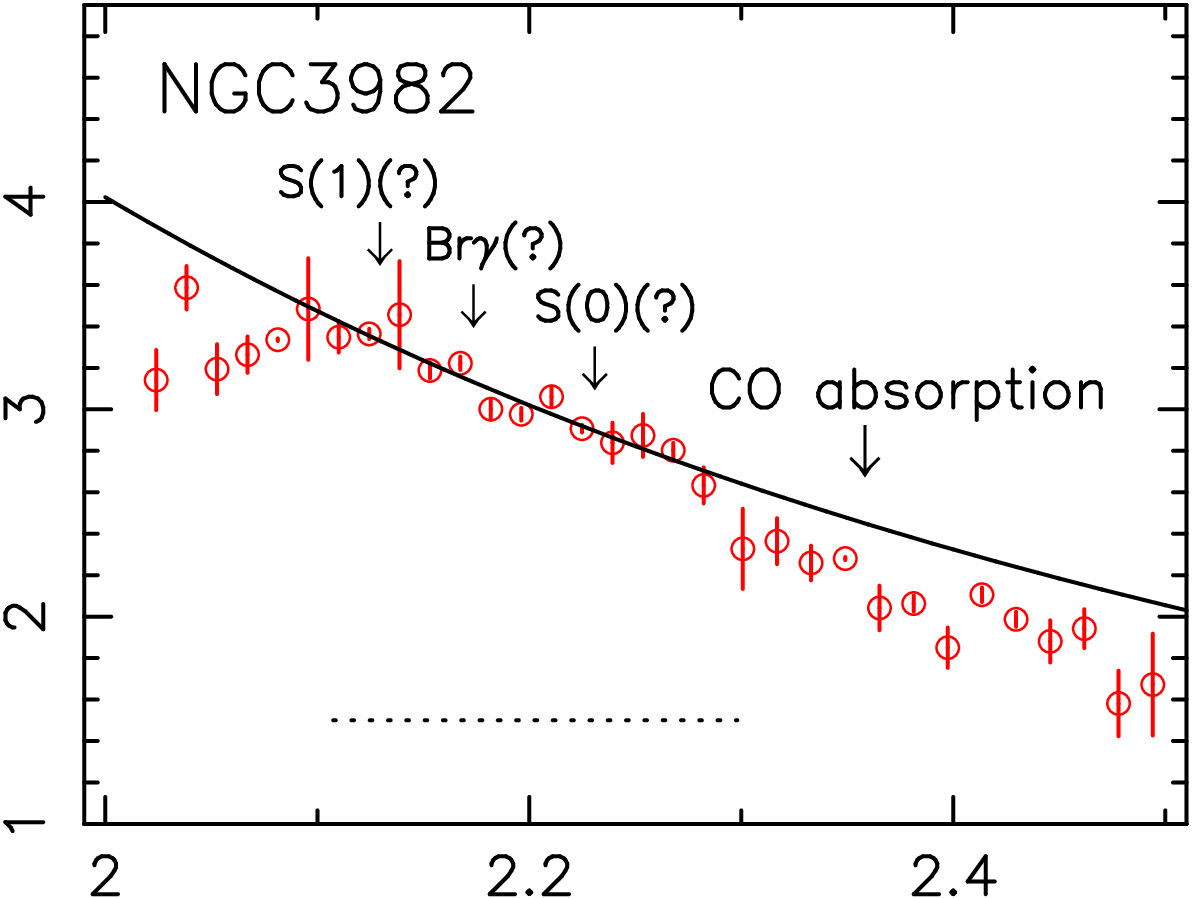}    
\includegraphics[scale=.4, angle=-90]{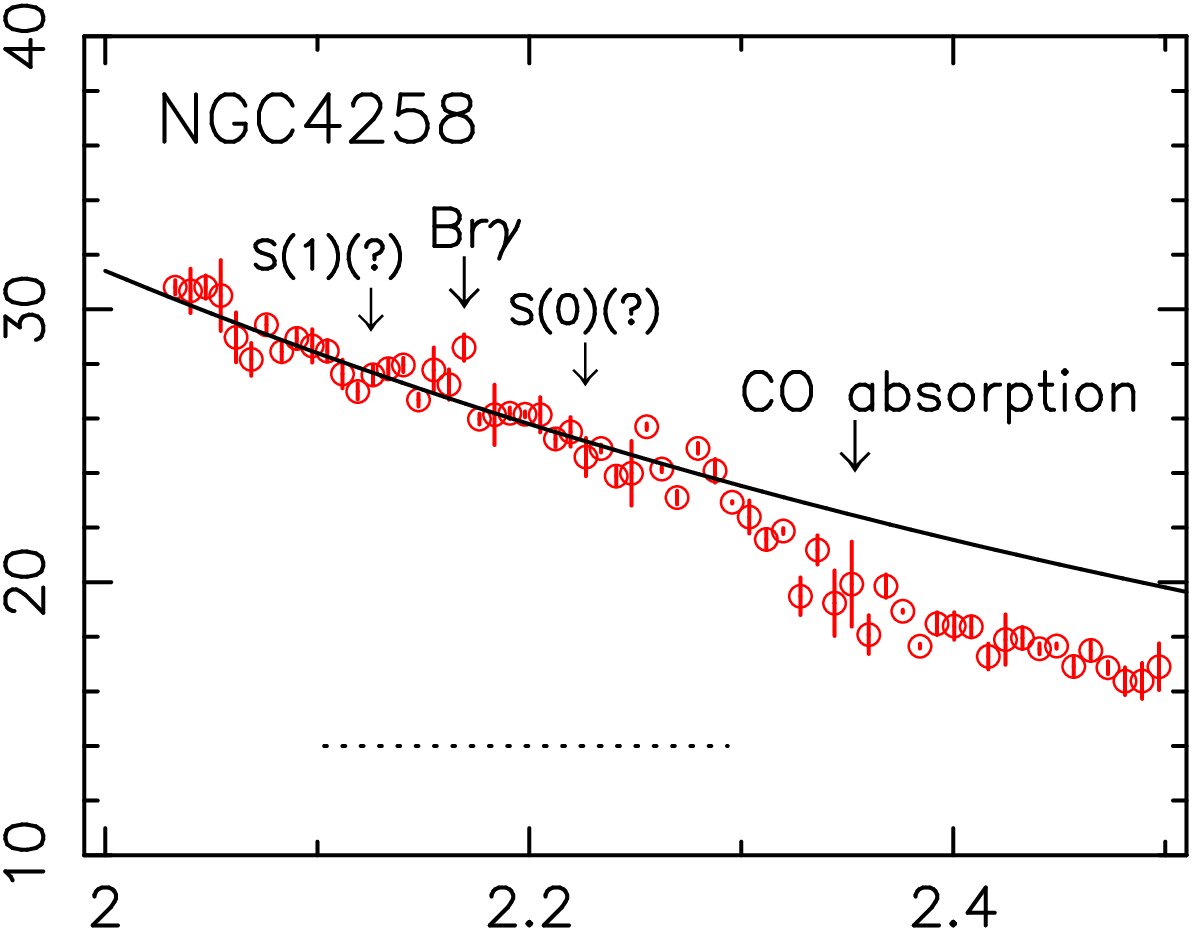}    \\
\caption{Infrared 2.0-2.5$\mu$m slit spectra (K-band) of 22 Seyfert nuclei.
 The abscissa and ordinate are the same as in Figure \ref{fig:L-band_spectra}.
 The phrases "S(1)", "Br$\gamma$", and "S(0)" show the excess emission
 lines of H$_{2}$ 1-0 S(1) ($\lambda_{rest}=2.122\mu$m),
 Br$\gamma$ ($\lambda_{rest}=2.166\mu$m) and H$_{2}$ 1-0 S(0) 
 ($\lambda_{rest}=2.223\mu$m), respectively.
 Indistinct lines are marked as "S(1)(?)", "Br$\gamma$(?)", and "S(0)(?)". 
 The symbol "CO absorption" indicates a clear absorption feature at
 $\lambda_{rest}=2.3-2.4 \mu$m.}
\label{fig:K-band_spectra}
\end{center}
\end{figure*}

\setcounter{figure}{1}
\begin{figure*}
\begin{center}
\includegraphics[scale=.4, angle=-90]{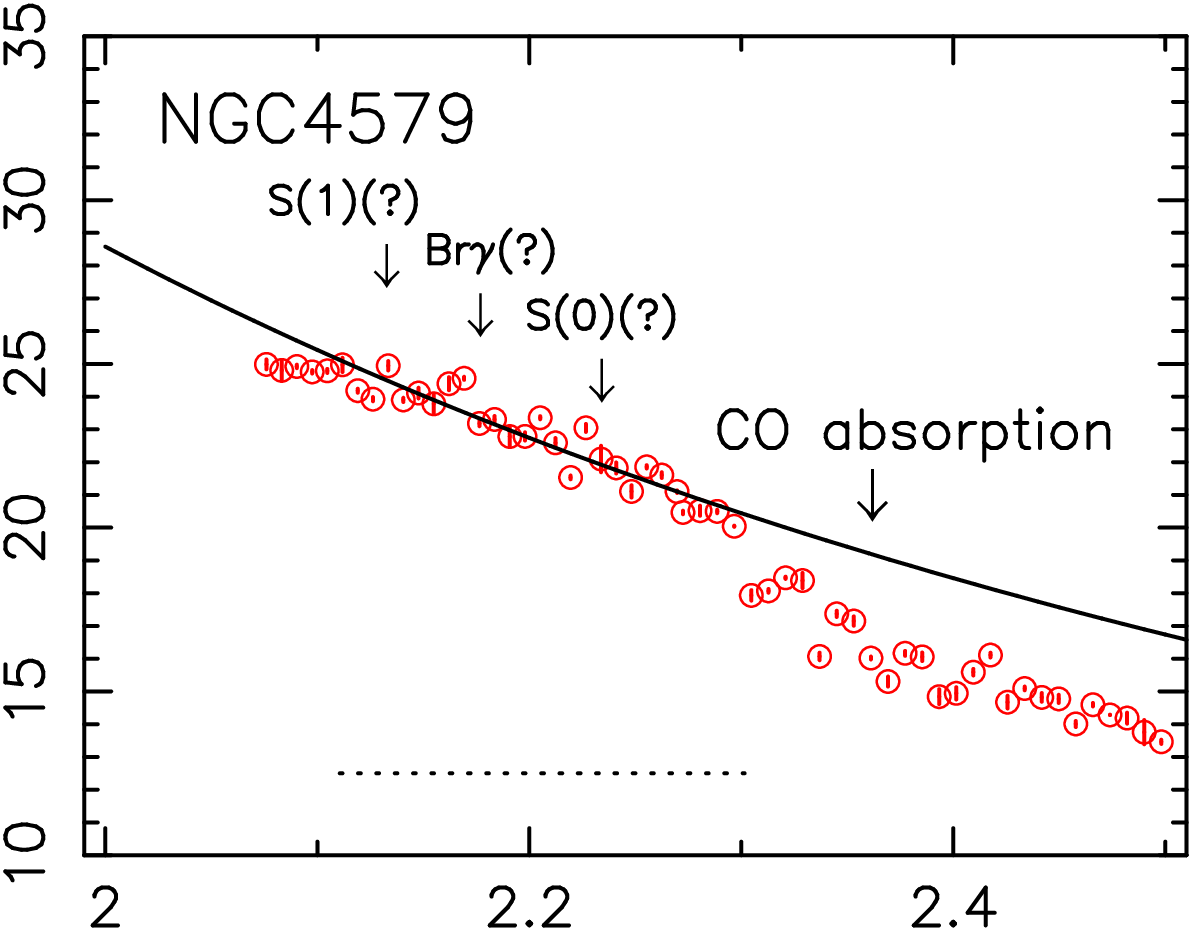}    
\includegraphics[scale=.4, angle=-90]{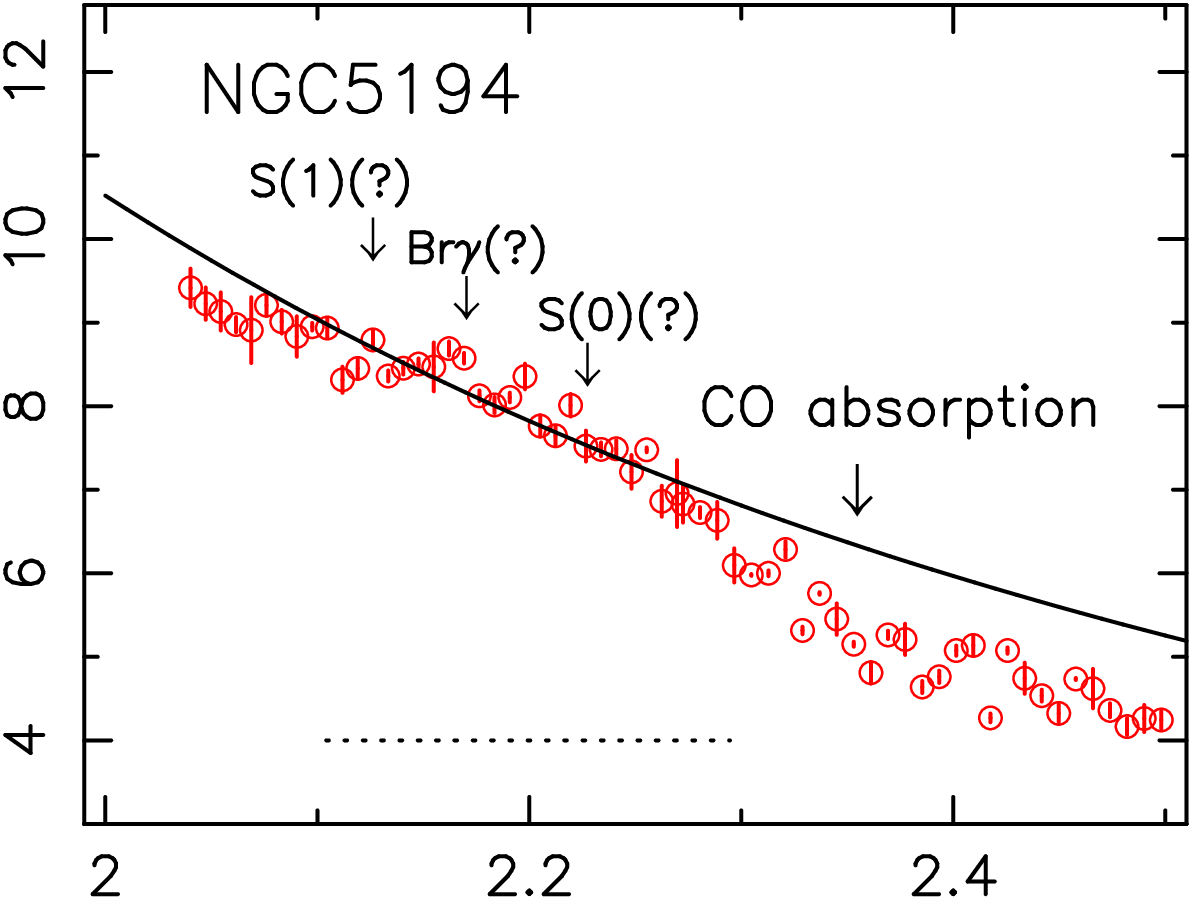}    
\includegraphics[scale=.4, angle=-90]{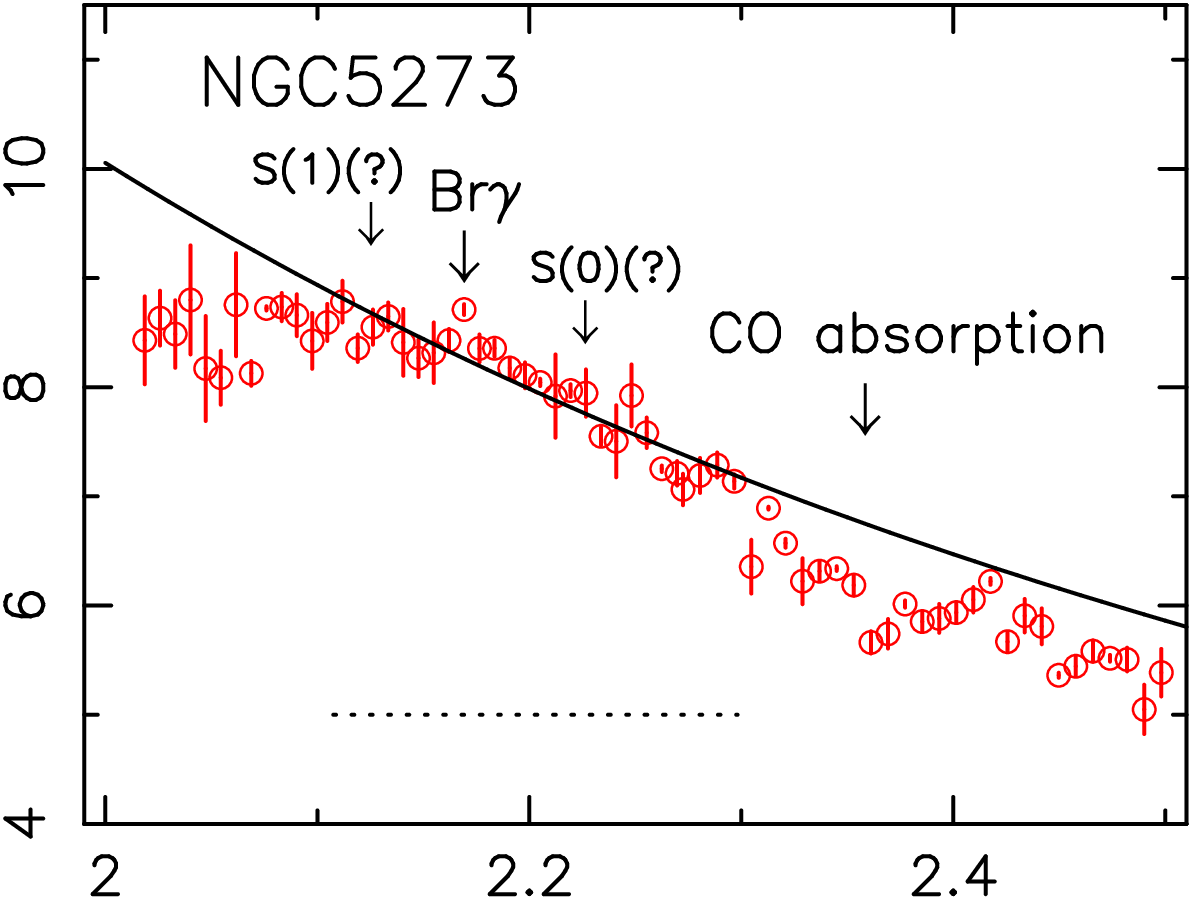}    
\includegraphics[scale=.4, angle=-90]{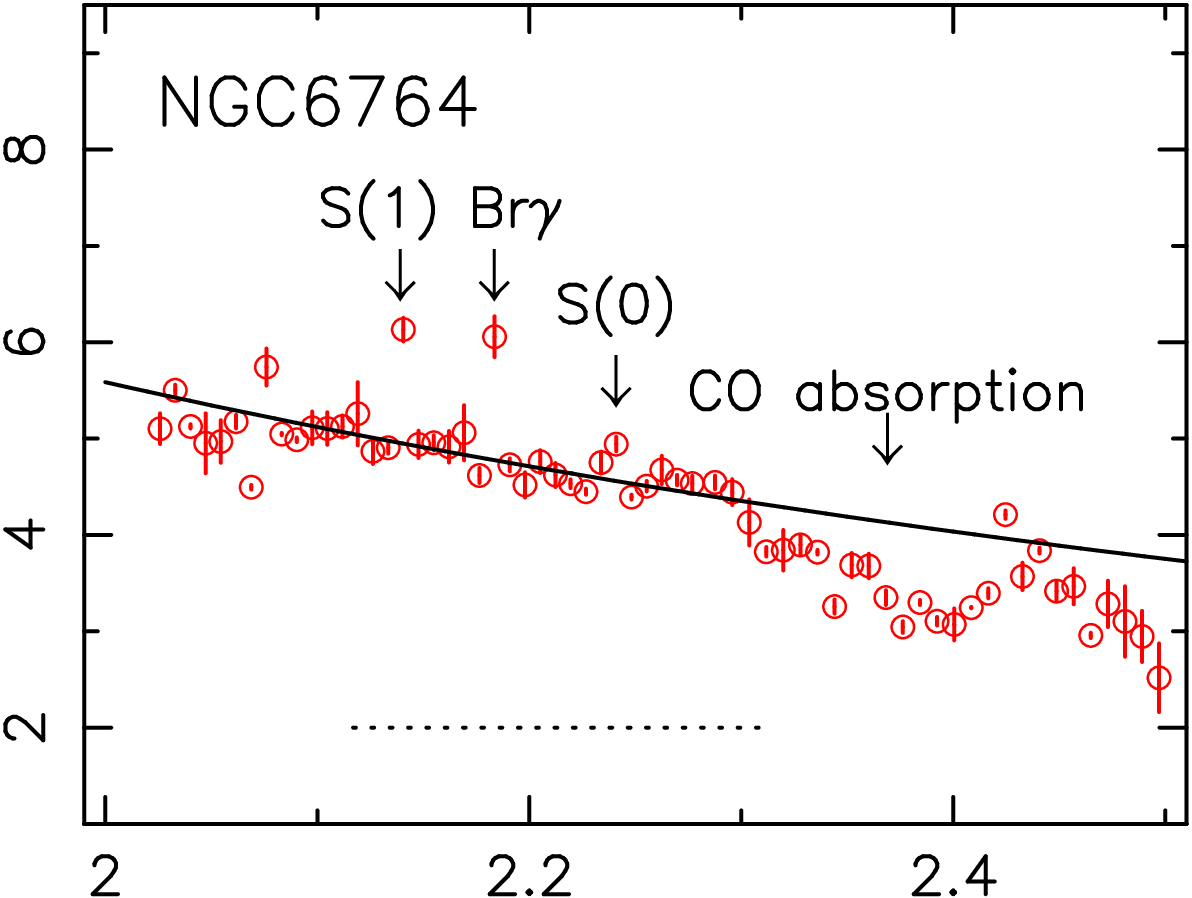}    
\includegraphics[scale=.4, angle=-90]{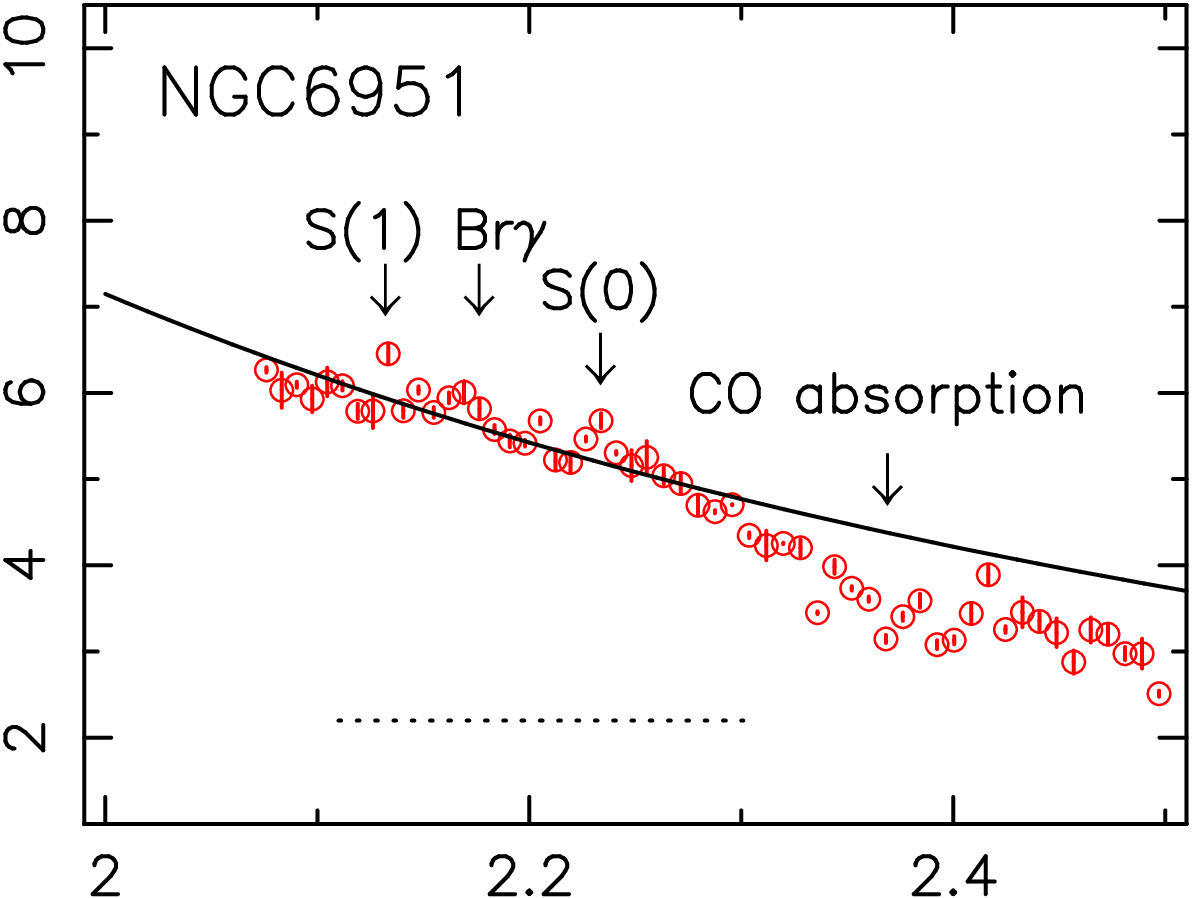}    
\includegraphics[scale=.4, angle=-90]{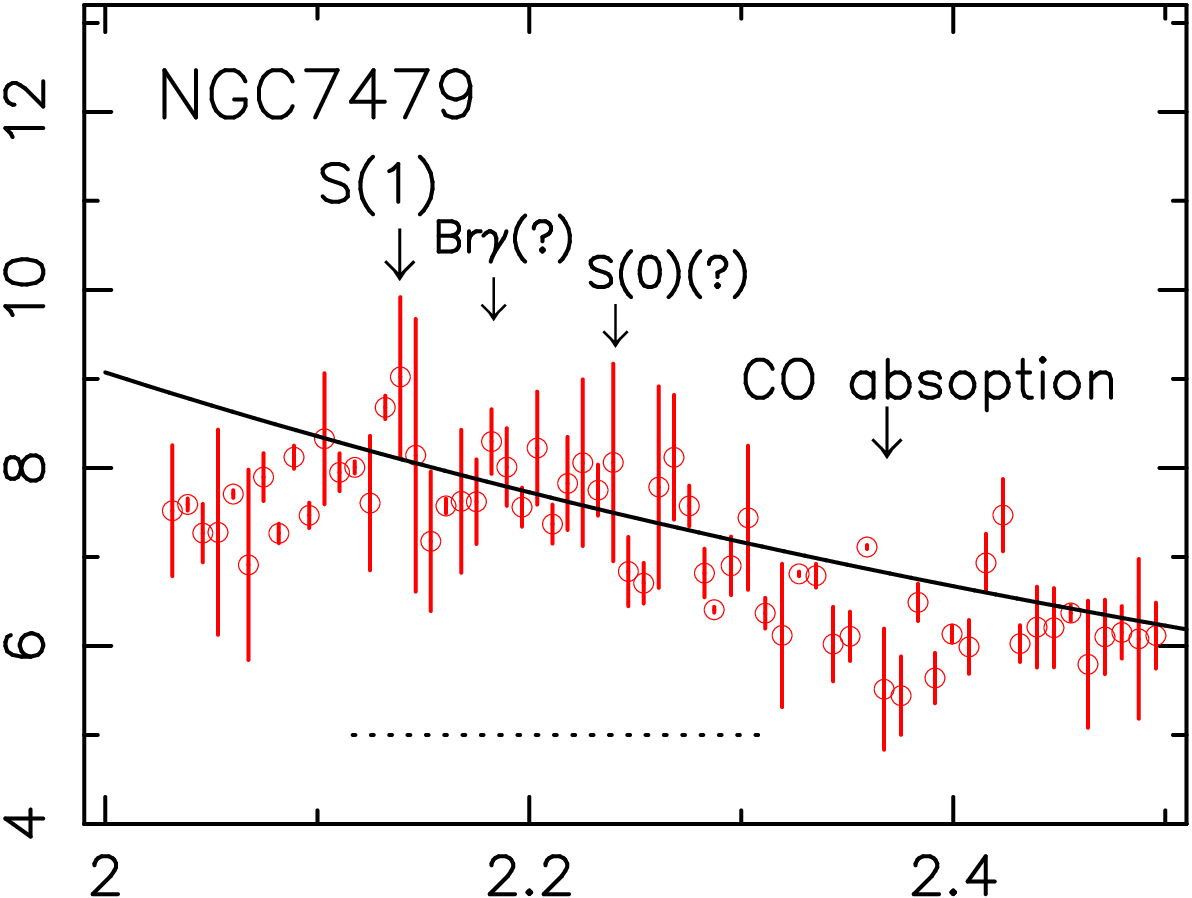}    
\includegraphics[scale=.4, angle=-90]{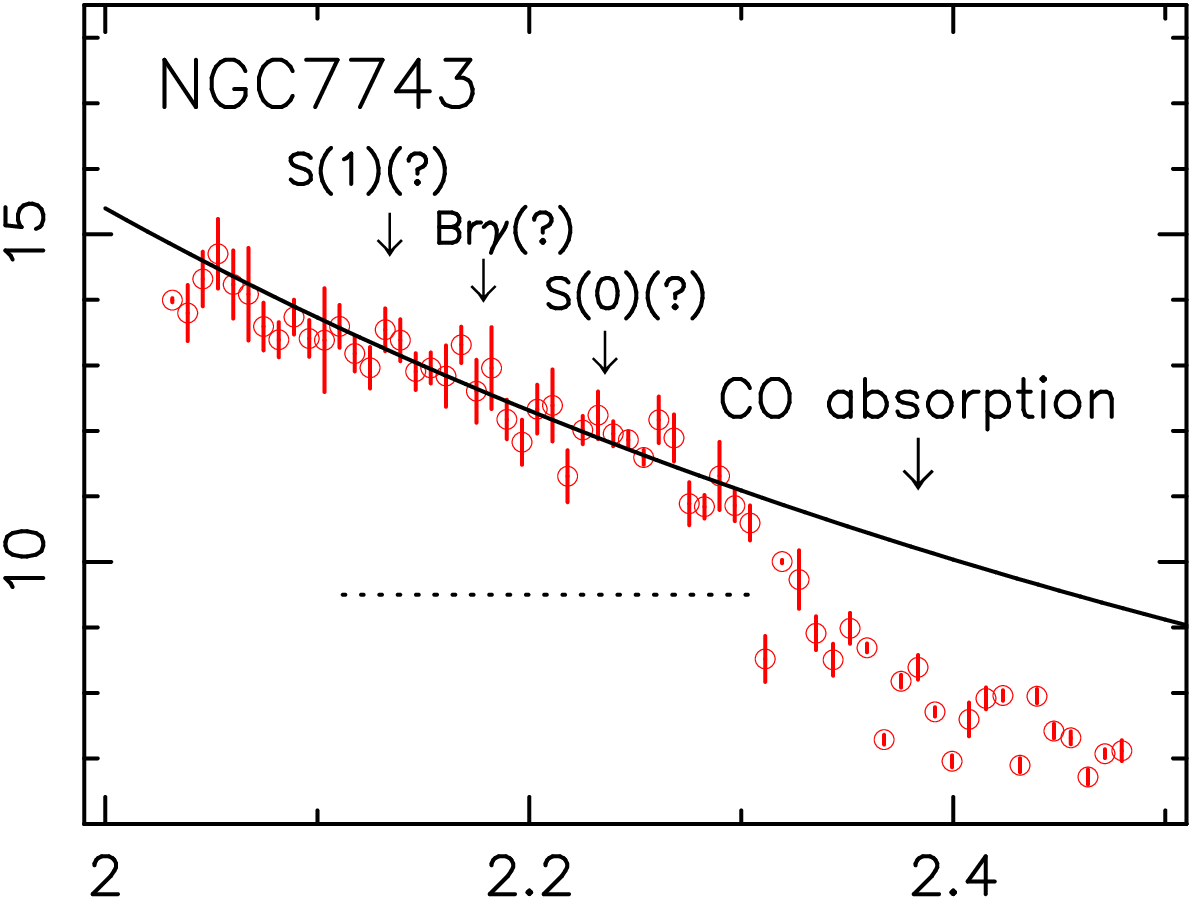}    
\includegraphics[scale=.4, angle=-90]{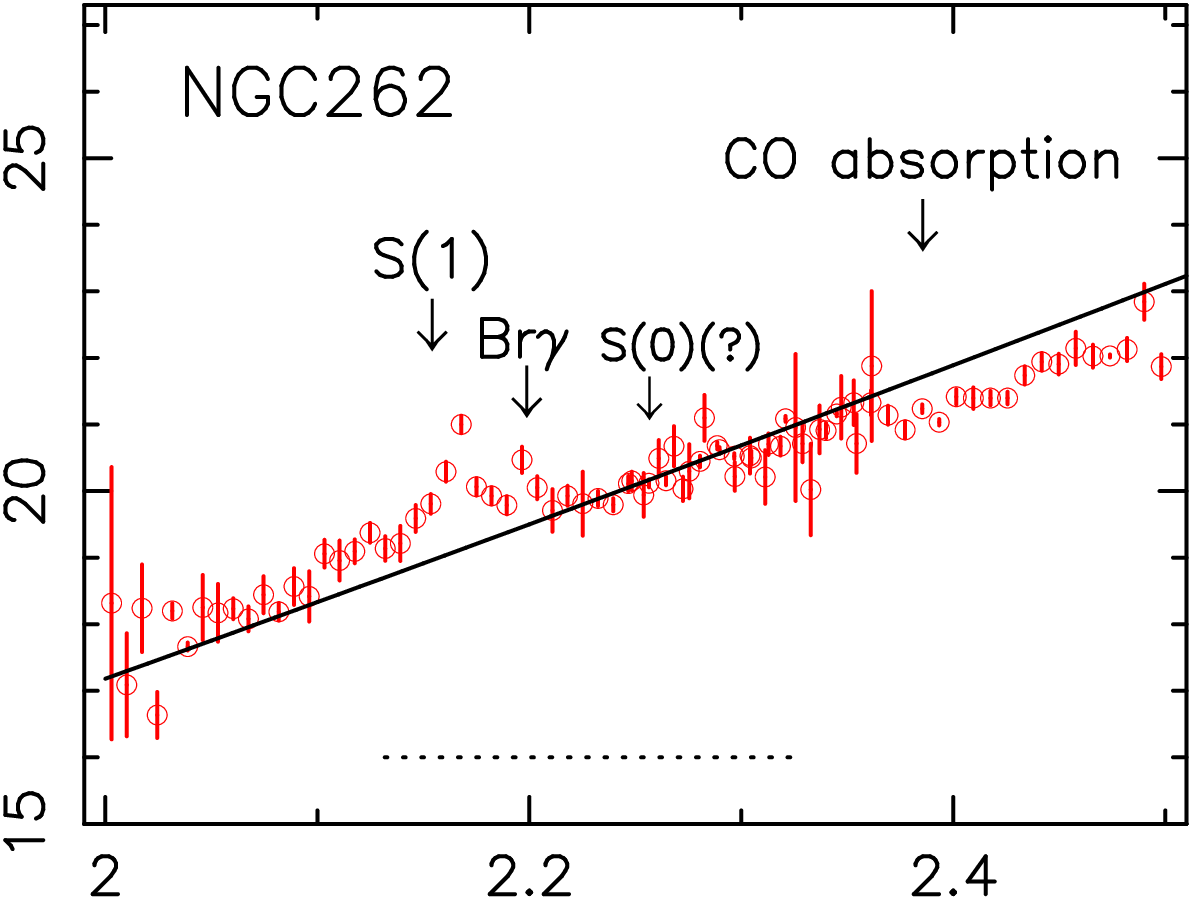}    
\caption{--Continued}
\end{center}
\end{figure*}

\setcounter{figure}{1}
\begin{figure*}
\begin{center}
\includegraphics[scale=.4, angle=-90]{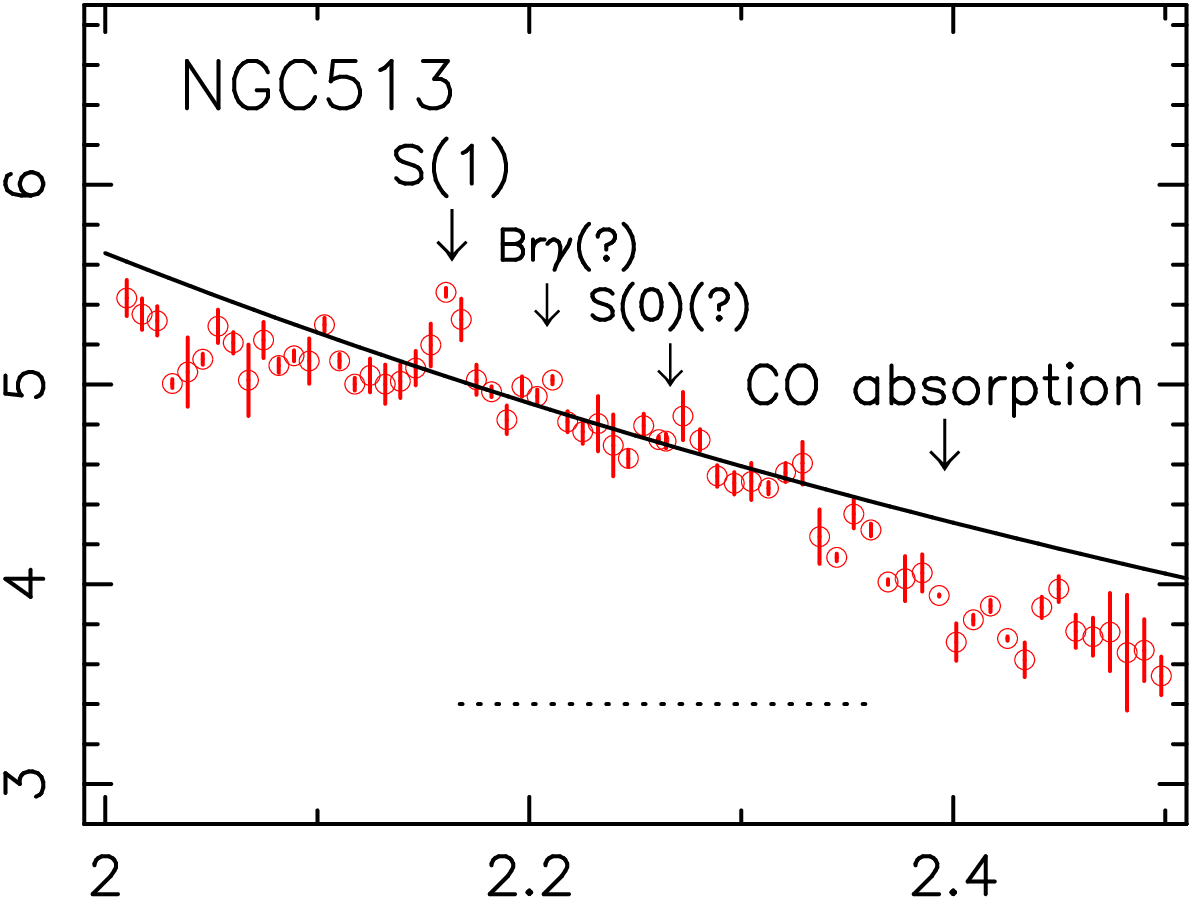}    
\includegraphics[scale=.4, angle=-90]{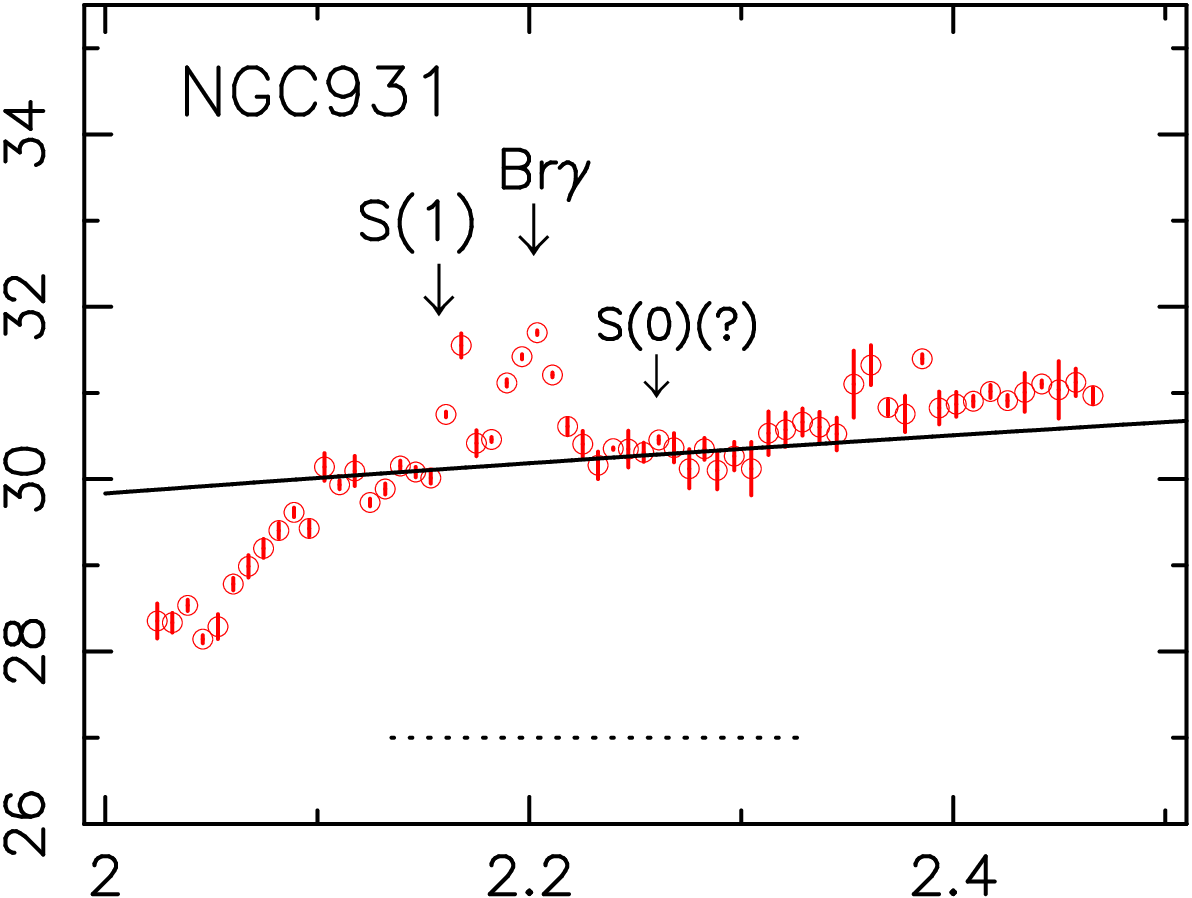}    
\includegraphics[scale=.4, angle=-90]{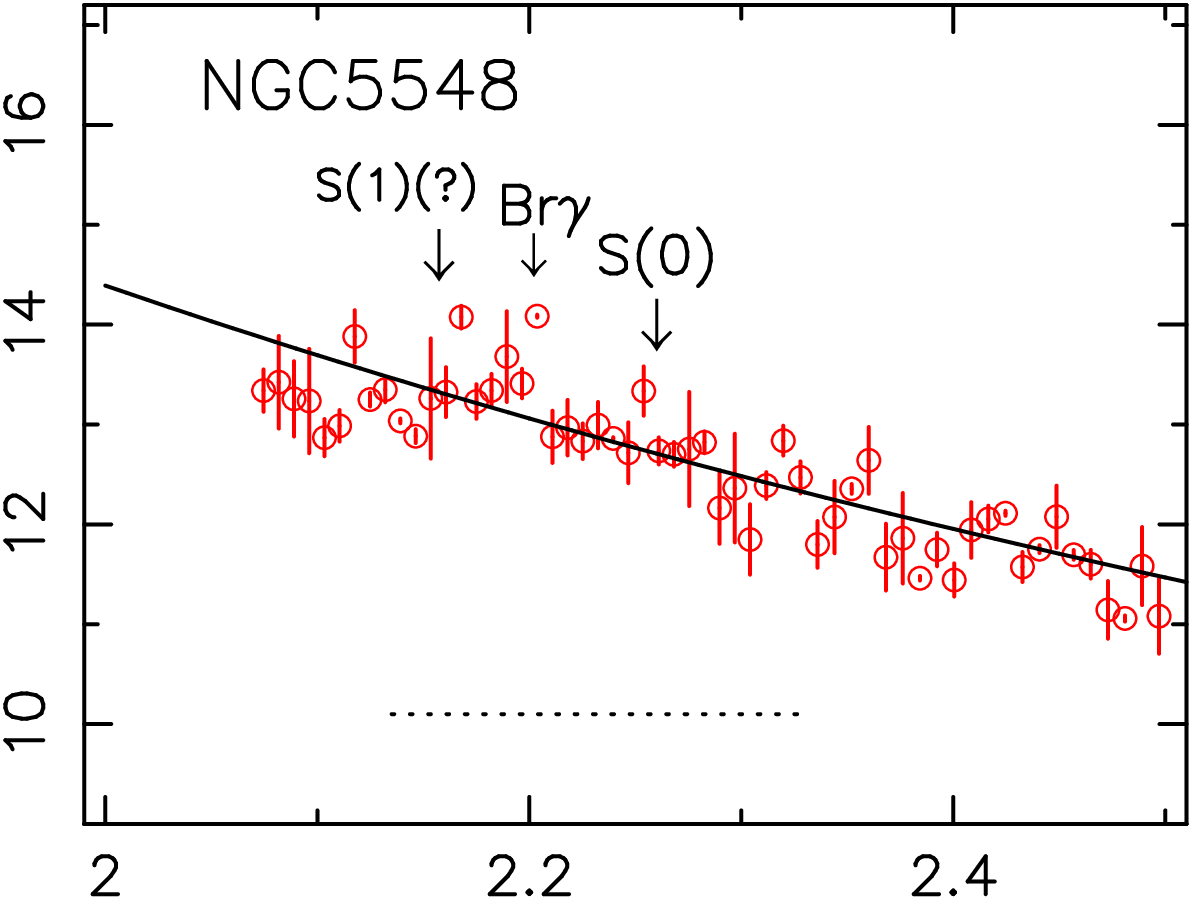}    
\includegraphics[scale=.4, angle=-90]{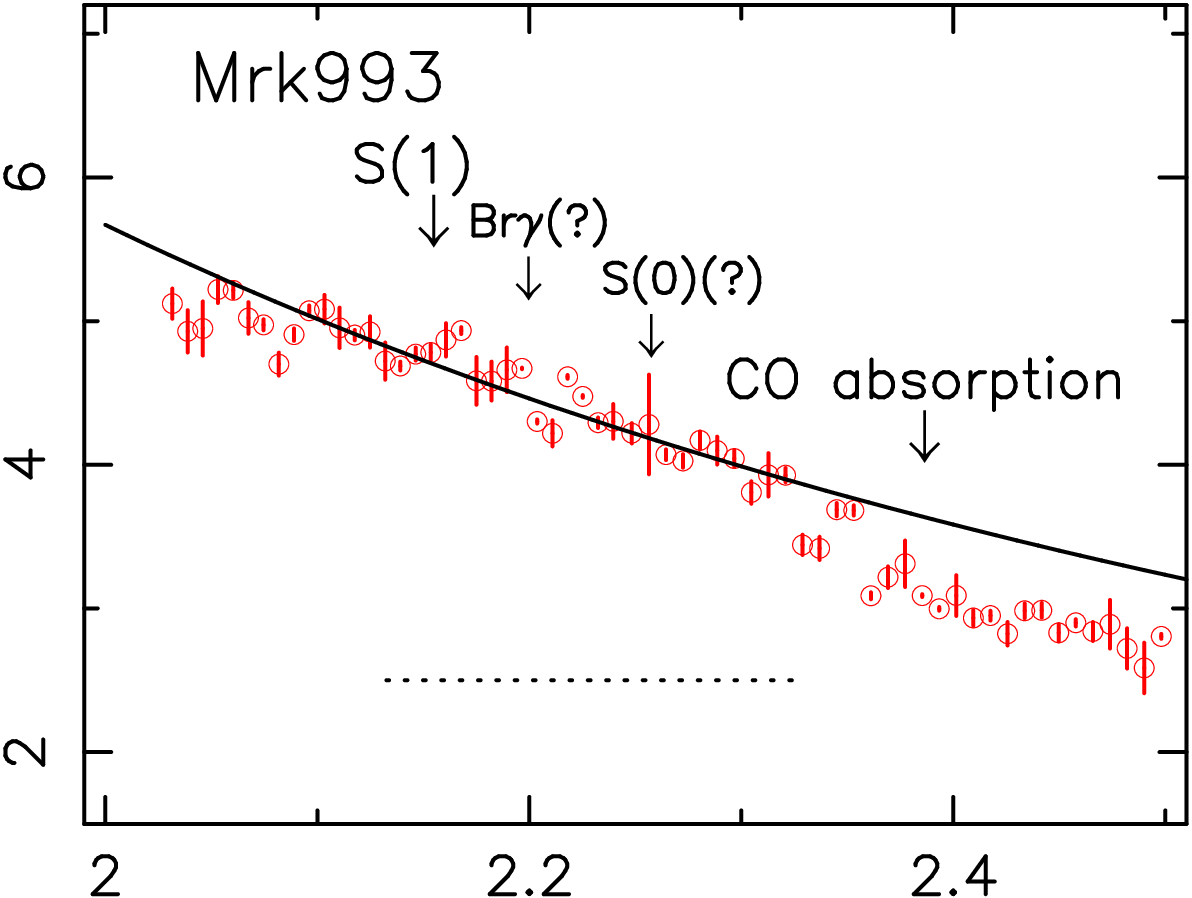}    
\includegraphics[scale=.4, angle=-90]{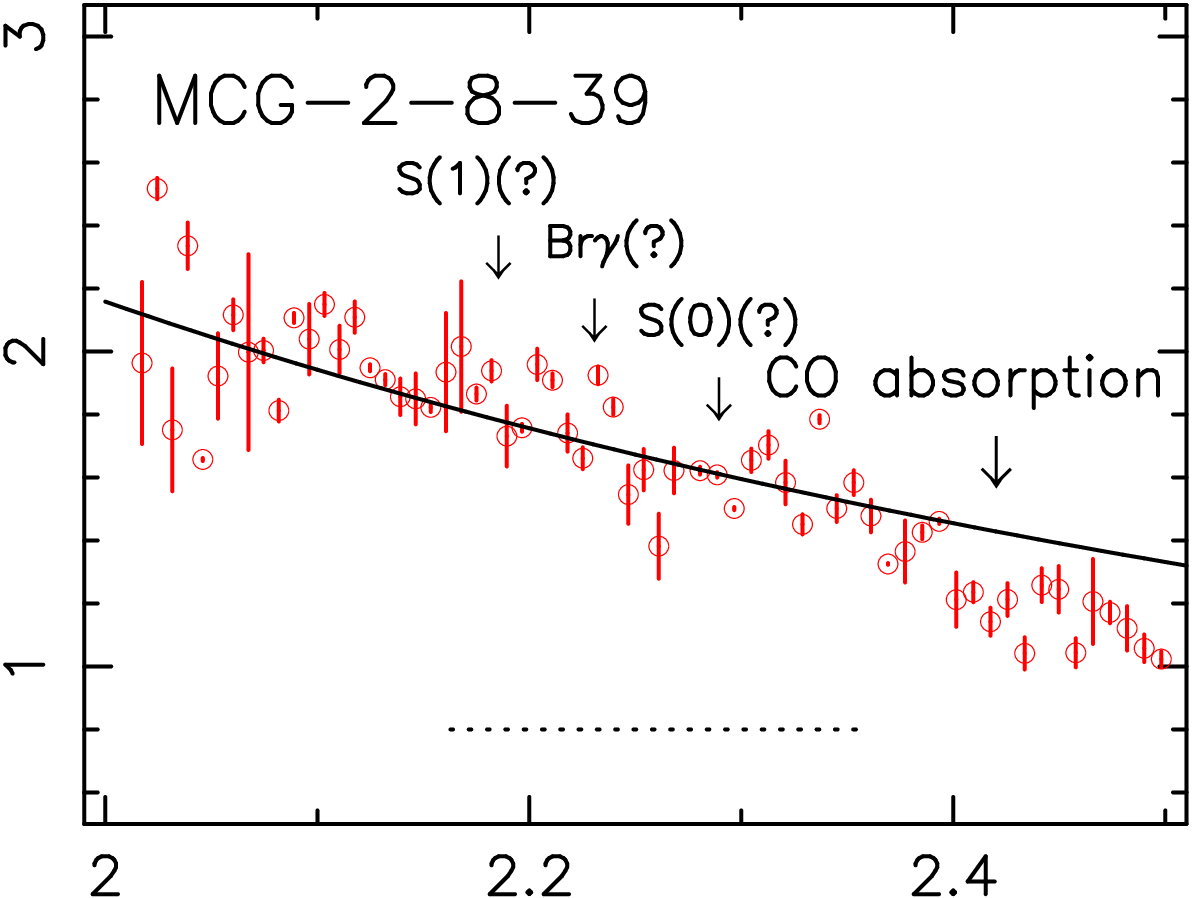}    
\includegraphics[scale=.4, angle=-90]{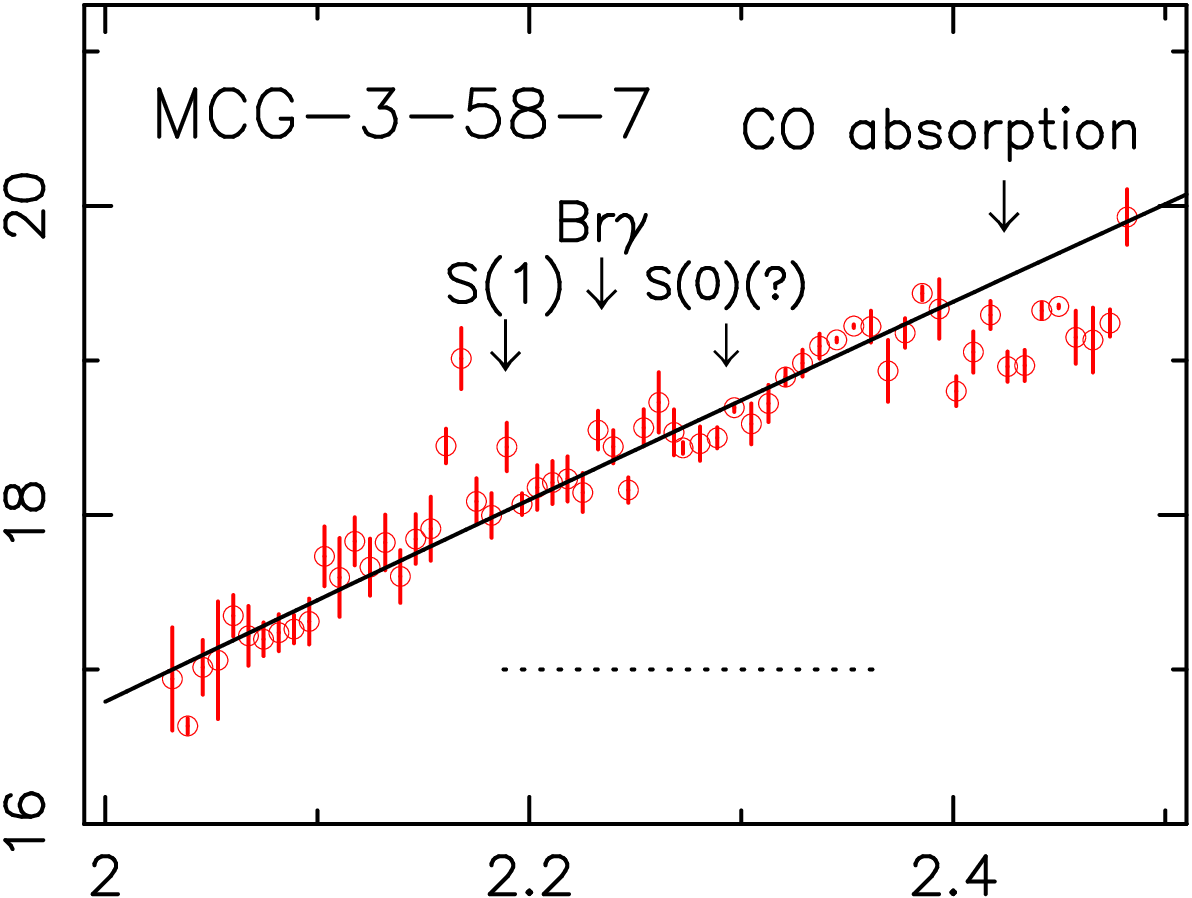}    
\includegraphics[scale=.4, angle=-90]{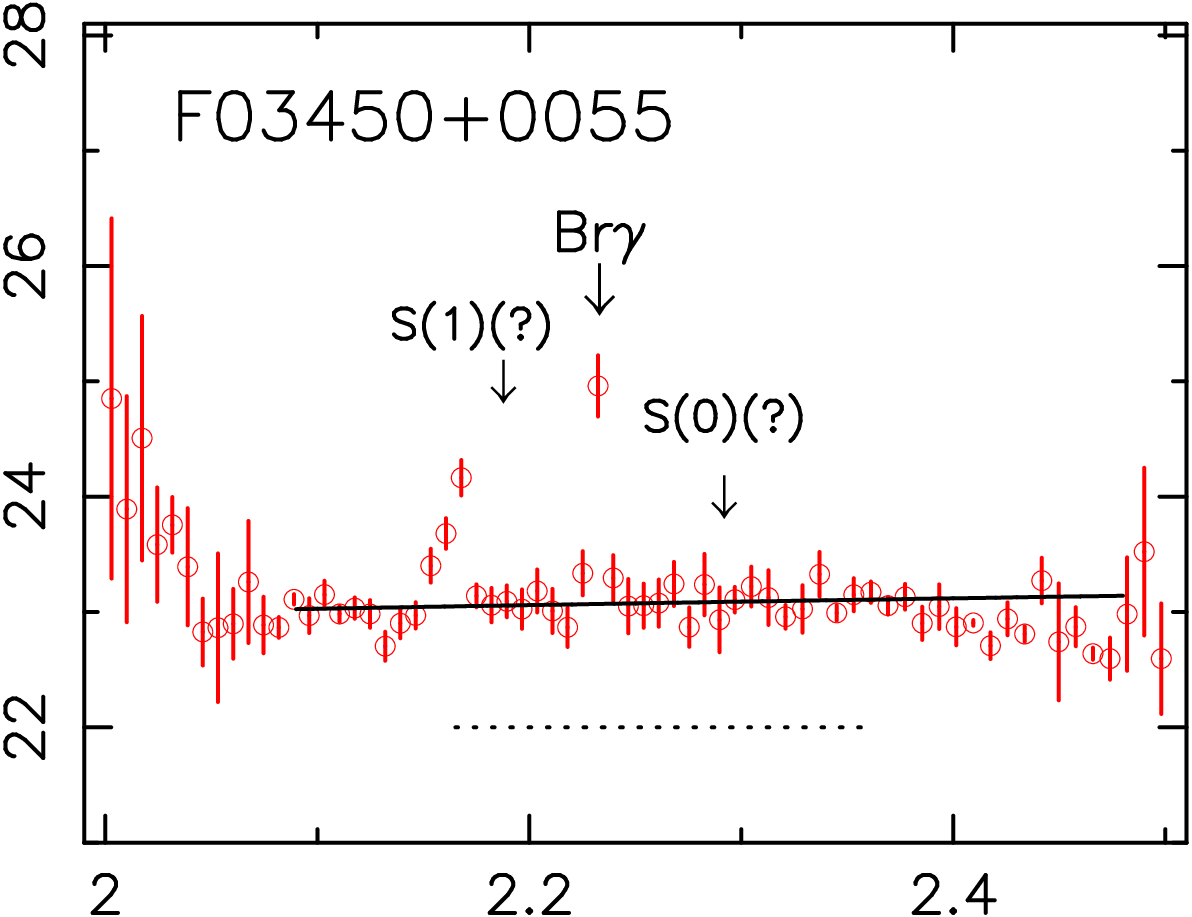}    
\includegraphics[scale=.4, angle=-90]{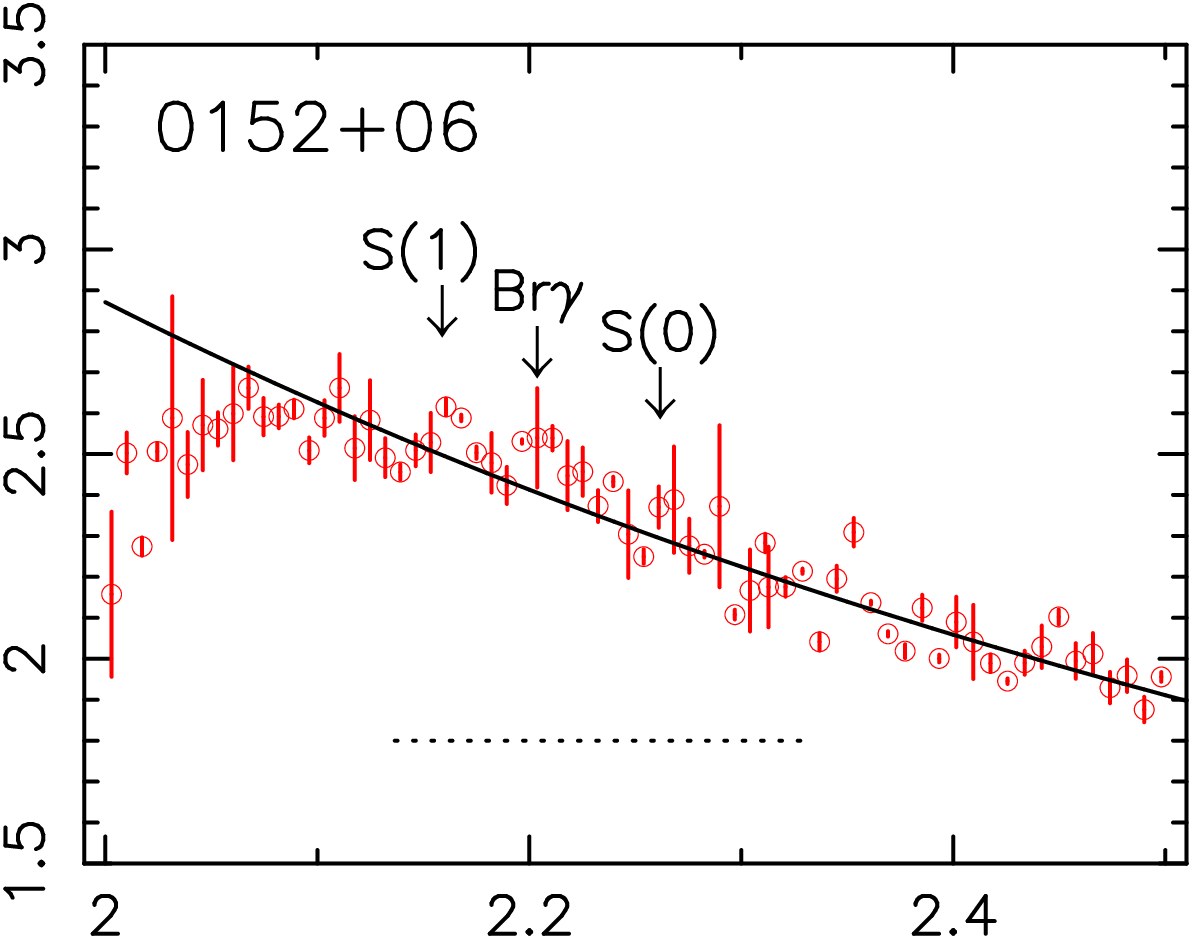}    
\caption{--Continued}
\end{center}
\end{figure*}

\clearpage
\begin{figure}
\begin{center}
\includegraphics[scale=.6, angle=-90]{figure3.ps}
\caption{Histogram of the $CO_{spec}$ index of LIRGs \citep{ridg94}.
The five objects with $CO_{spec}$ smaller than 0.1 (Mrk231, IRAS05189-2524,
 UCG5101, Arp299B1, and Arp299C) showed strong AGN signs in previous
 studies \citep{ID00, Ima01, della02, zezas03, IT03, Ballo04, Braito04, IN06}.
}
\label{fig:COhist_lirgs}
\end{center}
\end{figure}


\begin{figure}
\begin{center}
\includegraphics[scale=.6, angle=-90]{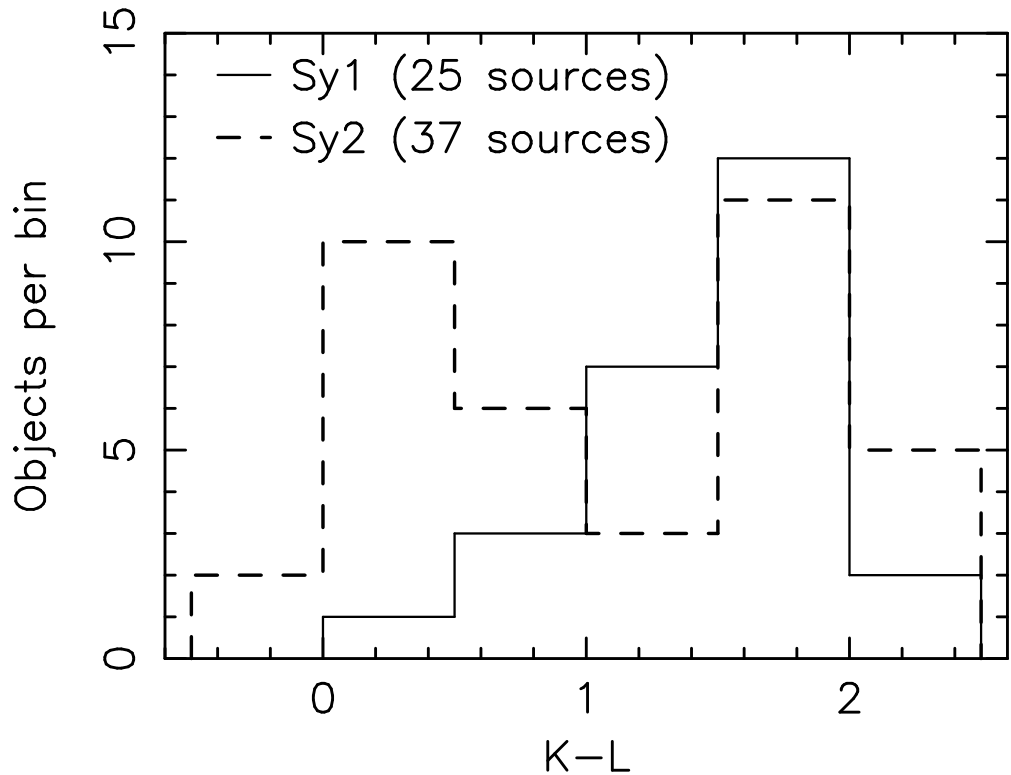}
\caption{Histogram of the $K-L$ color of Seyfert galaxies.
 In addition to the objects studied in this paper, data from \citet{IA04} and
 \citet{IW04} are included.
 The solid line represents Seyfert 1 galaxies, and the dashed line represents
 Seyfert 2 galaxies.
}
\label{fig:K-Lhist}
\end{center}
\end{figure}

\clearpage

\begin{figure}
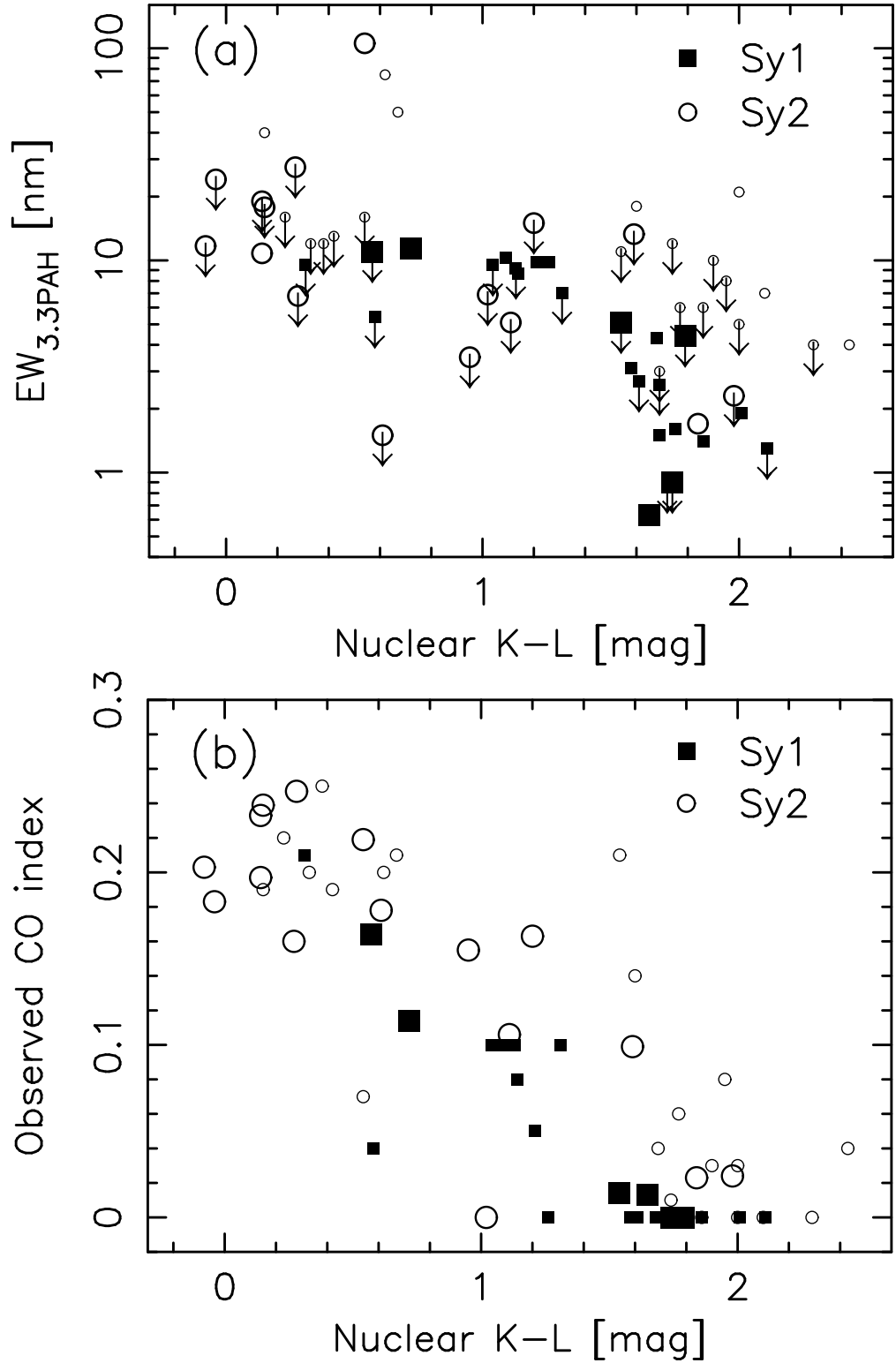

  \begin{center}
\includegraphics[scale=.6, angle=-90]{figure5a.ps}
\includegraphics[scale=.6, angle=-90]{figure5b.ps}
\caption{(a):Comparison of the $K-L$ color (abscissa) and equivalent width
 of the 3.3 $\mu$m PAH emission detected inside our slit spectra (ordinate).
 (b):The ordinate is the observed $CO_{spec}$ index.
 Filled squares represent the data for Seyfert 1 galaxies, and open circles for
 Seyfert 2 galaxies.
 Large symbols represent our sample, and small symbols represent previously
 gathered data \citep{IA04, IW04}.}
\label{fig:KL-PAHCO}
\end{center}
\end{figure}

\clearpage


\begin{figure}
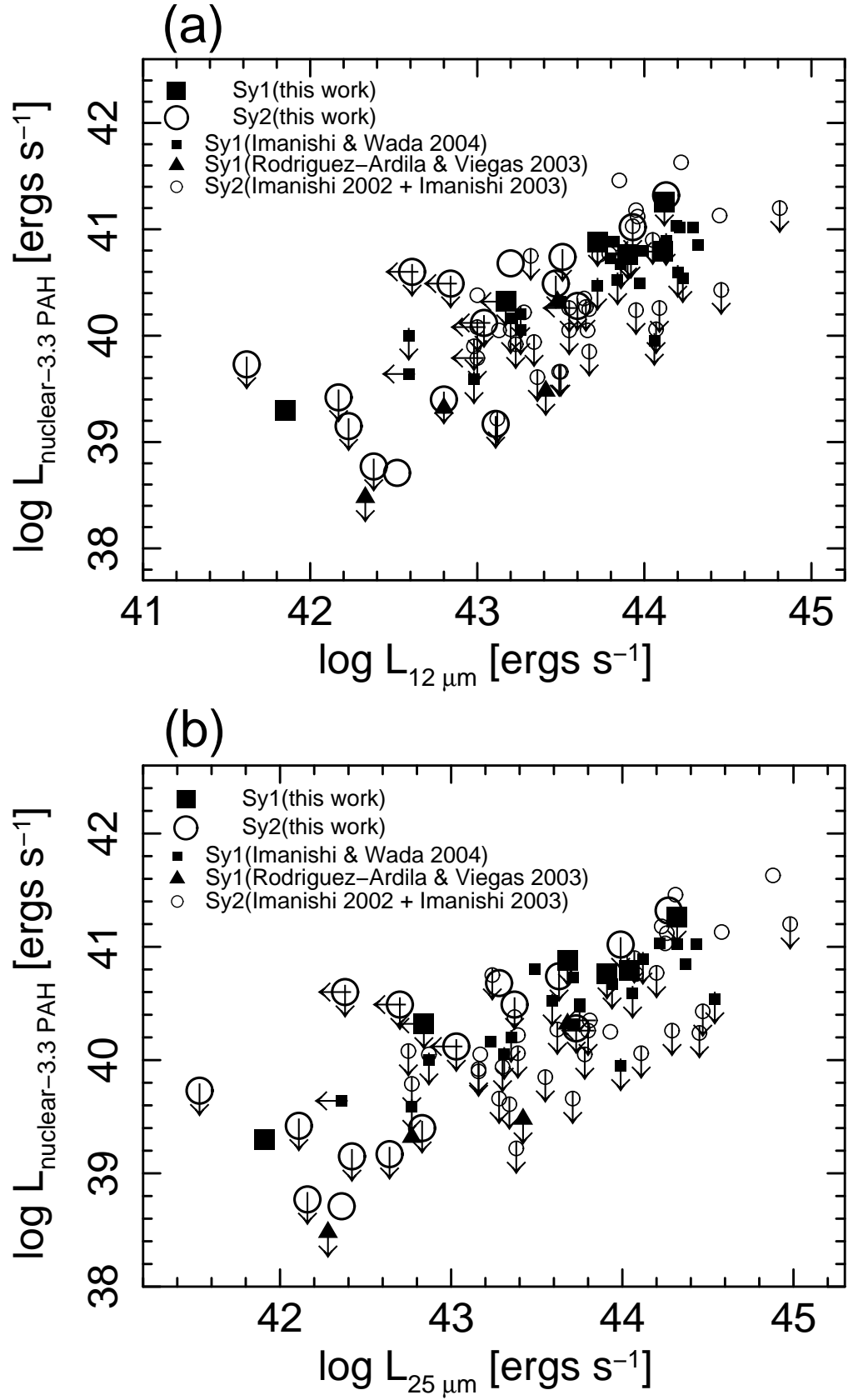

\begin{center}
\includegraphics[scale=.6, angle=-90]{figure6a.ps} \hspace{0.3cm}
\includegraphics[scale=.6, angle=-90]{figure6b.ps} \\
\caption{(a): $IRAS$ 12 $\mu$m luminosity, defined as $\nu$L$_{\nu}$ (abscissa)
 versus the 3.3 $\mu$m PAH emission luminosity detected inside our slit spectra
 (ordinate). Indices are the same as in Fig. \ref{fig:KL-PAHCO}.
 Seyfert 1 galaxies studied by other groups (IH1934-063, Mrk766, NGC3227, and
 NGC4051; \cite{Rod03})
 are plotted as filled triangles.
 (b): Same as (a), but the abscissa is $IRAS$ 25 $\mu$m luminosity.}
\label{fig:AGNpowervsPAHlumin}
\end{center}
\end{figure}

\clearpage

\begin{figure}
\begin{center}
\includegraphics[scale=.6]{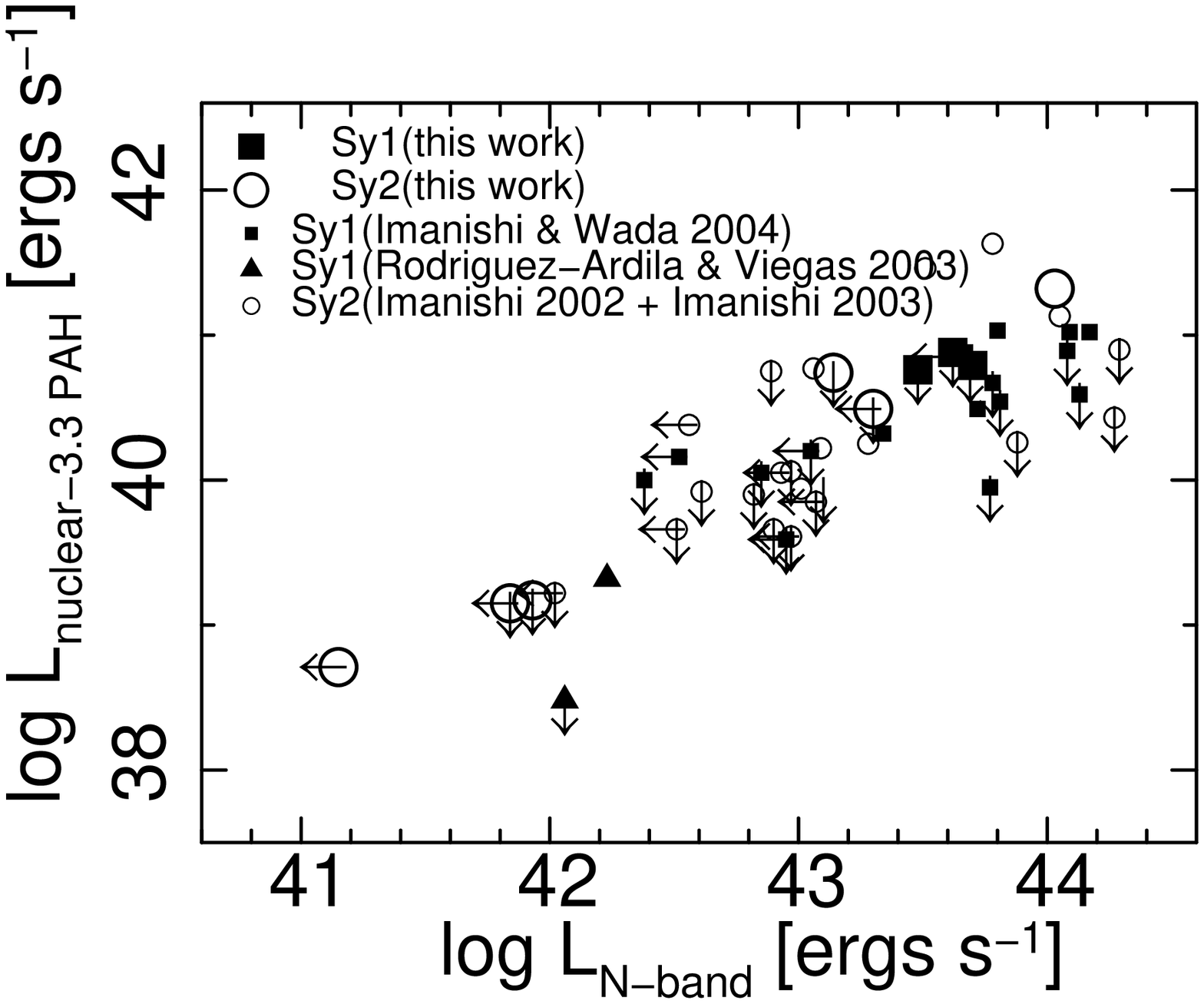} \\
\caption{Same as Fig. \ref{fig:AGNpowervsPAHlumin}, but the abscissa is
  ground-based N-band luminosity measured with small aperture.
 Almost all N-band data in the abscissa are 10.8 $\mu$m luminosity measured
  with a 1$\farcs$5 aperture \citep{gorjian04}.
 \citet{Gal05} observed 10.4 $\mu$m luminosity with 2-arcsec aperture for
  NGC2992 and NGC7469 using ESO 3.6-m telescope/TIMMI2, and \citet{horst09}
  observed 10.49 $\mu$m luminosity with 1.27-arcsec aperture for NGC4579,
  Mrk509 and MCG -3-34-64 using VLT/VISIR. 
 We adopted the new values for the five objects.
 Among the four sources studied by \citet{Rod03}, only two objects 
  (NGC3227 and NGC4051) are plotted because of the availability of measured
  10.8 $\mu$m luminosity \citep{gorjian04}.
 The abscissa is a good indicator of AGN power, and the
  ordinate probes nuclear starburst luminosities.}
\label{fig:N-band_AGNpowervsPAHlumin}
\end{center}
\end{figure}

\begin{figure}
\begin{center}
\includegraphics[scale=.6]{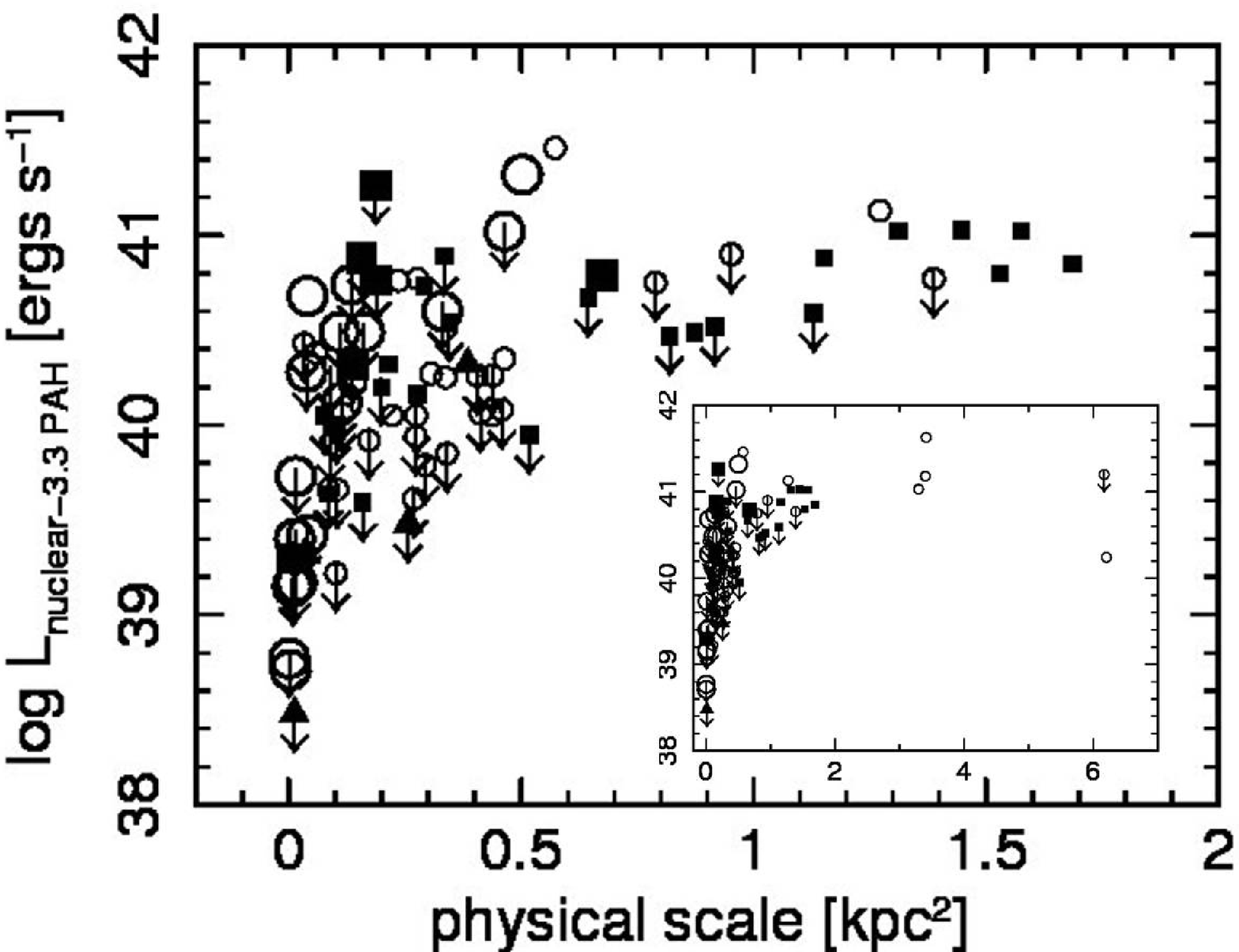} \\
\caption{Relationship between physical scale we probed with slits and the 3.3
 $\mu$m PAH emission luminosity inside the slits (same symbols as in Figure 6).
 Since all physical areas of objects except 5 objects (Mrk34, Mrk78, Mrk273,
 Mrk463, Mrk477) plotted in the figure are smaller than 2 square kpc, larger
 figure focus on the physical area from 0.001 to 1.827 kpc$^2$ to show the
 distribution. A figure including these 5 objects are pasted small on the
 lower right of larger figure.}
\label{fig:surfacevsL3.3}
\end{center}
\end{figure}

\clearpage

\begin{figure}
\begin{center}
\includegraphics[scale=.6, angle=-90]{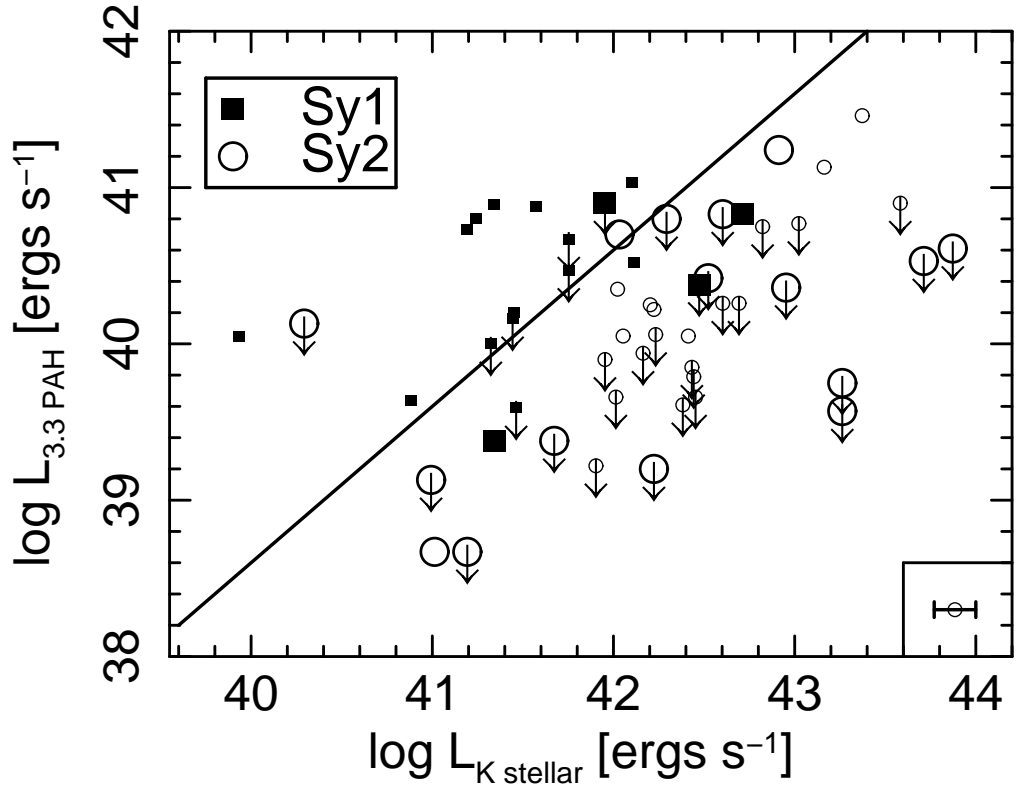} \\
\caption{Comparison between the K-band stellar luminosity (abscissa) and the
 nuclear 3.3 $\mu$m PAH  luminosity detected inside our slit spectra
 (ordinate).
 Symbols are the same as in Figure \ref{fig:KL-PAHCO}.
 The length of the bar at the lower right of the figure represents the range in
 uncertainty of $L_{Kstellar}$ due to $CO_{spec}=0.2-0.3$.
 The solid line shows the predicted ratio of $L_{3.3PAH}/L_{Kstellar}$
 ($\sim 10^{-1.4}$) for starbursts as seen in LIRGs.}
\label{fig:LkstellarvsPAHlumin}
\end{center}
\end{figure}

\end{document}